\newif\ifanswers
\documentclass[fleqn,10pt]{wlscirep}
\usepackage[utf8]{inputenc}
\usepackage[T1]{fontenc}
\usepackage{pdflscape}

\ifanswers
\usepackage{xr}
\makeatletter
\newcommand*{\addFileDependency}[1]{% argument=file name and extension
  \typeout{(#1)}
  \@addtofilelist{#1}
  \IfFileExists{#1}{}{\typeout{No file #1.}}
}
\makeatother
\newcommand*{\myexternaldocument}[1]{%
    \externaldocument{#1}%
    \addFileDependency{#1.tex}%
    \addFileDependency{#1.aux}%
}
\myexternaldocument{supplement}
\fi
\title{SANA: Cross-Species Prediction of Gene Ontology GO Annotations via Topological Network Alignment}

\author[1]{Siyue Wang}
\author[1]{Giles R. S. Atkinson}
\author[1,*]{Wayne B. Hayes}
\affil[1]{Department of Computer Science, University of California, Irvine, CA 92697-3435, USA}

\affil[*]{whayes@uci.edu}

\begin{abstract}

Topological network alignment aims to align two networks node-wise in order to maximize the observed common connection (edge) topology between them. The topological alignment of two Protein-Protein Interaction (PPI) networks should thus expose protein pairs with similar interaction partners allowing, for example, the prediction of common Gene Ontology (GO) terms. Unfortunately, no network alignment algorithm based on topology alone has been able to achieve this aim, though those that include sequence similarity have seen some success. We argue that this failure of topology alone is due to the sparsity and incompleteness of the PPI network data of almost all species, which provides the network topology with a small signal-to-noise ratio that is effectively swamped when sequence information is added to the mix.
Here we show that the weak signal can be detected using multiple stochastic samples of ``good'' topological network alignments, which allows us to observe regions of the two networks that are {\em robustly} aligned across multiple samples.
The resulting {\it Network Alignment Frequency} (NAF) strongly correlates with GO-based Resnik semantic similarity and enables the first successful cross-species predictions of GO terms based on topology-only network alignments. Our best predictions have an AUPR of about 0.4, which is competitive with state-of-the-art algorithms, even when there is no observable sequence similarity and no known homology relationship.
While our results provide only a ``proof of concept'' on existing network data, we hypothesize that predicting GO terms from topology-only network alignments will become increasingly practical as the volume and quality of PPI network data increase.

\end{abstract}

\begin{document}
\flushbottom
\maketitle
\thispagestyle{empty}
%The main text (not including abstract, Methods, References and figure legends) is limited to 5,000 words. Figure legends are limited to 350 words each. As a guide, references should not exceed 70. Footnotes are not used.
%Display items: maximum about 1 per 500 words of text, max 10 for 5,000 word paper.

\section*{Introduction and Motivation}

While much effort is devoted to prediction of protein function by mapping sequence and structure to function, not all proteins have analogs to ones with known function, and the sequence-function relationship is far from 1-to-1: there can be functional similarity in the absence of sequence similarity \cite{furuse1998claudin,fisher2006conservation,schlicker2006new}, and conversely identical sequences can possess multiple, completely different functions\cite{kabsch1984use,morrone2011denatured,schlicker2006new}. Confusing matters further, minor changes in sequence can result in significant changes to function \cite{kimchi2007silent,zhao2014determining}, and similar structure does not always imply similar function \cite{madsen1999psoriasis}. However, one thing is Fcertain: a protein’s function is intimately tied to its set of interaction partners. Since protein-protein interaction (PPI) networks can be measured directly, they potentially provide a road map to function that avoids the complexities of relating sequence and structure to function.

Given that all life on Earth is related, and that proteins derived from genes that have even a remote a common ancestor often share not only sequence but also functional similarity\cite{kachroo2015systematic}, it is reasonable to hypothesize that proteins in different species that share common function might be aligned together by a {\em network} alignment driven to maximize the number of common interactions observed in an alignment. Stated in terms of graph theory, we expect that nodes in two different PPI networks that share common function should also share similar topology among their network interactions. More succinctly, we expect {\em network topology} and {\em protein function} to be related. Importantly, the statement that proteins with similar function are likely to share similar interaction partners does not {\em require} any sequence relationship between the proteins claimed to have similar function; similar network connectivity may be sufficient. This is the basis on which we can hypothesize that topological network alignment may be able to discover inter-species functional orthology even in the absence of sequence similarity.

%Orthologous genes across species arise from the existence of a common ancestor gene. Since proteins arise from genes, the significant genetic similarity across species implies that orthologous proteins will tend to share high similarity in both sequence and function. Since the ``function’’ of a protein is largely defined by its set of interaction partners in its protein-protein interaction (PPI) network, we expect orthologous proteins to have significant similarity in network interaction topology. This common interaction topology suggests that aligning the PPI networks of two species by aligning common {\em edges} should result in the alignment of functionally similar {\em nodes}. This suggests that network topology {\em alone} should allow cross-species prediction of protein function.

Unfortunately, PPI networks for most species are noisy \cite{wodak2013protein}, incomplete\cite{vidal2016much} and biased\cite{Han2005,luck2017proteome}. 
Such data make it difficult to detect common network topology, so that ``failure to find network conservation [between] species [is] likely due to low network coverage, not evolutionary divergence.''\cite{ideker2012differential} %Rather than discouraging research on finding common network topology between species, this sparsity has instead spurred the creation of over forty algorithms attempting to solve the PPI network alignment problem; several reviews and comparisons exist\cite{GRAAL,MamanoHayesSANA,clark2014comparison,optnetalign,crawford2015fair,faisal2015post,guzzi2017survey,balomenos_tracking_2015}. Unfortunately, due to this low and uneven coverage, barely a handful of network alignment algorithms have been able to use topology {\em alone} to infer functional similarity, and then only for a handful\cite{Berg2006} or at most a few dozen\cite{GRAAL,HGRAAL,faisal2014global} proteins. The {\em vast} majority of network alignment algorithms instead add information external to the networks---usually protein-pair sequence similarities---in order to recover functionally similar proteins across species. In fact, several studies have explicitly demonstrated the existence of a so-called ``sequence-topology trade-off'': the functional similarity of network alignments is positively correlated with the weight given to sequence similarity in the objective function, and negatively correlated with the weight given to topological similarity.\cite{Ulign,gligorijevic2015fuse,NATALIE,NATALIE2,kalecky2018primalign,alberich2019alignet}.
For example, the most recent human PPI network from BioGRID \href{https://downloads.thebiogrid.org/BioGRID/Release-Archive/BIOGRID-3.5.184}{(version 3.5.184, released April 2020)} contains 368,005 unique interactions amongst 17,815 unique human proteins; for comparison, the next most complete mammal in the same release is mouse, which contains barely 6\% of the interactions of human, at only 22,903 interactions amongst 7,543 unique mouse proteins. (Note that the numbers given on the BioGRID website for each species include {\em interactions with proteins outside the named species}. These must be removed in order to extract the PPI network of the desired species. We also remove self-interactions, to simplify the graph theory.) Given that the number of edges in the human BioGRID network has consistently grown by about 30\% each year for the past decade and shows no signs of leveling off, both networks must be considered incomplete. %The next most complete PPI networks are S. cerevisiae (116,000 edges among 7,000 nodes) and D. melanogaster (52,400 interactions among 9,000 proteins).

Given the highly disparate levels of PPI network completeness between species, it may come as no surprise that, among the more than fifty attempts in the literature at aligning PPI networks, very few have been able to demonstrate a statistically significant relationship between topological and functional or semantic similarity, with most successes involving local network topology as described by {\it graphlets}  \cite{Przulj2004,milenkovic2008uncovering,GRAAL,MIGRAAL,HGRAAL,faisal2014global,DavisPrzulj2015TopoFunction,gaudelet2018higher,malod2019functional}. Instead, most authors understandably augment the objective function for {\em network} alignments with {\em sequence} similarity of aligned proteins, and such methods met with early success\cite{kelley2003conserved} and continue to meet with success. The problem with this approach is one of signal to noise: any novel functional information hidden in the weak signal that may exist in the common topology between today’s (highly incomplete) networks is likely to be ``drowned out’’ by the much stronger---and already well-understood---signal that exists between proteins of similar sequence. Thus, {\em network} alignments driven by an objective function that includes {\em sequence} similarity may lose the opportunity to learn from any weak signal that may exist in the topology of PPI networks but is obscured by little or no sequence similarity.

What has been lacking in topology-driven network alignments to date is a way to cut through the noise and incompleteness of existing PPI network data to find the functional information hidden in the noisy and incomplete topological data. Our solution is to ``fight fire with fire’’, and utilize intentionally generated randomness to separate signal from noise. Given two networks whose topological similarity we wish to explore, we randomly walk through the alignment search space, eventually converging on a network alignment that exposes a near-optimal amount of topological similarity. Since each random walk takes a different path towards optimality, nodes that share the greatest amount of topological similarity have the greatest chance of becoming aligned across independent paths taken towards a near-optimal solution. Our random walk through search space is generated using simulated annealing, which has a rich history of success in optimizing NP-complete problems \cite{mitra1985convergence,romeo1986efficient,SA1,szu1987fast,sekihara1992details,szykman1997improving,Strens2003,suman2006survey,mcmullen1999determination,meise1998convergence,AguiareOliveiraJunior2012,dowsland2012simulated,larsen2016simulated}. Its {\em randomness} is key: each run of our {\em Simulated Annealing Network Aligner}, or SANA \cite{MamanoHayesSANA,hayes2020introductory}, follows a different, randomized path towards an alignment that uncovers close to the maximum amount of common topology that can be discovered between two networks\cite{WeBeat}. Since each path to a near-optimal alignment is different, each run of SANA produces a {\em different alignment}—but all alignments have nearly the same, close-to-optimal score. SANA effectively produces a {\em random sample} from the frontier of near-optimal alignments. If there is any weak signal of true common topology between a pair of PPI networks, we would expect such common topology to re-appear across these independently generated, near-optimal alignments with a frequency above random. In other words, the alignment of truly similar regions is {\em repeatable}. For example, if SANA independently generates 100 alignments, the better-than-random re-alignment of regions with better-than-random topological similarity manifests as a better-than-random chance that individual {\em pairs of proteins} embedded in these regions will appear at frequencies that are higher than random chance would allow. Those pairs of proteins that appear most frequently will tend to lie in regions with the greatest amount of topological similarity, and consequently we would expect such aligned pairs of proteins to have the highest functional similarity among our aligned protein pairs.

We dub the result {\it Network Alignment Frequency}, or NAF. The NAF of a pair of proteins $(p,q)$ from different species measures the propensity that they will align repeatedly across multiple independently generated near-optimal alignments. We find that NAF strongly correlates with Resnik’s Semantic Similarity (cf. Figure \ref{fig:Resnik-vs-NAF+Pearsons}).

\subsection*{Contribution}
In this paper, our network alignments are driven by network topology alone: the only input is two lists of protein-protein interactions (PPIs)---one PPI network for each species. We demonstrate that SANA's {\it Network Alignment Frequency} (NAF) %not only correlates with Resnik similarity, but is able to {\em predict} novel GO annotations of proteins.
%Since simulated annealing is a {\em random} search algorithm, each run of our algorithm (SANA) on the same two networks produces a different alignment. These two facets---(1) random search that (2) eventually produces alignments with near-optimal scores---means that each time we run SANA, it effectively produces a {\em random sample} on the frontier of alignments with near-optimal scores.  If there are regions in the two networks that contain sufficiently similar topology according to the objective being optimized, then these regions will have a {\em propensity} to repeatedly align together despite the randomness injected by simulated annealing. In other words, some regions of the alignment are so similar that their alignment together is {\em repeatable}---meaning robust to random noise.
%On each run, SANA produces a 1-to-1 alignment between protein pairs. Here we show that if we run SANA 100 times on the same two networks optimizing the same objective function to near-optimality, then the number of times (out of 100) that the same pair of proteins is aligned together---called the {\it Network Alignment Frequency} or {\it NAF}---is positively correlated with Gene Ontology (GO)-based Resnik similarity. This correlation is independent of sequence similarity, and is sufficiently strong that we have produced the first-ever {\it bona fide} novel cross-species predictions of the function of individual proteins which have been 
not only correlates with Resnik similarity, but is able to predict novel GO annotations, even in the absence of detectable sequence similarity.
Our results are validated in two ways: with predictions made in 2010 validated {\it en masse} by comparison with GO terms available in 2020 (10 years later); and on a smaller scale, with predictions made using data available in later 2018 manually validated by literature search today. The latter predictions, based on high NAF scores, were made by transferring GO annotations from a mouse protein that was annotated with GO terms, to a human protein that lacked such annotations and had no detectable sequence similarity according to NCBI PSI-BLAST, nor any known homology relationship using the latest available orthology databases (see Methods).

%Cilia-related GO terms form just a tiny subset of all possible GO terms. While Table \ref{tab:cilia} has just 16 predictions, with 100\% precision in the top 10 predictions and 81\% across all 16---our NAF-based scores produce tens of thousands of protein pairs with similar NAF scores to those in Table \ref{tab:cilia} across the largest of recent BioGRID networks.

Finally, we note that it is not merely the increase in data volume over the past decade, but our {\em method} that has enabled these results, since our 2010-based predictions used {\em only} data that was available as of April 2010, and none of the network alignment algorithms published in the intervening decade has successfully leveraged topology alone to predict a significant number of GO annotations with acceptable accuracy. %To prove this point, we have re-performed our experiments using BioGRID networks from 10 years ago---BioGRID 3.0.64 (April 2010)---and made NAF-based functional predictions based only on GO terms available at that time. We eliminate from consideration all pairs of proteins that (1) have detectable sequence similarity according to BLAST or PSI-BLAST with their default parameters, (2) all known orthologs from NCBI Homologene, InParanoid 8\cite{sonnhammer2015inparanoid}, the 2019 release of EggNog \cite{huerta2019eggnog}.

The outline of our paper is as follows: we describe Gene Ontology annotations including which evidence codes we deem as ``involving sequence’’ (cf. Table \ref{tab:BioGRID+NOSEQ}(bottom)), and introduce network alignment (cf. Figure \ref{fig:NetAlign}) and the various measure of topological similarity that we employ. We then define NAF---{\it Network Alignment Frequency}---which is a measure of confidence for the alignment of each protein pair output by our alignment algorithm SANA \cite{MamanoHayesSANA}. Figure \ref{fig:Resnik-vs-NAF+Pearsons} then demonstrates that NAF correlates with Resnik semantic similarity, while the large middle table of Fig \ref{fig:Resnik-vs-NAF+Pearsons} shows that the correlation is especially strong when restricted to proteins that are well-annotated. One of our most important results is demonstrated in Figure \ref{fig:Resnik-Know}: the distribution of Resnik similarity scores of network-aligned protein pairs is {\it\bf independent of whether the pair possess sequence similarity}. In other words, NAF uncovers semantic similarity that is invisible to sequence-based methods. Supplementary Table \ref{tab:CCS-mean-degree} lists the most dense regions of our network alignments, sorted by mean degree, while Tables \ref{tab:HS-precision} and \ref{tab:NAF-precision} demonstrate that prediction precision correlates strongly with NAF in the regions with highest mean degree. Figure \ref{fig:PR-2010-by-Evidence+Pred-vs-aligQual}(bottom) presents AUPR curves for all 2010-based predictions of human GO annotations validated in 2020; Table \ref{tab:Fstar} and Supplementary Table \ref{tab:Category} provide the associated $F^*$ measures. Finally, Tables \ref{tab:cilia} and \ref{tab:Fancd2-TRIM25} detail novel predictions of human GO terms based on information available in 2018 and manually validated by literature search.

\section*{Results}

\subsection*{Global network alignment}
\begin{figure}
  \centering
  \includegraphics[width=0.7\linewidth]{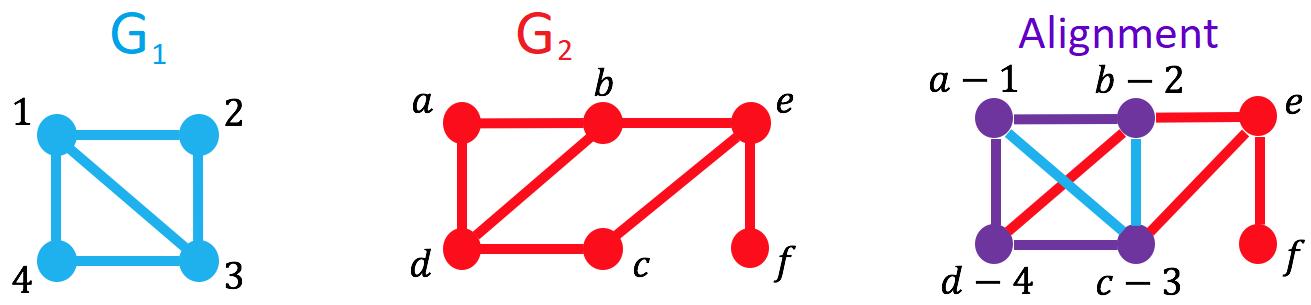}
  \caption{A schematic depiction of a 1-to-1 Pairwise Global Network Alignment (PGNA). The input graphs are $G_1$ (blue, with fewer nodes), and $G_2$ (red). The network alignment can be depicted itself as a network with two types of nodes (purple and red) and three types of edges (purple, blue, and red). Aligned nodes and edges are purple, depicting a mix of red and blue. Unaligned nodes and edges retain the color of the graph they came from. Note that in the aligned network, two common measures of topological network similarity can easily be interpreted visually: EC$=|$purple edges$|$/$|$purple+blue edges$|$, while $S^3=|$purple edges$|/|$edges of all colors between purple nodes$|$.}
  \label{fig:NetAlign}
\end{figure}
We focus on the {\it Pairwise Global Network Alignment} (PGNA) problem: {\it pairwise}, because we align exactly two networks $G_1$ and $G_2$ that have $n_1$ and $n_2$ nodes, and we assume without loss of generality that $n_1\le n_2$; {\it global}, because we aim to find a mapping from {\em every} node in $G_1$ to some node in $G_2$; and {\it network} (as opposed to sequence) alignment because we aim to use {\em only} the network connectivity information (aka global network topology) to guide creation of the network alignment. (See Methods for a formal definition.)
Figure \ref{fig:NetAlign} depicts a schematic diagram of a small PGNA.

\subsection*{Network alignment quality measures}

To demonstrate a relationship between network topology and semantic similarity, we start by elaborating on how each is measured.

\subsubsection*{Semantic similarity between pairs of individually aligned proteins}

Given a pair of proteins $p\in G_1, q\in G_2$, we measure their semantic similarity using the ``maximum'' variant of Resnik Semantic Similarity
\cite{resnik1995using,resnik1999semantic} applied to Gene Ontology (GO) terms\cite{GO} as implemented by the Python package FastSemSim\cite{guzzi2012semantic}. Every GO term that annotates a gene or protein has an associated {\em evidence code} describing the evidence that backs the annotation. Most evidence codes are either based directly on experiment, or inferred through some mechanism. Some mechanisms for inferring GO annotations include sequence analysis. Since one of our main goals is to demonstrate that NAF can highlight Resnik similarity in the absence of sequence similarity, we distinguish between Resnik values that allow all types of evidence (``allGO'') vs. those that disallow any evidence based on sequence (``NOSEQ''). Table \ref{tab:BioGRID+NOSEQ}(bottom) lists the evidence codes we disallow in the latter case.

\subsubsection*{Topological similarity of a global alignment between two networks}
The topological similarity of an alignment between two networks can be scored in many ways, including quantifying edge overlap \cite{GRAAL,MAGNA,WAVE}, node ``importance'' \cite{HubAlign}, graphlet similarity \cite{milenkovic2008uncovering,LGRAAL,GREAT}, graph edit distance \cite{GEDEVO,CytoGEDEVO}, and graph spectra \cite{GHOST}. While some work has been conducted to compare how alignment strategies and objective functions each independently affect the biological relevance of an alignment\cite{WAVE,Milenkovic2013}, our companion paper\cite{WeBeat} performs the first comprehensive, level-playing field study to compare a large number of topological measures for their ability to recover biological information.
Figure \ref{fig:NetAlign} provides a schematic example of two purely edge-based measures: EC\cite{GRAAL} (variously called Edge Correctness, Coverage, Correspondence, or Conservation by various authors), and $S^3$ (the {\it Symmetric Substructure Score}\cite{MAGNA}). %, which has the same numerator as EC, but a denominator that counts all edges between aligned nodes (cf. Figure \ref{fig:NetAlign}), so that it is unchanged by swapping networks. Maximizing the value of either EC or $S^3$ is NP-complete, since achieving a value of 1 corresponds to solving the subgraph isomorphism problem\cite{GareyJohnson}. HubAlign\cite{HubAlign} introduced the {\it Importance} measure, which is essentially a recursive measure of degree: a node has high Importance if it has high degree, and its neighbors also have high degree, etc. The {\it Importance Similarity} between two nodes is the lesser of the two Importance values\cite{HubAlign}.

%Another popular method for measuring local substructure uses graphlets\cite{Przulj2004Graphlets}, which are small {3--5 node} subgraphs of a larger graph. %Below we will use two slightly different variants of graphlets, which we refer to simply as {\it graphlets}\cite{milenkovic2008uncovering}, or GRAAL-type graphlet since they were first used in the global network alignment context by that algorithm\cite{GRAAL}.
%Though there are more methods of measuring network topological similarity, we will focus on EC and {\it Graphlet Degree Vector Similarity} (GDV-graphlet) the former because we recently demonstrated EC's superiority in among edge-based measures, and the latter due to its historical success. (Results from other measures are in the Supplementary.)
%the second we call type {\it L-GRAAL}\cite{LGRAAL}. Both are based on orbit degree vectors as computed by ORCA\cite{ORCA}, and have been shown to correlate with protein function\cite{GRAAL,Przulj2014HiddenLanguage,DavisPrzulj2015TopoFunction}.

\subsection*{Statistical sampling of stochastically-generated network alignments using simulated annealing}
Anybody who shakes a box of loose items in an attempt to make the contents ``settle'' already intuitively understands annealing: vigorous shaking re-initializes the system to a new random state, while more refined shaking hones the solution towards a ``settled'' state which is typically different each time. Crucially, all settled states found by the same ``shaking schedule'' tend to end with roughly equal energy, even though the final positions of the package contents are different each time. In its essence, our {\it Network Alignment Frequency} (NAF) detects pairs of proteins whose alignment is {\em repeatable} across multiple, independent, stochastically generated, near-optimal alignments.

\begin{table}
\centering
\begin{tabular}{|l|l|l|r|r|r|r|}
\hline
     Species  & ShortName & Common name &   nodes  &       edges    &    mean degree  &   max degree \\
\hline
{\it H. sapiens}    &     HS &    human       &    17200 &         282181 &        32.8 &         2385 \\
{\it S. cerevisiae} &     SC &    baker's yeast &   5984 &         104962 &        35.1 &         3603 \\
{\it D. melanogaster}&    DM &    fruit fly   &     8728 &          46364 &        10.6 &          266 \\
{\it A. thaliana}   &     AT &    water cress &     9364 &          34725 &         7.42 &         1341 \\
{\it M. musculus}    &     MM &    mouse       &    6777 &          18108 &         5.34 &         1671 \\
{\it S. pombe}      &     SP &    fission yeast&    2811 &           8931 &         6.36 &          298 \\
{\it C. elegans}    &     CE &    round worm  &     3194 &           5572 &         3.49 &          181 \\
{\it R. norvegicus} &     RN &    rat         &     2391 &           3554 &         2.97 &          808 \\
\hline
\end{tabular}

\vspace{3mm}
\begin{tabular}{|l|l|}
\hline
Code & Description of sequence-based evidence (ie., disallowed in our predictions) \\
\hline
IBA & curated transfer amongst related sequences Based on common Ancestry (derived by sequence comparison) \\
IEA & Electronic Annotation (strong sequence-based evidence not directly traceable to experimental evidence) \\
ISM & Inferred from sequence model \\
ISA & Inferred from sequence alignment\\
ISO & Inferred from Sequence Orthology\\
IGC & Inferred from Genomic Context \\
RCA & Inferred from Reviewed Computational Analysis \\
ISS & Inferred from Sequence or Structural Similarity \\
\hline
\end{tabular}

\caption{{\bf TOP:}
    \href{https://downloads.thebiogrid.org/BioGRID/Release-Archive/BIOGRID-3.4.164}{BioGRID (version 3.4.164, downloaded Sept. 2018)}, sorted by number of edges. The graphs are undirected; duplicate edges, self-loops and all interactions with proteins outside the specified species were removed.
    {\bf BOTTOM:}     Sequence-based GO evidence codes disallowed in ``NOSEQ'' cases: Note that we are rather more Draconian in our interpretation of ``sequence-based'' than is the norm: we disallow any code in which sequence could have had {\em any} influence, including manually curated sequence comparison. This supports our hypothesis that NAF discovers semantic similarity ``in the absence of sequence similarity''.
    }
\label{tab:BioGRID+NOSEQ}
\end{table}

\subsubsection*{Network Alignment Frequency (NAF)}

We say that a pair of proteins that appears frequently in well-scoring topological alignments have a high {\it propensity} to align together.
%We define the {\it propensity for alignment} between two proteins $p\in V_1$ and $q\in V_2$ as $\pi_{uv}=\ldots$.
For %each of the 5 topological objective functions above, and for
each of the 28 pairs of BioGRID networks from Table \ref{tab:BioGRID+NOSEQ}(top), we independently generate 100 alignments, each driven to optimize the same objective function for a one hour run of SANA. (All runs used a 24-core Intel X5680 CPU running at 3.33GHz with 96GB of RAM.) We chose 1 hour because that was the shortest run that produced objective function values within a few percent of the asymptotic value of much longer runs \cite{WeBeat}. Once the 100 runs are finished, we count the frequency (minimum zero, maximum 100) that each pair of aligned nodes appears across the 100 network alignments. The result is NAF: a node-by-node {\em output} measure $\phi_{pq}$, which is the frequency, or {\em propensity}, of alignment between proteins $p\in G_1, q\in G_2$. The higher the frequency, the higher the propensity for alignment between $p$ and $q$. Note that while many measures exist\cite{BergLassig06,milenkovic2008uncovering,GRAAL,LGRAAL,GHOST,DavisPrzulj2015TopoFunction,WAVE,GREAT} for computing topological similarity between two nodes $p\in G_1,q\in G_2$, they are all pre-computed and provided as {\em input} to the alignment process, remaining constant throughout the process. Ours is the first topology-only network alignment method to produce a pair-by-pair score as {\em output}.

The network alignment frequencies generated above by multiple runs of SANA are a generalization of {\it core alignments}, that were introduced by GRAAL\cite{GRAAL} and developed further by H-GRAAL\cite{HGRAAL}. GRAAL used randomness only to break ties while building an alignment greedily using graphlet measures, while H-GRAAL used the Hungarian Algorithm to exhaustively enumerate {\em all} optimal solutions to a given graphlet-based local measure. In both cases, it was observed that there were a subset of aligned protein pairs (the ``core'') that appeared in all optimal alignments, and that the mean semantic similarity measured across this core of always-aligned protein pairs was higher than protein pairs whose alignment partners changed between alignments. Network alignment frequency (NAF) simply generalizes this idea to stochastically generated network alignments that have been optimized to maximize some measure of global topological similarity.

We note that even though SANA produces only 1-to-1 network alignments on each individual run, the merged output of $N$ such alignments effectively produces many-to-many network alignments, with the added value of an output score for each possible pair of nodes. This merging of multiple network alignments also alleviates a potential problem called ``low alignment coverage.'' In particular, it has been noted\cite{Ulign} that 1-to-1 network alignment algorithms do not provide alignment suggestions for all possible nodes in both networks. Their solution was to combine the outputs of several algorithms in order to improve this coverage. However, our network alignment frequency measure makes this unnecessary, since every possible pair of nodes can be assigned an approximate propensity value; pairs that never appear are simply assigned an approximate propensity of zero. %In addition, in the process of {\em creating} any particular one-hour network alignment, SANA typically performs $10^5-10^6$ iterations per second, making it highly likely that the vast majority of pairs are considered multiple times. Thus, any pairs that do not appear in the 100 final network alignments are very likely ``covered'', but appropriately assigned an approximate alignment propensity of zero.

\begin{figure}\small
%     $\;\;\;\;\;\;\;\;\;\;\;\;\;\;\;\;\;\;\;\;\;\;\;\;\;\;\;\;\;\;\;\;\;\;\;\;\;\;\;\;\;\;$EC-seq
%     $\;\;\;\;\;\;\;\;\;\;\;\;\;\;\;\;\;\;\;\;\;\;\;\;\;\;\;\;\;\;\;\;\;\;\;\;\;\;$EC-allGO
%     $\;\;\;\;\;\;\;\;\;\;\;\;\;\;\;\;\;\;\;\;\;\;\;\;\;\;\;\;\;\;$EC-NOSEQ
%     $\;\;\;\;\;\;\;\;\;\;\;\;\;\;\;\;\;\;\;$graphlets-seq
%     $\;\;\;\;\;\;\;\;\;$graphlets-allGO
%     $\;\;\;\;\;\;$graphlets-NOSEQ$\;\;\;\;\;\;\;$    
\centering
{\large $\longleftarrow$ \hspace{1.5cm}  EC-Optimized       \hspace{2.3cm} $\longrightarrow$ $||$
        $\longleftarrow$ \hspace{0.7cm} graphlet-GDV-optimized \hspace{1cm} $\longrightarrow$}\\
\rotatebox{90}{\small $\;\;\;\;\;\;\;$MM-HS Resnik}
\includegraphics[width=0.155\linewidth]{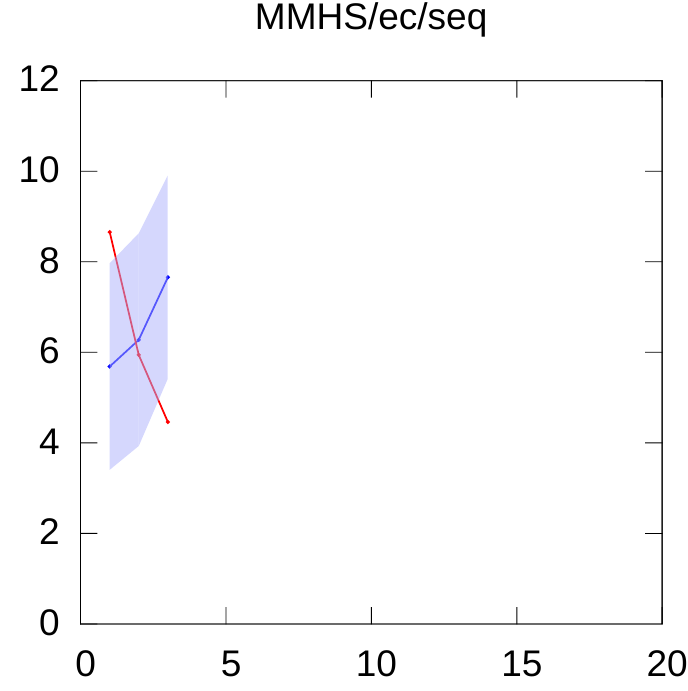}
\includegraphics[width=0.145\linewidth]{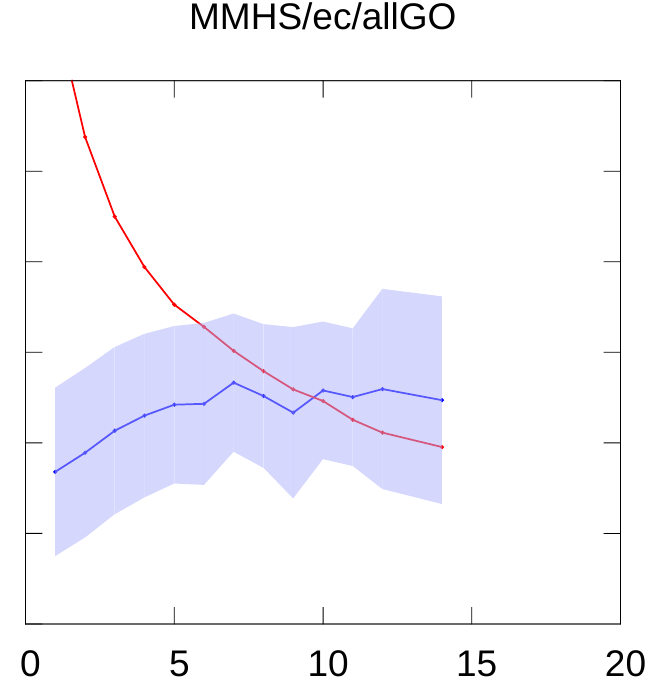}
\includegraphics[width=0.17\linewidth]{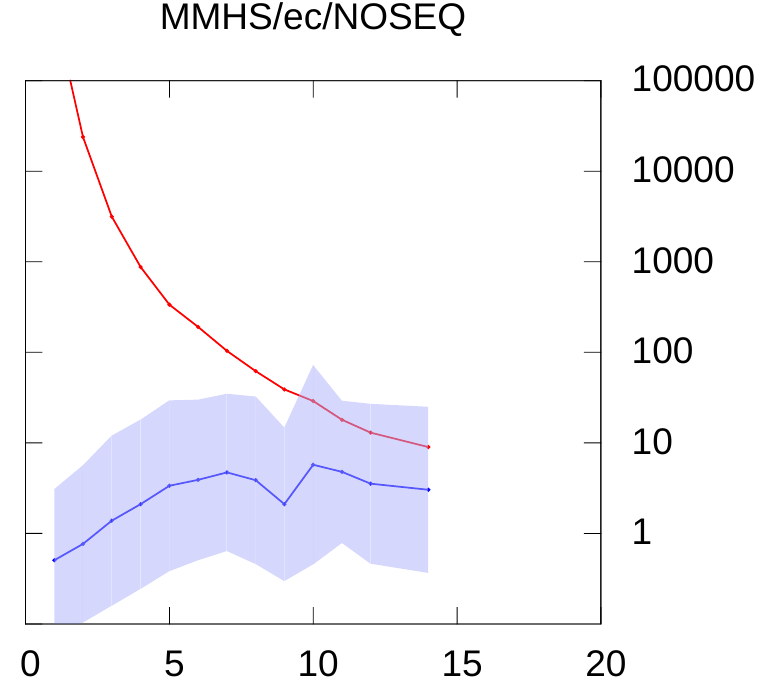}
\includegraphics[width=0.155\linewidth]{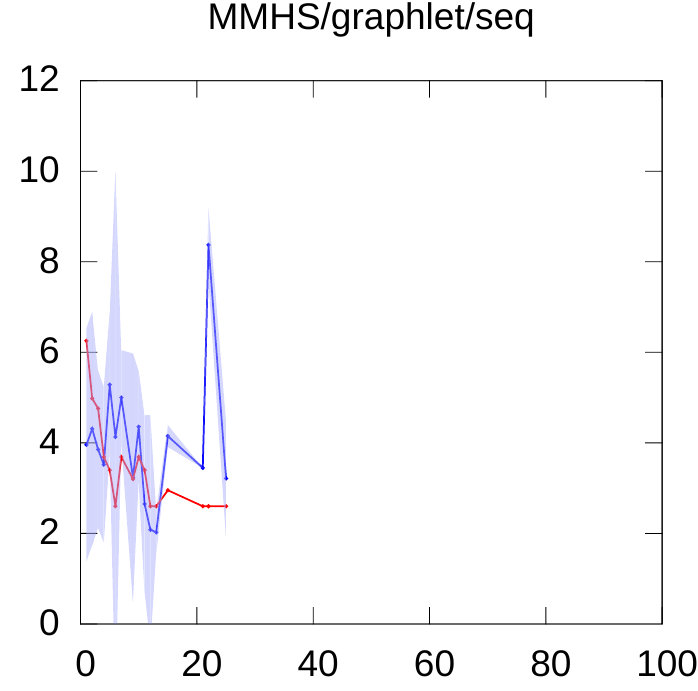}
\includegraphics[width=0.145\linewidth]{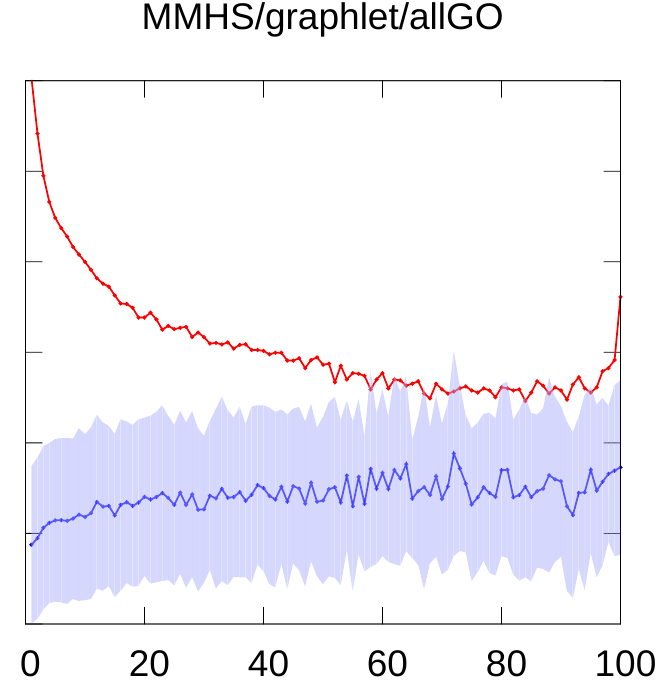}
\includegraphics[width=0.17\linewidth]{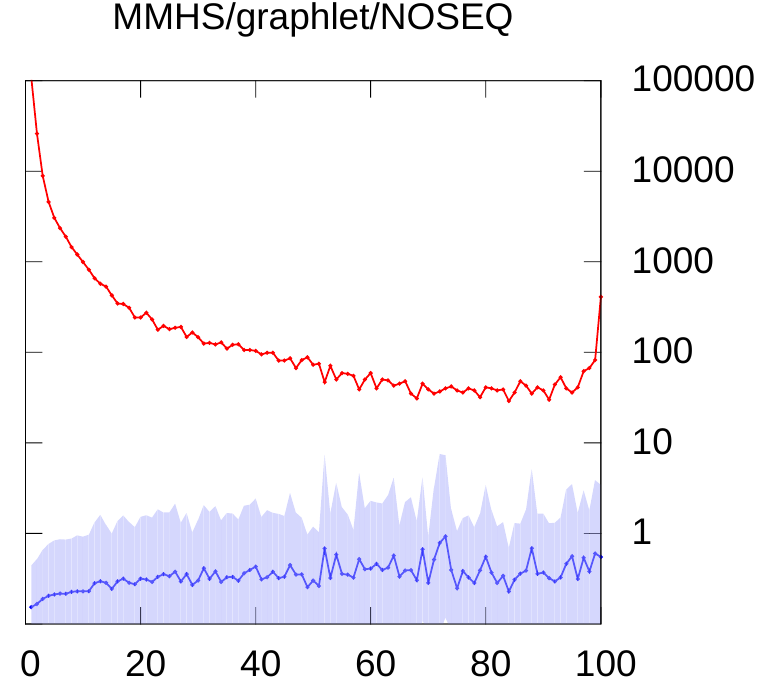}
\\
\rotatebox{90}{\small $\;\;\;\;\;\;\;$SC-HS Resnik}
\includegraphics[width=0.155\linewidth]{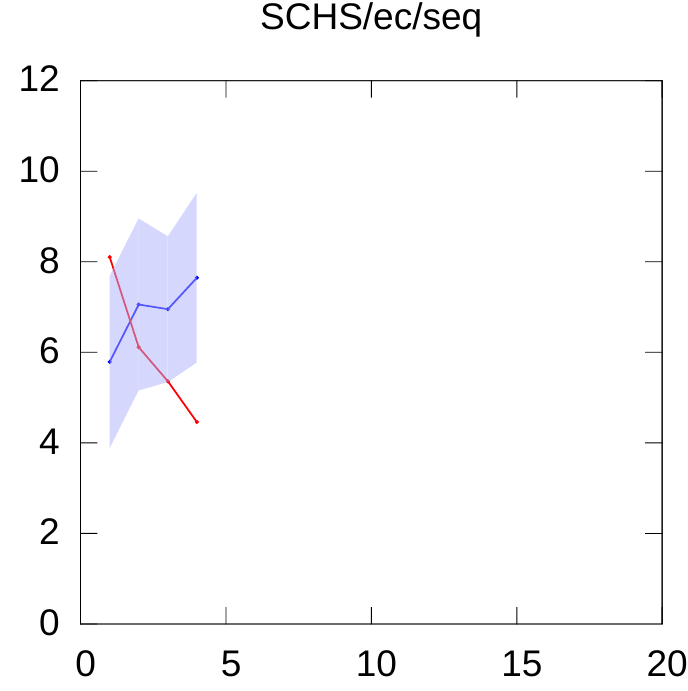}
\includegraphics[width=0.145\linewidth]{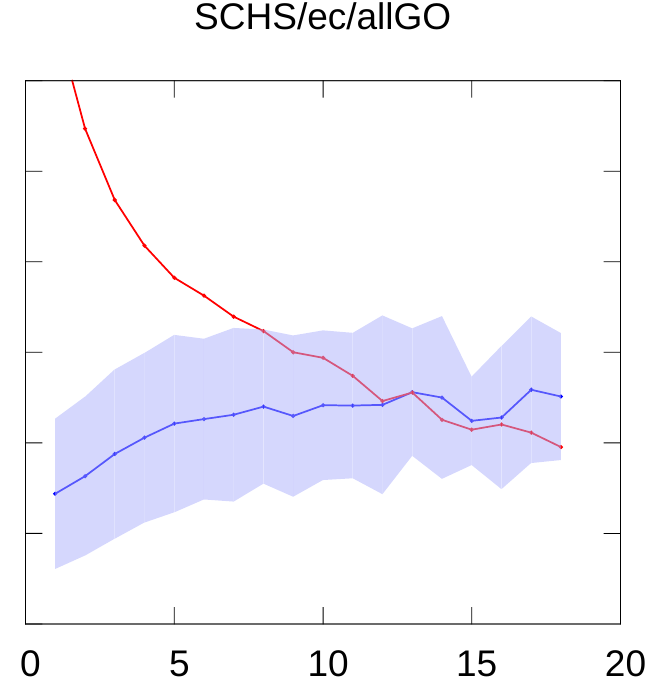}
\includegraphics[width=0.17\linewidth]{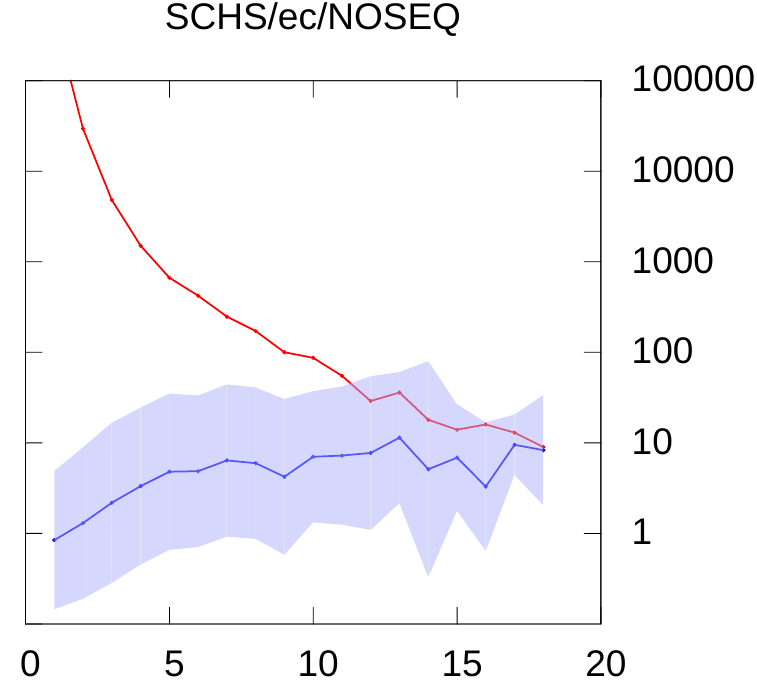}
\includegraphics[width=0.155\linewidth]{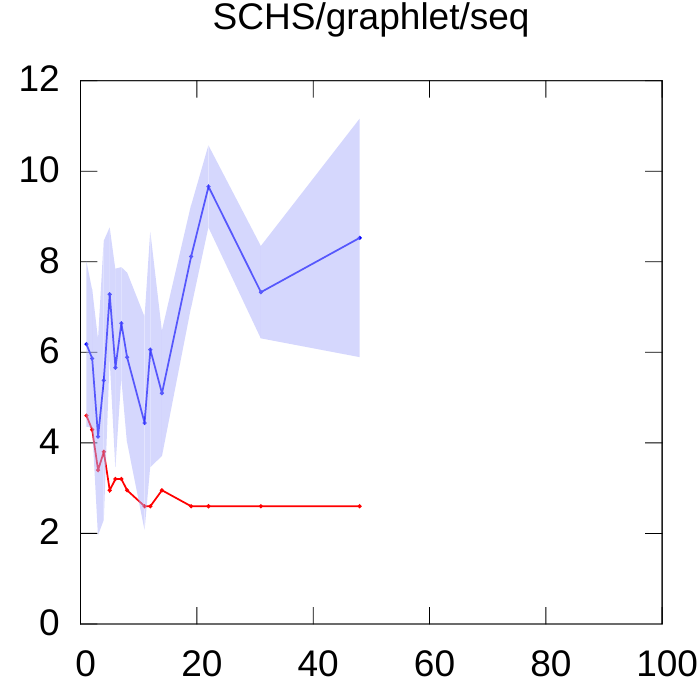}
\includegraphics[width=0.145\linewidth]{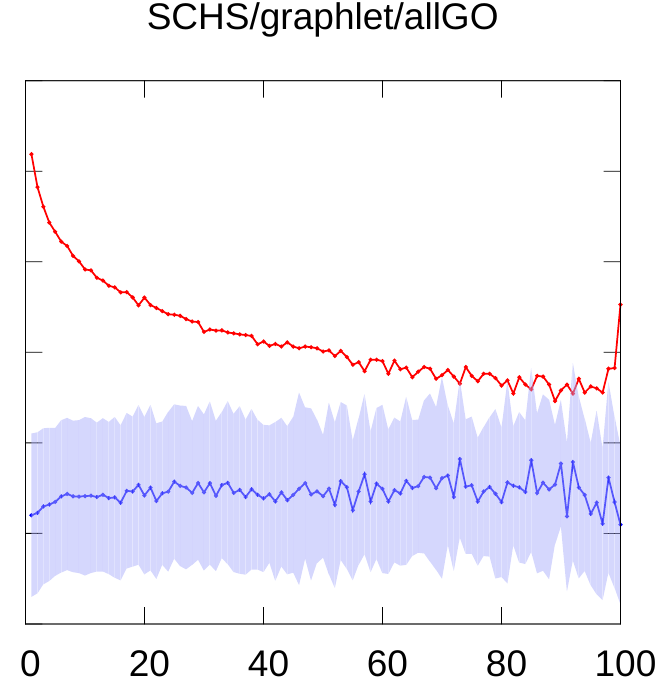}
\includegraphics[width=0.17\linewidth]{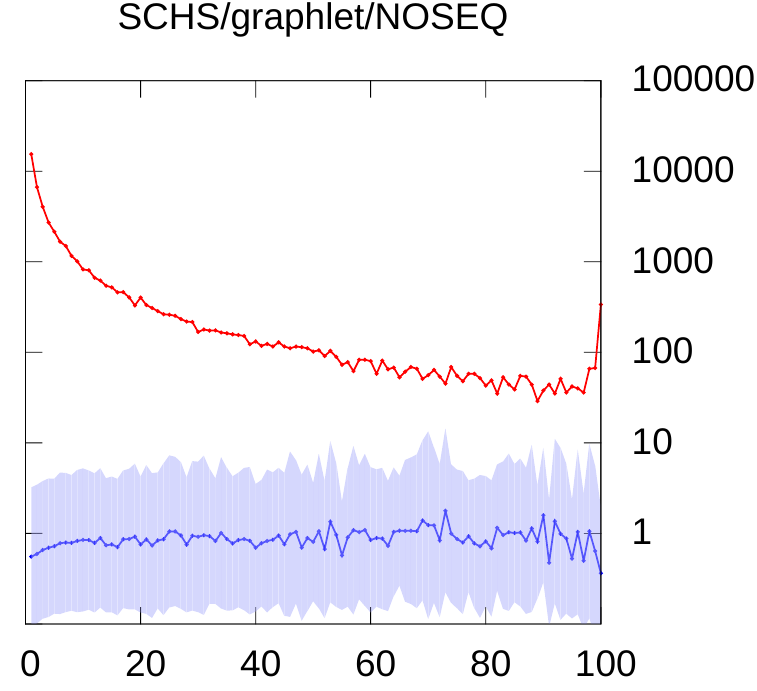}\\
 \hspace{1mm} seqSim/allGO \hspace{5mm} allPairs/allGO \hspace{4mm} allPairs/GO-NOSEQ
 \hspace{4mm} $||$ \hspace{2mm}
 \hspace{1mm} seqSim/allGO \hspace{5mm} allPairs/allGO \hspace{4mm} allPairs/GO-NOSEQ
 
 \vspace{5mm}
    \begin{tabular}{|lll|ccc|ccc|}
    \hline
   &&&  \multicolumn{3}{c|}{All evidence codes}        & \multicolumn{3}{c|}{No sequence-based evidence} \\
    \hline
Species & Obj.&  \#aligs &           $\rho$ & $p$-value       &$\sigma$'s& $\rho$ & $p$-value       &$\sigma$'s \\
    \hline
MM-HS   &EC    & 28779 & 0.06 & 1.66$\times 10^{-28}$  & 11.7 & 0.08   & 2.05$\times 10^{-43}$ & 14.4  \\
      &graphlet& 61469 & 0.13 & 2.57$\times 10^{-248}$ & 34   & 0.10   & 3.66$\times 10^{-145}$ & 26  \\
    \hline
SC-HS   &EC     & 37785 & 0.15 & 2.18$\times 10^{-193}$ & 30   & 0.13 & 4.59$\times 10^{-154}$ & 26.8 \\
    &graphlet  & 35450 & 0.06 & 2.27$\times 10^{-32}$  & 12.5 & 0.03 & 8.58$\times 10^{-09}$ & 6.87 \\
    \hline
    \end{tabular}

    \vspace{5mm}
    \centering
    \begin{tabular}{|r|rrlr|ccr|}
    \hline
    row & $N$ & $\rho(S,M)$ & $p$-value & $\sigma$  & species & $S$ & $M$ \\
    \hline
    1&  23&     0.661&  0.00645&       4.04&   CE-DM&  16&      1      \\
    2&  84&     0.488&  $7.11^{-5}$&    5.06&   CE-MM&  32&      4      \\
    3&  210&    0.431&  $3.22^{-9}$&    6.89&   CE-SC&  16&      4      \\
    4&  93&     0.389&  0.00384&       4.03&   MM-DM&  8&       1      \\
    5&  366&    0.373&  $1.14^{-11}$&   7.68&   CE-SC&  8&       4      \\
    6&  506&    0.364&  $2.37^{-15}$&   8.77&   CE-MM&  16&      4      \\
    7&  591&    0.308&  $2.07^{-12}$&   7.87&   CE-MM&  8&       4      \\
    8&  222&    0.294&  0.000505&      4.57&   CE-MM&  16&      2      \\
    9&  385&    0.245&  0.000103&      4.94&   MM-HS&  16&      4      \\
    10& 277&    0.244&  0.00261&       4.18&   CE-SC&  8&       2      \\
    11& 462&    0.241&  $1.59^{-5}$&    5.33&   CE-MM&  8&       2      \\
    12& 4613&   0.187&  $6.94^{-35}$&   12.93&  MM-HS&  8&       4      \\
    13& 2653&   0.185&  $3.08^{-19}$&   9.69&   MM-HS&  8&       2      \\
    14& 1254&   0.180&  $3.07^{-8}$&    6.46&   MM-HS&  8&       1      \\
    15& 2609&   0.141&  $1.45^{-10}$&   7.29&   CE-HS&  8&       2      \\
    16& 6509&   0.132&  $6.07^{-24}$&   10.77&  CE-HS&  8&       4      \\
    17& 1275&   0.128&  0.000794&      4.59&   CE-HS&  16&      4      \\
    18& 1547&   0.119&  0.000477&      4.72&   MM-DM&  8&       4      \\
    19& 1930&   0.115&  $9.79^{-5}$&    5.07&   DM-HS&  8&       4      \\
    \hline
    \end{tabular}

    \vspace{3mm}
    \begin{tabular}{|c|crlc|}
    \hline
    Independent variable & $N$ & Pearson$^2$&    $p$-value    &    $\sigma$ \\
    \hline
    $M$ & 1599&    -0.244&   $2.6\times 10^{-33}$&  12.68 \\
    $S$ & 1599&     0.417&   $2.5\times 10^{-108}$& 23.17 \\
    $S/M$ &1599&    0.564&   $2.4\times 10^{-225}$& 34.53 \\
    \hline
    \end{tabular}
\caption{\small
    {\bf Two rows of figures at top} plot the Resnik-based similarity vs. NAF between mouse-human (top row) and yeast-human (bottom row).
    {\bf Top Table:} Pearson correlation ($\rho$) and statistical significance of the plots. 
    {\bf Middle Table}: Filtering for well-annotated proteins, we see higher Pearson correlations between NAF and Resnik score (allowing all evidence codes) that result when filtering for well-annotated protein pairs in EC-driven alignments; $N$ is the number of aligned protein pairs for which {\em both} proteins are annotated with at least $S$ GO terms that are each annotate at most $M$ proteins per species. We exhaustively list every pair of BioGRID species for which the Pearson $p$-value is less than $10^{-2}$ for $S\ge 8$ and $M\le 4$; the table is sorted by $\rho(S,M)$. {\bf Bottom Table}: Pearson correlation between $M$, $S$, and $\rho(S,M)$ above, across all species and values $M$ and $S$ for which $\rho(S,M)$ was statistically significant.
    (See text for further discussion.)
}
\label{fig:Resnik-vs-NAF+Pearsons}
\end{figure}

\subsection*{Correlation between semantic similarity and Network Alignment Frequency (NAF)}
For each value $\phi$ of NAF, the mean Resnik similarity was computed across all aligned protein pairs with at least frequency $\phi$. We then plotted the Resnik values of various subsets of pairs allowing various subsets of GO evidence codes.
We will depict our results split across three ``axes'': (a) which topological objective was being optimized (our two examples here being EC\cite{GRAAL} and graphlet-GDV\cite{milenkovic2008uncovering}); (b) whether or not the aligned node pair possess sequence similarity; and (c) whether we allowed the use of sequence-based GO evidence codes when computing the Resnik score. Before studying the details, we first draw attention to our primary conclusion: when the aligned pair of proteins possess sequence similarity, then sequence-based evidence codes provide a ``boost'' to the Resnik score; conversely, this boost is impossible for aligned pairs of proteins that do {\em not} possess sequence similarity, resulting in a potential bias towards a low Resnik score for such pairs. We stress that the separation of aligned protein pairs into those that do, or do not, possess sequence similarity is done {\em after the fact}: sequence plays {\em absolutely} no role in creating our alignments or computing NAF. The sequence of events is (1) create 100 alignments by optimizing a topology-only objective function; (2) compute NAF for each pair of aligned proteins observed in the 100 alignments; (3) compute two Resnik scores for each pair of aligned proteins: one that allows the use of sequence-based evidence codes, and one that does not; (4) finally, once all scores are fixed (both NAF and Resnik), separate the aligned protein pairs into two groups: those that possess sequence similarity, and those that do not.

Figure \ref{fig:Resnik-vs-NAF+Pearsons} plots Resnik similarity {\it vs.} NAF in 12 ``postage-stamp'' sub-figures, arranged with the top row of postage stamps depicting alignments between mouse (MM) and human (HS), and the bottom row between yeast (SC) and human.
In each row, left three postage stamps (which we call a ``column-triplet'') depict alignments that were driven to optimize EC, while the right column-triplet were driven to optimize {\it Graphlet Degree Vector Similarity} \cite{milenkovic2008uncovering}.
Each ``postage stamp'' displays the mean (blue line) and standard deviation (blue shaded area) of Resnik semantic similarity (measured on the left axis with scores from 0 to 12) between pairs of individually aligned proteins as a function of NAF. The red line (right axis, logarithmic from 1 to $10^5$) depicts the {\em number} of pairs that aligned with that NAF or higher.
Within each column-triplet, the three columns depict 
\begin{itemize}
    \item[(left)] {\em only} those aligned protein pairs that possess sequence similarity, and for which we allowed sequence-based evidence codes in the Resnik score (column labelled at the bottom with ``seqSim/allGO'');
    \item[(mid)] {\em all} aligned protein pairs, again allowing sequence-based evidence (column ``allPairs/allGO'');
    \item[(right)] {\em all} aligned protein pairs, but disallowing sequence-based evidence codes (column ``allPairs/GO-NOSEQ'').
\end{itemize}
Note that the latter two columns of each column-triplet in Figure \ref{fig:Resnik-vs-NAF+Pearsons} depict the {\em same set of aligned node pairs}, the only difference being that the former allows sequence-based evidence codes while the latter does not. Conversely, the first column of each triplet lists only those pairs that actually possess sequence similarity (see Methods).

In each column-triplet of \ref{fig:Resnik-vs-NAF+Pearsons}, comparing the three postage stamps reveals, respectively, that (1) allowing sequence-based evidence significantly enhances the measured Resnik similarity, but obviously only for that minority of pairs that actually {\em possess} sequence similarity; (2) the sequence-similar pairs and their sequence-based evidence enhance the mean Resnik similarity across {\em all} aligned pairs, over the Resnik value obtained when (3) no sequence-based evidence is allowed.
(In the cases that the semantic similarity trend reverses and starts to decrease with alignment frequency, it is usually when the number of aligned pairs is below 30, which can be attributed to statistical noise.)
Comparing the two objective functions, we see that EC achieves maximum NAF frequencies of about 15--20 with mean Resnik scores of about 4--8 (depending on whether we allow sequence-based evidence). In contrast, the graphlet-GDV objective provides hundreds of aligned pairs with very high NAF (up to 100), though their Resnik scores are significantly lower on average. We will see below that even with these apparent low scores, graphlet-based objectives still retain significant predictive power.
Supplementary Figure \ref{fig:Resnik-vs-NAF-BP-CC-MF} shows that NAF correlates well with Resnik similarity even when we separate GO terms based on biological process (BP), cellular component (CC), and molecular function (MF), across all aligned pairs and allowing all GO terms.

We move now to the tables below the postage stamps of Figure \ref{fig:Resnik-vs-NAF+Pearsons}. The top table lists the Pearson correlations and statistical significance of the plots. For the species pairs mouse-human (top two rows) and yeast-human (bottom two rows), we list the number of aligned protein pairs (``\#aligs'') with NAF score 2\% or more, and compute the Pearson correlation between NAF and Resnik using either all GO terms (middle section) or including only GO terms not based on sequence (right section). In each section we list the Pearson correlation $\rho$, the $p$-value computed using Fisher's $r$-to-$z$ transformation, as well as the number of standard deviations ($\sigma$'s) from random represented by that $p$-value.

%As in Figure \ref{fig:Resnik-vs-NAF+Pearsons}[top], we see that the edge-based measures (EC, $S^3$, Importance) perform well on more distantly related species fly-human and yeast-human (top row, left 3 and right 3 columns, respectively), as well as on pairs of the mammals rat, mouse, and human (second row). The graphlet-based measures perform poorly on these distant species (Supplementary) but perform very well on any pair of the mammals.

The correlations ($\rho$ values) listed in the top table of Figure \ref{fig:Resnik-vs-NAF+Pearsons} are on the low side. The primary reason for this is due to lack of GO information: the majority of proteins have few GO annotations, or only very vague ones. The mathematical formulation of the Resnik score requires that {\em both} proteins be well-annotated to achieve a high score \cite{resnik1995using,resnik1999semantic}. For example, if only a small fraction $\varepsilon$ of proteins are well-annotated by some criterion, then only a fraction $\approx\varepsilon^2$ of protein {\em pairs} will be well-annotated by the same criterion. Luckily, our 100 alignments provide us with about half a million pairs of aligned proteins for any given pair of species---more than enough to allow us to filter for well-annotated pairs. (If both proteins are well-annotated but with very different annotations, then they will have a {\em meaningful} low score, as opposed to a low score due to lack of information.)

To account for this, we will filter protein pairs for annotation quality. First, note that a GO term’s specificity is inversely proportional to how many proteins it annotates: GO terms that annotate only a few proteins tend to provide more specific information than vague GO terms that annotate thousands of proteins. Furthermore, proteins annotated with highly specific GO terms tend to be better understood than those that are not. In the large middle table of Figure \ref{fig:Resnik-vs-NAF+Pearsons}, each row displays the correlation between NAF and Resnik after filtering for well-annotated protein pairs. In particular, for a given row labeled with $(S,M)$ in the last two columns, each protein must independently be annotated by at least $S$ distinct GO terms each of which annotates at most $M$ proteins per species. The table exhaustively lists every statistically significant ($p$<0.01) correlation observed for $S\ge 8$ and $M\le 4$ optimizing the EC objective, sorted by $\rho(S,M)$. For example, the top row depicts alignments between the species pair CE-DM (worm {\it C. elegans} and fruit fly {\it D. melanogaster}); although not depicted, the 100 CE-DM alignments contained exactly 302,169 distinct protein pairs with non-zero NAF; among these, there were only $N=23$ in which {\em both} proteins were annotated by at least $S=16$ distinct GO terms each of which annotated at most $M=1$ proteins in its respective species. In other words, these 23 protein pairs are {\em very} well understood---they each possess least 16 GO terms that {\em uniquely} annotate {\em that} protein and no other in its species. In this case, we see that correlation between NAF and Resnik is $\rho(S,M)=0.661$---much higher than the correlations seen among the unrestricted pairs in Table in Figure \ref{fig:Resnik-vs-NAF+Pearsons}.

The large middle table of Figure \ref{fig:Resnik-vs-NAF+Pearsons} lists only a small subset of ($S,M$) values we tested, which included all pairs where $S$ and $M$ independently ranged from 1 to 1024 in powers of 2 (10 values each), for both the $EC$ and graphlet measures---200 rows per species---across all ${5 \choose 2}=10$ pairs of the 5 best-annotated BioGRID species: {\it C. elegans, D. melanogaster, M. musculus, S. cerevesia, and H. sapiens} (CE, DM, MM, SC, and HS, respectively). Merging all of these cases gives a table with 2,000 rows, each one with a NAF-Resnik Pearson correlation $\rho(S,M)$ and $p$-value.
Of particular interest is what happens when we compute the Pearson correlation between $\rho(S,M)$ and either $M$ or $S$. More formally: Given a pair of species $s_1,s_2$ and values of $M$ and $S$ each ranging from 1 to 1024 in powers of 2, let $\rho(S,M)$ refer to the Pearson correlation between NAF and Resnik restricted to protein pairs satisfying the $S,M$ requirements. Out of the 2,000 rows, there are 1599 in which the correlation $\rho(S,M)$ is statistically significant ($p<5\times 10^{-6}$, chosen to ensure a statistical significance of at least $p<0.01$ after Bonferroni correction across 2,000 rows), we find that $\rho(S,M)$ is itself correlated with each of $M$ and $S$, independently. Since this is a correlation of correlations, we refer to it as a Pearson$^2$. Observing the bottom table of Figure \ref{fig:Resnik-vs-NAF+Pearsons}, we see that there is a strong and highly significant correlation with $M$ (negative because specificity increases as M decreases), and separately a strong and highly significant positive correlation with $S$ (the number of such GO terms possessed by both proteins). The correlation becomes even stronger if we use $S/M$ as the independent variable. In English, the bottom table of Figure \ref{fig:Resnik-vs-NAF+Pearsons} demonstrates that
    {\it the more we know about two proteins that have been aligned, the better the correlation between their alignment frequency (NAF) and their mutual Resnik score.}
This observation suggests that high NAF scores tend to uncover protein pairs with genuine high similarity, even if that similarity is not (yet) well-documented with GO terms; in turn, this suggests that NAF can be used as a measure of confidence that two proteins possess GO-based semantic similarity.

\subsection*{The NAF-function postulate}
In each column-triplet of Figure \ref{fig:Resnik-vs-NAF+Pearsons}, the second and third columns (``allPairs'') show significantly lower Resnik scores than the first, which plots {\em only} pairs that possess sequence similarity according to BLAST (bitscore threshold 13, E-values allowed from 0 to 1000).
%If alignment frequency scales with semantic similarity equally well for pairs with and without sequence similarity, this suggests that there may be far more semantically similar proteins that do {\em not} share sequence similarity than than those that do.
Since NAF aligns protein pairs based {\em only} on similar network topology, and Tables \ref{fig:Resnik-vs-NAF+Pearsons}--\ref{fig:Resnik-vs-NAF+Pearsons}(3rd Table) strongly support the hypothesis that NAF correlates with Resnik semantic similarity, we propose the following:

\begin{quote}
{\bf NAF-function Postulate}: {\it protein pairs aligned at or above a given Network Alignment Frequency (NAF) are drawn from a single distribution of functional similarities, regardless of whether or not they possess significant sequence similarity.}
\label{eq:NFP}
\end{quote}
We provide evidence for the NAF-function Postulate below, but if true, it suggests that, compared to the first column of each column-triplet in Figure \ref{fig:Resnik-vs-NAF+Pearsons}, the lower scores of the second and third columns is spurious, because allowing GO terms derived from sequence-based evidence will only benefit that minority of protein pairs that actually {\em possess} sequence similarity; those pairs that do not possess sequence similarity cannot benefit from sequence-based evidence that does not exist. Of course, we are not claiming that sequence-based evidence is untrustworthy; it is simply {\em inapplicable} to protein pairs that do not possess sequence similarity. If one assumes that the Resnik scores in the left column (``seqSim/allGO'') are indicative of {\em true} similarity for protein pairs at a particular NAF, then the NAF-Function Postulate asserts that the Resnik scores in the second and third columns are artificially low.
In essence, the NAF-function postulate states: {\it sequence-based evidence doesn't help when it doesn't exist---but absence of evidence is not evidence of absence.} This, combined with the already-known fact that functional similarity can exist despite little or no detectable sequence similarity\cite{furuse1998claudin,fisher2006conservation,schlicker2006new}, makes the NAF-function postulate a plausible extension of existing knowledge.

\begin{figure}
\centering
\includegraphics[width=0.4\linewidth]{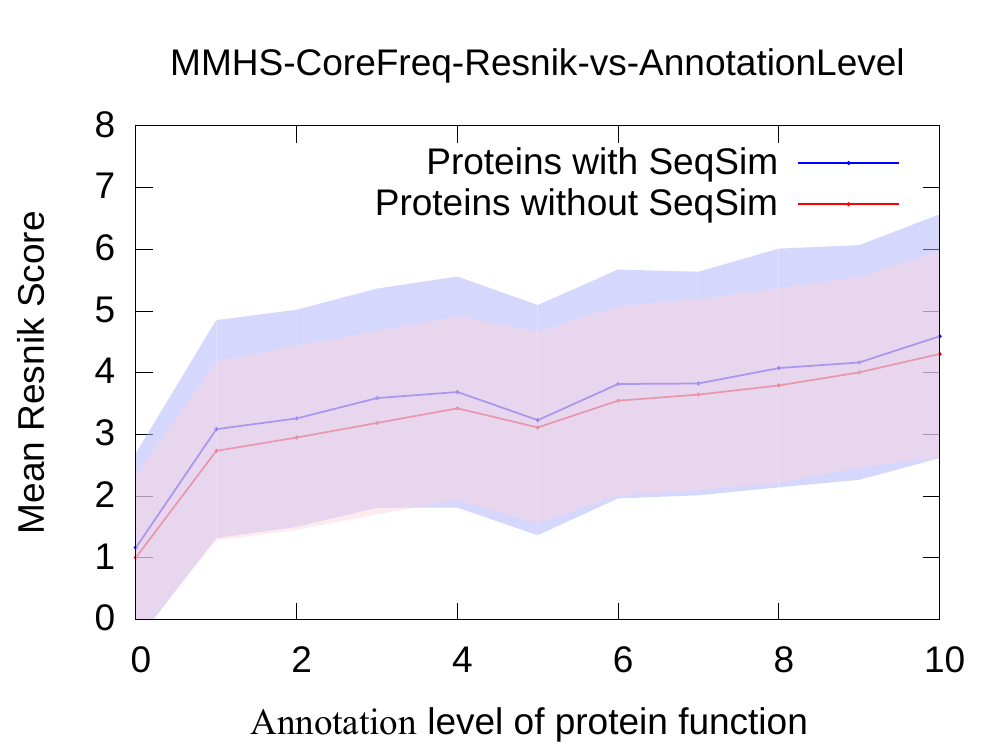}
\includegraphics[width=0.4\linewidth]{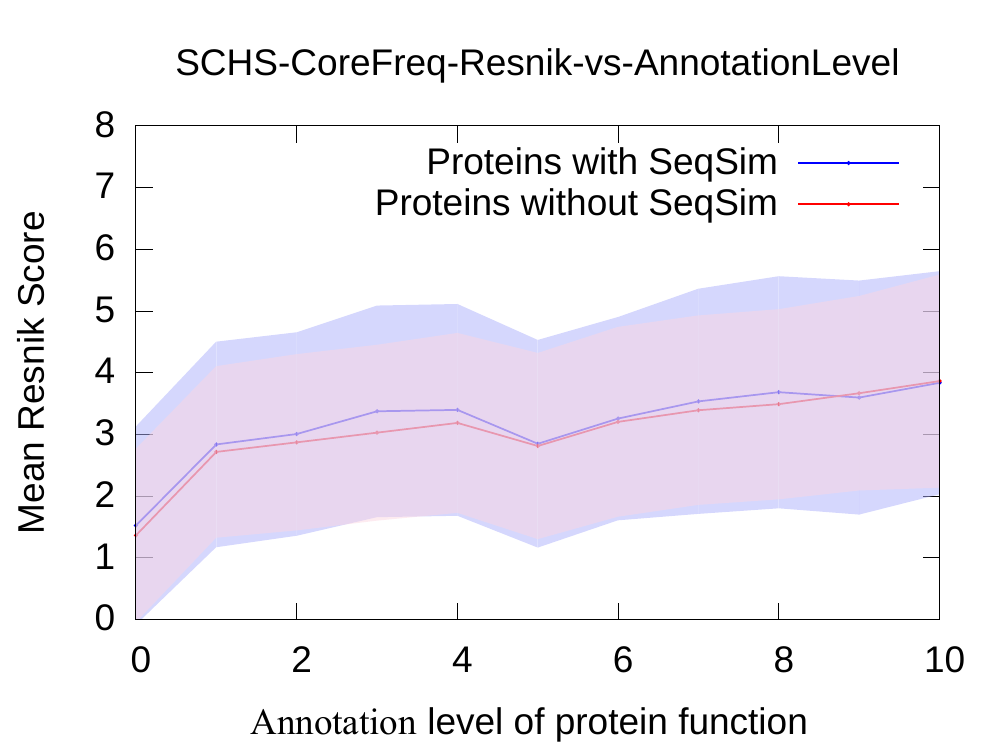}
\caption{{\bf Protein pairs aligned by network topology alone have equal Resnik similarity---not including sequence-based evidence---independent of whether they possess sequence similarity.}
Note the horizontal axis here is no longer NAF, it is {\em annotation level} of aligned pairs across all those with NAF 2\% or higher.
We plot mean Resnik score as a function of annotation level for MMHS (left) and SCHS (right). In each plot, aligned protein pairs $(p,q)$ are binned along the horizontal axis into the integer part of the NetGO-based annotation detail \cite{NetGO2017} of the lesser understood protein.  The vertical axis is mean Resnik score, with shading out to $1\sigma$ standard deviation of the pairs in that bin. Blue is protein pairs {\em with} sequence similarity, red is those without. In all cases, the Pearson correlations are above 0.35 and have $p$-values below $10^{-300}$ before binning to take the mean, while the $p$-value of Pearson correlation of the binned means are about $2\times 10^{-3}$; the {\em difference} between the means has $p$-value above 0.4---ie., far from statistical significance.
The Resnik scores here are significantly lower than those in Figure \ref{fig:Resnik-vs-NAF+Pearsons} because (a) we have, of necessity, removed all sequence-based evidence, and (b) the mean is dominated by the high number of low-NAF (NAF=2\%) pairs.
}
\label{fig:Resnik-Know}
\end{figure}

We now provide evidence for the NAF-function Postulate. First, to apply a level-playing-field comparison of Resnik similarity between pairs of nodes that may or may not share sequence similarity, we disallow the use of sequence-based evidence in computing the Resnik score (cf. Table \ref{tab:BioGRID+NOSEQ}(bottom)). Surprisingly, even after removing sequence-based evidence, sequence-similar proteins retain a significant Resnik advantage at fixed NAF. Careful investigation reveals that proteins with sequence similarity tend to be better-annotated even with {\em non}-sequence evidence than those without (Supplementary Figure \ref{fig:seq-vs-noseq-GOk}). While the reason behind this bias in annotation levels is beyond the scope of this paper (popularity\cite{luck2017proteome}?), the effect is easily removed by accounting for level of GO annotation.
Figure \ref{fig:Resnik-Know} plots the mean Resnik score as a function of {\em annotation level} (ie., number of GO terms, disallowing sequence-based evidence), across all aligned protein pairs with NAF 2\% or more. After separating those aligned protein pairs with, and without, sequence similarity, we observe that the two curves are statistically indistinguishable, suggesting that sequence similarity plays little or no role in the NAF-Function Postulate.
%First, assume that the function of a protein is effectively {\em defined} by its interaction partners and the pathways it participates in. Thus, proteins with more similar interaction partners---and thus similar topological neighborhoods---will tend to have more similar function. %We further note, with emphasis, that the words ``sequence similarity'' do not play a role in the assumption that network topology defines function, except indirectly in that similar sequence leads to similar binding domains, similar 3D structure, which leads eventually to similar interaction partners---but that traditional route from sequence to function is {\em far} longer, more circuitous, and more error-prone than the statement ``network neighborhood defines function''.
%We now offer a further hypothesis: that all protein pairs aligned with equal NAF---sequence-similar or not---have functional (or at least semantic) similarities drawn from the same distribution.
%More specifically, Figure \ref{fig:Resnik-Know} supports supports the NAF-Function Postulate by showing that once we remove the effect of disparate levels of annotation detail and compare Resnik similarity of proteins with equal annotation detail and without allowing sequence-based evidence, the difference in Resnik similarity between pairs with and without sequence similarity {\em vanishes}.
In other words, while high sequence similarity is often {\em sufficient} to infer functional or semantic similarity, it is by no means {\em necessary}: removing sequence-based evidence and comparing the Resnik similarity between protein pairs at equal annotation level, the impact of sequence similarity on semantic similarity is negligible. More to the point, this suggests that when two proteins {\em without} sequence similarity are aligned with NAF at or above some threshold $\phi$, their semantic similarity tends to be about the same as equal-NAF pairs {\em with} sequence similarity. While Figure \ref{fig:Resnik-Know} only demonstrates this for $\phi=2$, the previous sentence equates to the NAF-Function Postulate.

Finally, we note the obvious fact that protein pairs with high sequence similarity are rare among the space of all protein pairs, which is why---when it occurs---sequence similarity correlates well with semantic similarity. Similarly, protein pairs with high topology-based network similarity (as quantified by NAF) are also rare in the space of all protein pairs, and that network similarity correlates equally well with semantic similarity. Figure \ref{fig:Resnik-Know} establishes that topological network similarity also correlates with functional and semantic similarity, {\em independent} of whether the topologically-aligned protein pairs share sequence similarity.

\subsection*{NAF predicts common GO terms even in the absence of sequence similarity}

The bottom two tables in Figure \ref{fig:Resnik-vs-NAF+Pearsons} show than when both proteins are well-annotated, we observe a strong positive correlation between NAF and the demonstrable similarity between the pair of proteins aligned. This suggests that NAF can be used as a measure of confidence that two proteins share some common set of GO terms: if two proteins are aligned with high NAF but only one of them is well-annotated, there is a basis for using the GO terms possessed by one as predictions of GO terms possessed by the other, with NAF providing a measure of confidence of the predictions. Here we test this hypothesis in several ways.

\subsubsection*{Predictions from the year 2010, validated today}

To demonstrate that NAF's success is not simply due to the greater amount of network data available today than previously, we have performed the required 100 alignments on the same species as in Table \ref{tab:BioGRID+NOSEQ}(top), but using \href{https://downloads.thebiogrid.org/BioGRID/Release-Archive/BIOGRID-3.0.64}{BioGRID 3.0.64, released on 23 April 2010}. We then used the \href{http://archive.geneontology.org/full/2010-04-01/}{Gene Ontology release of the same month} to predict novel (as of April 2010) GO annotations between species, as follows:
Let $p_{g,e}$ represent the fact that protein $p$ is annotated with GO term $g$, supported by evidence code $e$. 
For each pair of proteins $(p,q)$ aligned by SANA with NAF $\ge\phi$, assume we wish to use the GO terms of $p$ (the ``source'') to predict those of $q$ (the ``target''). For each GO term $g$ from the source protein $p$, and for each evidence code $e$ relating $p$ to $g$, we increment a counter $q_{g,e}$ by $\phi$. Note that this allows GO terms and their evidence codes for target $q$ to accumulate across {\em different} proteins $p$ of the source species---essentially, if $q$ is aligned with multiple proteins $p$ and all of these alignment partners agree than $q$ should be annotated with GO term $g$, then the NAF value accumulates across all such $p$'s. For example, if a GO term appears among multiple non-human proteins each aligned with the same human one, all contribute to the score of the human protein being annotated with that GO term, with that evidence code. At the end, we have a cumulative score for $q$ being annotated with $g$, across various evidence codes $e$. If the cumulative score is above a pre-specified threshold $\Phi$ (used in precision-recall calculations, see Methods), it counts as a prediction.
We then validate these predictions by checking to see if the predicted GO term shows up as annotating the human protein in a later release of the GO database. 
%In the following analysis, we ensured that our predictions were orthogonal to those that could be produced using sequence analysis by eliminating all pairs, regardless of NAF, that had sequence similarity according to BLAST run on the protein sequences with its default parameters, as well as all pairs that were listed as even the most distantly related orthologs in InParanoid 8 or the 2019 release of EggNog.
We find that the validation rate depends heavily on the evidence code used to justify the annotation of the non-human protein. By far the evidence codes with the greatest predictive power (from 2010) are IPI (Inferred from Physical Interaction), EXP (experimentally determined), and IDA (Inferred from Direct Assay), in that order. (Keep in mind that these are the evidence codes for the {\em source} protein---the non-human one.) This resulted in over 3,000 novel annotations to almost 4,000 human proteins, including 137 human proteins that had {\em zero} GO annotations as of April 2010. 

We made every effort to exclude annotations that could have been either predicted, {\em or validated} using any form of sequence information. In particular, we eliminated from consideration (1) any pair of proteins that had sequence similarity according BLAST (used with its default parameters resulting in bit scores of 13 or higher); and (2) any pair of proteins listed as orthologs---even distant ones---in any of NCBI Homologene, InParanoid 8 \cite{sonnhammer2015inparanoid}, or the 2019 release of EggNog \cite{huerta2019eggnog}. Additionally, we excluded any GO annotation that was supported by {\em any} sequence-based evidence code, {\em even if it also had non-sequence-based evidence}. Finally, this procedure was applied {\em both} to the 2010 GO release from which GO term predictions were sourced, as well as 2020 GO release which was used to validate predictions. Though these conditions are likely more stringent than one would want in a production-level prediction pipeline, our goal here is to demonstrate that none of the predictions discussed below could have been made, or even validated, using any form of sequence information. In short, the predictions below should be largely {\em orthogonal} to predictions that are based on sequence analysis.

\begin{figure}
    \small \centering
    \includegraphics[width=0.4 \textwidth]{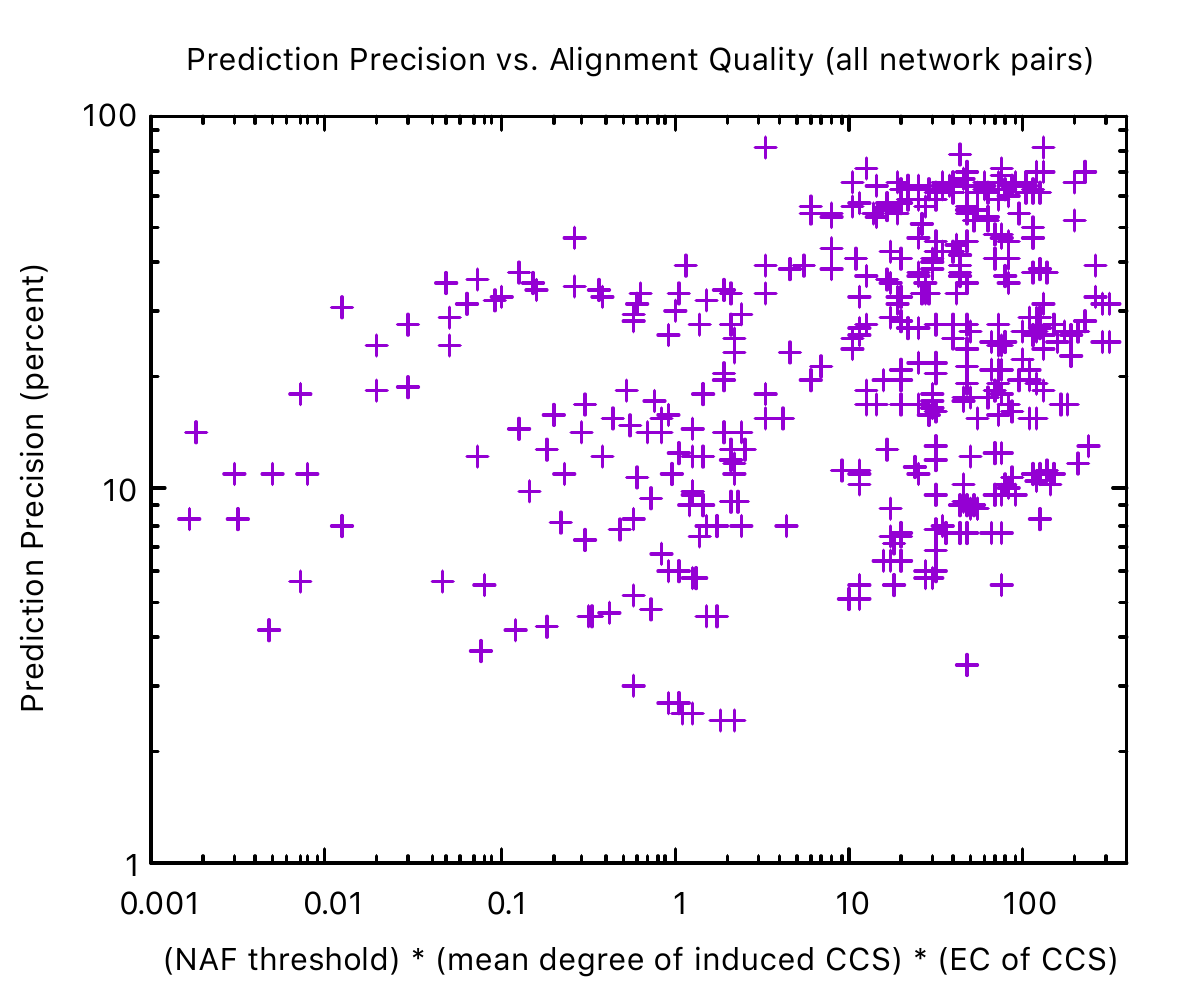}
    \includegraphics[width=0.4 \textwidth]{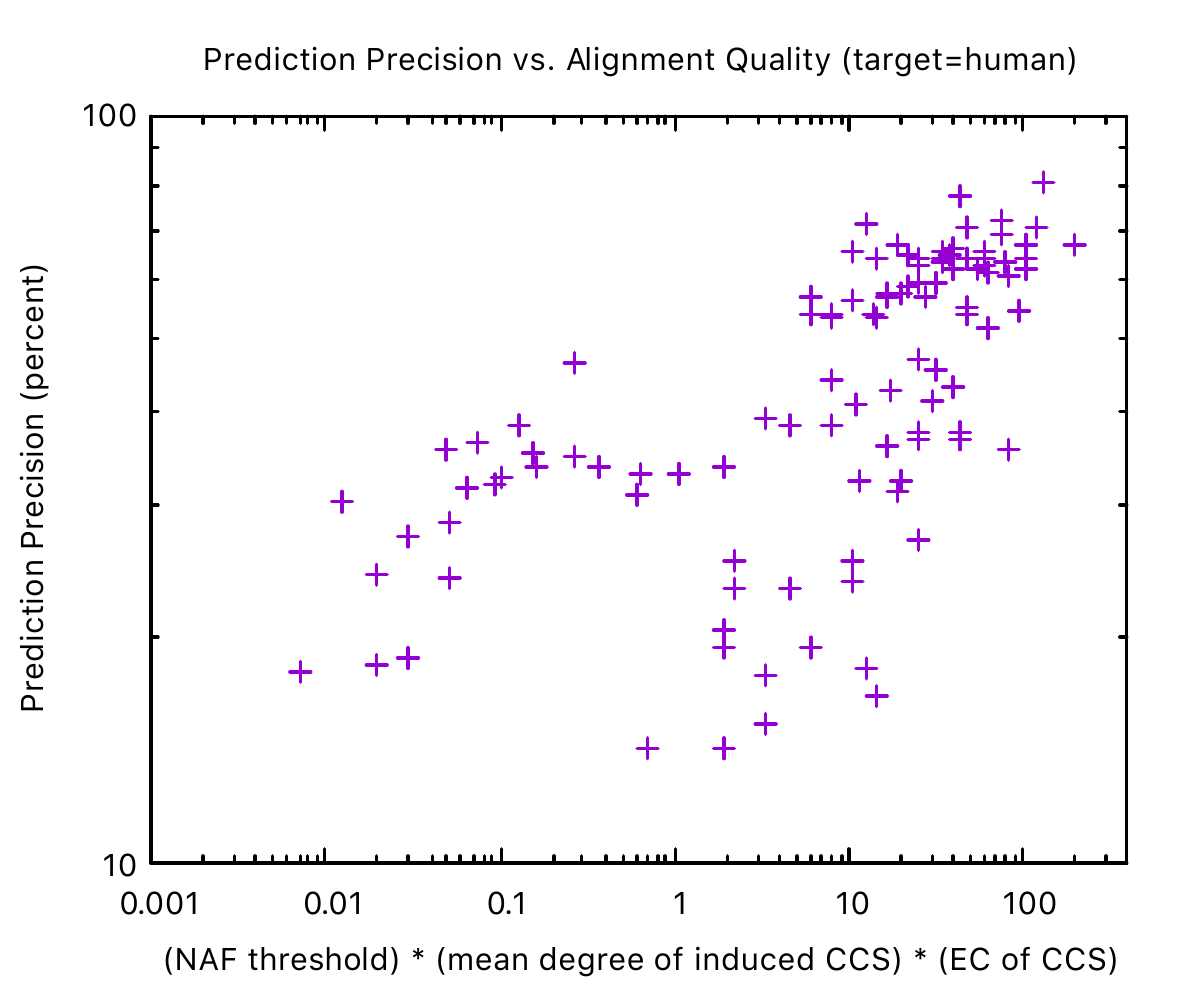}

    \vspace{3mm}
    \includegraphics[width=0.32\textwidth]{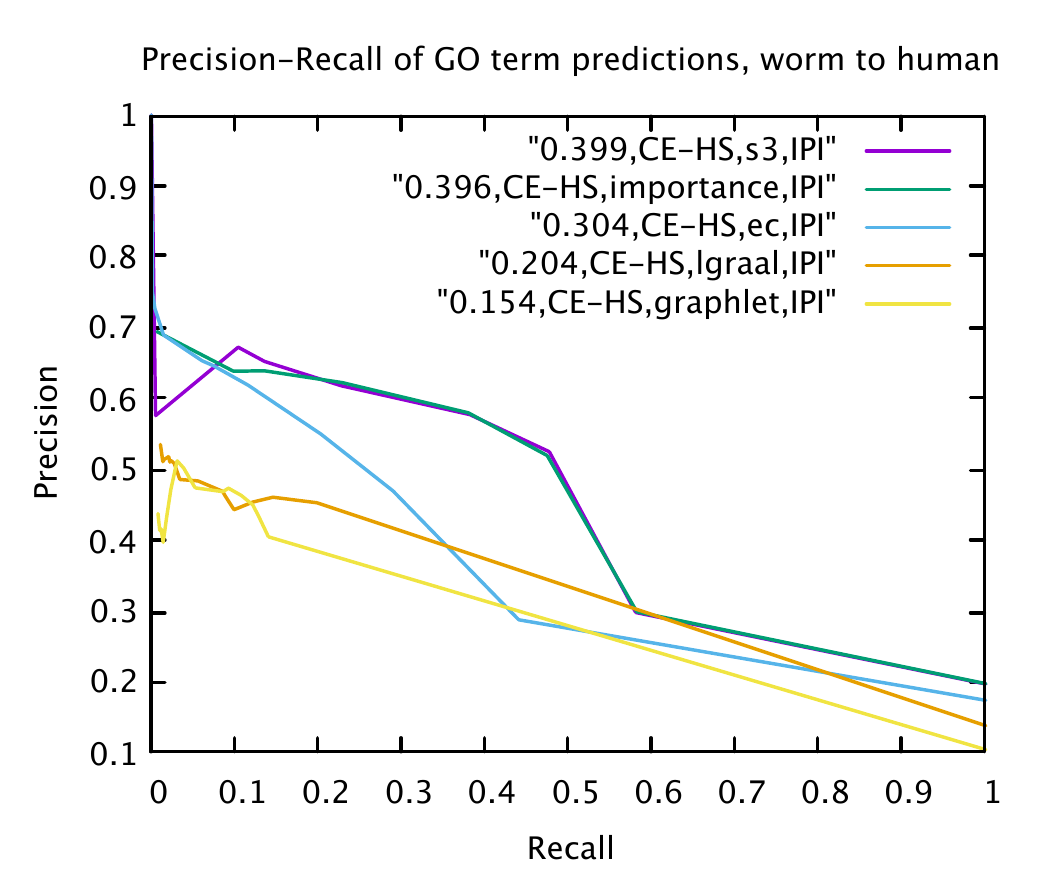}
    \includegraphics[width=0.32\textwidth]{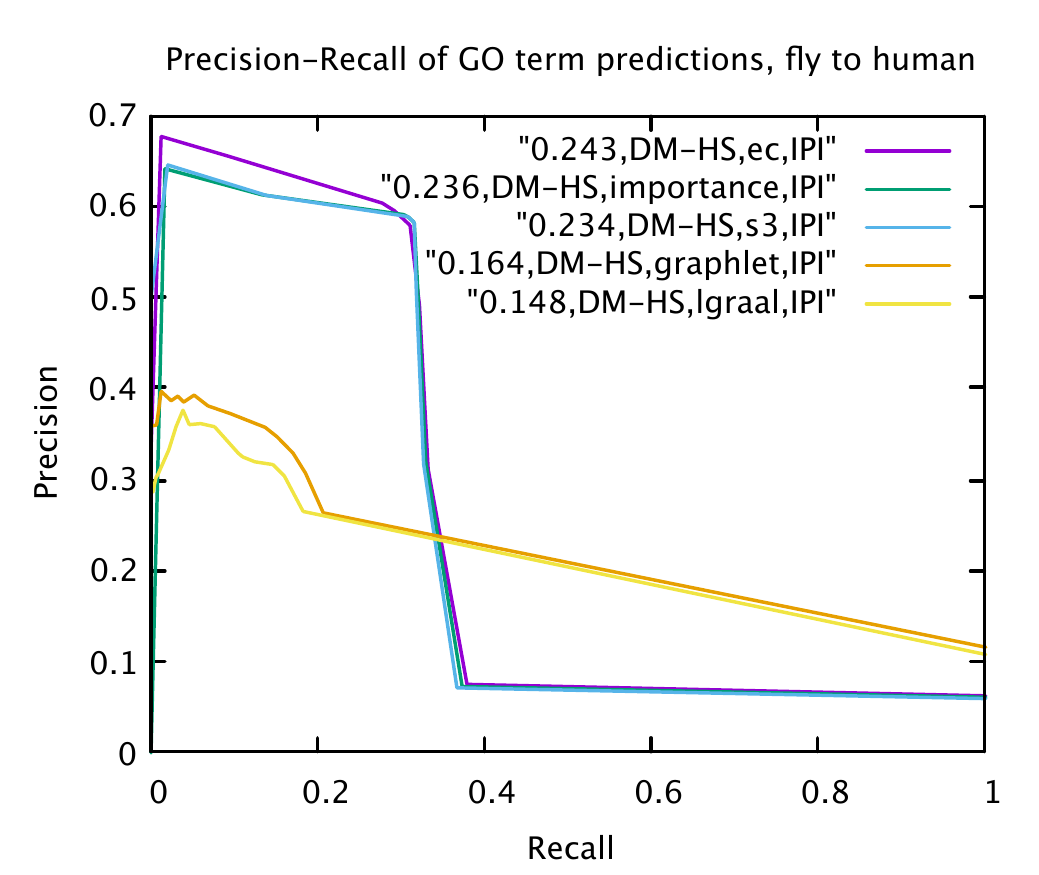}
    \includegraphics[width=0.32\textwidth]{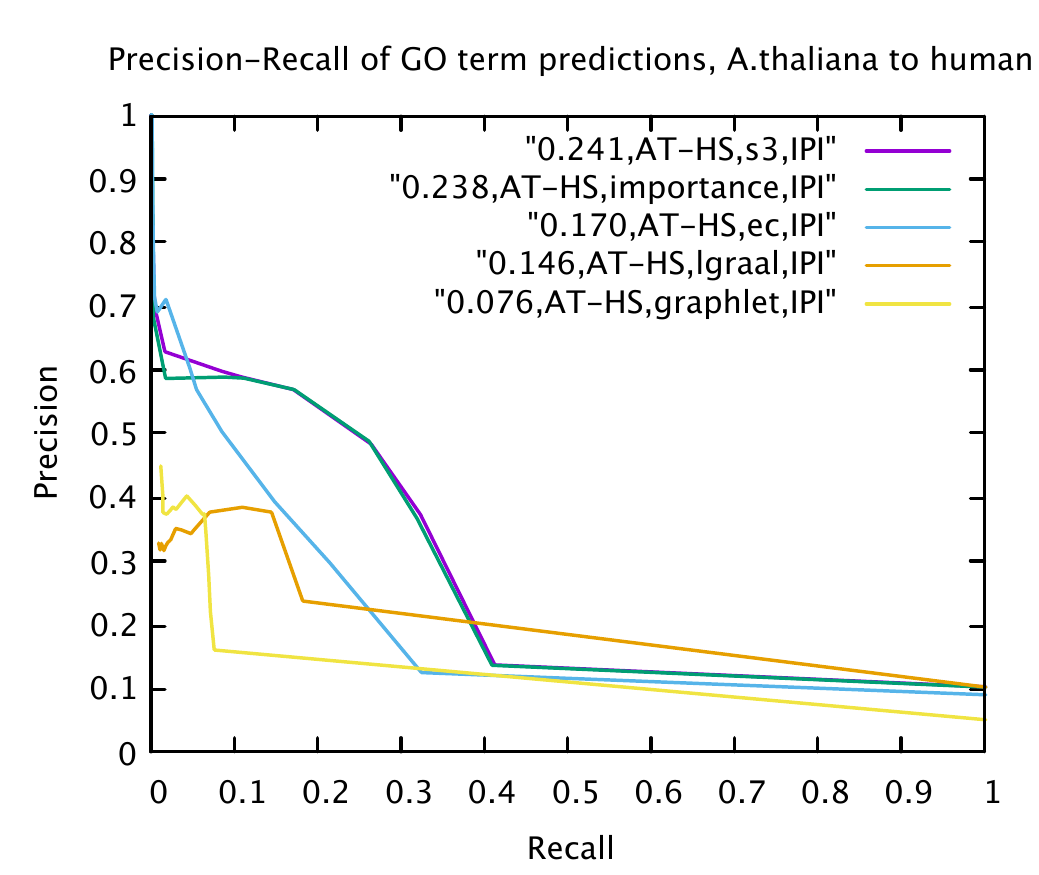}\\
    \includegraphics[width=0.32\textwidth]{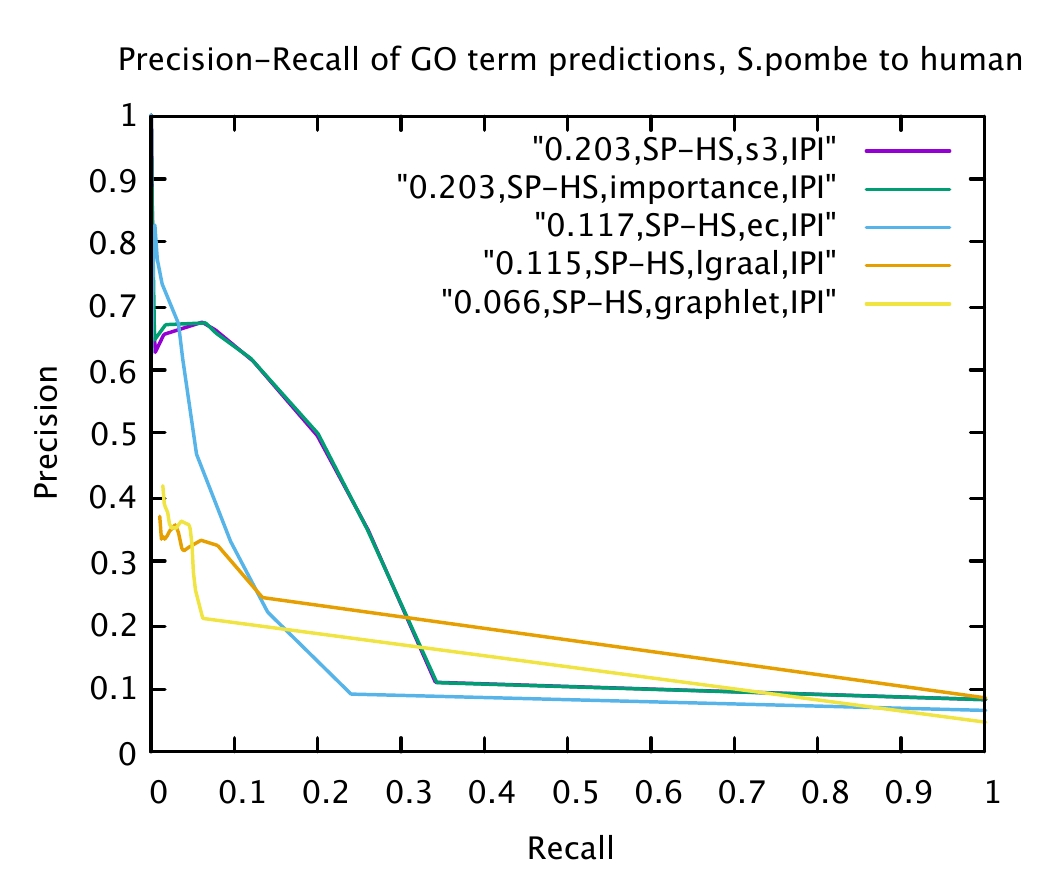}
    \includegraphics[width=0.32\textwidth]{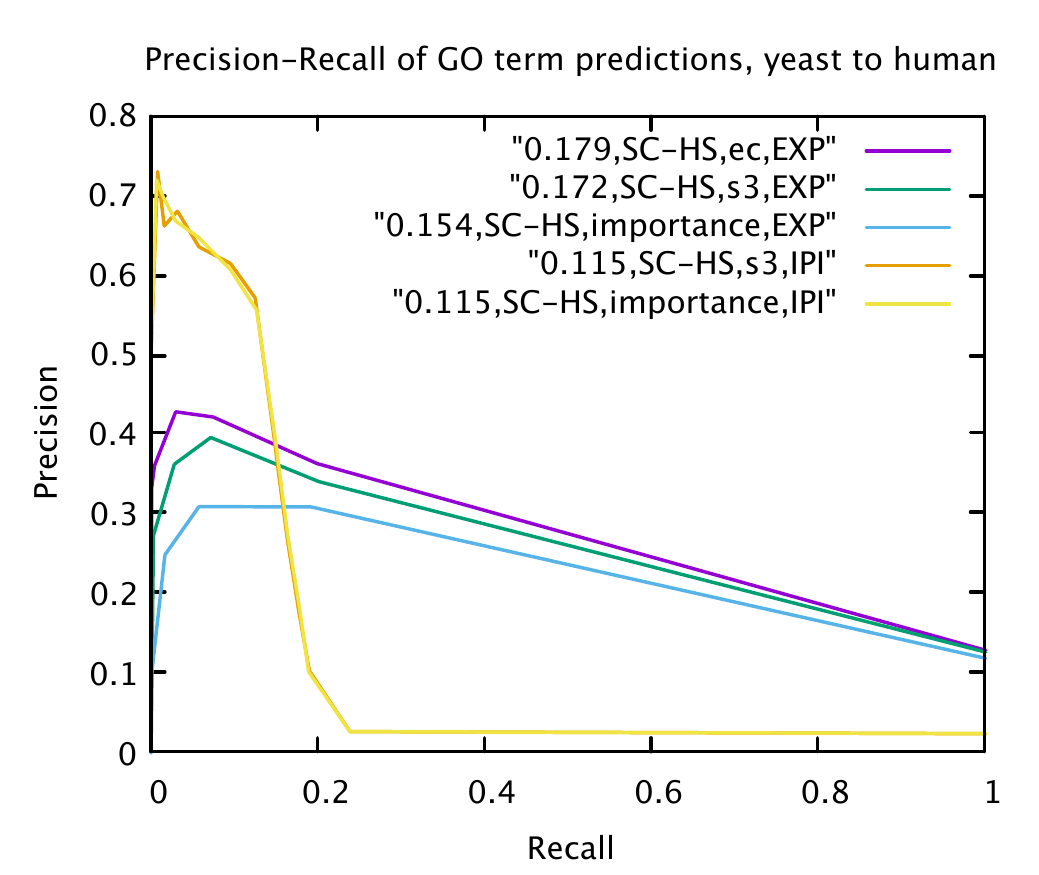}
    \includegraphics[width=0.32\textwidth]{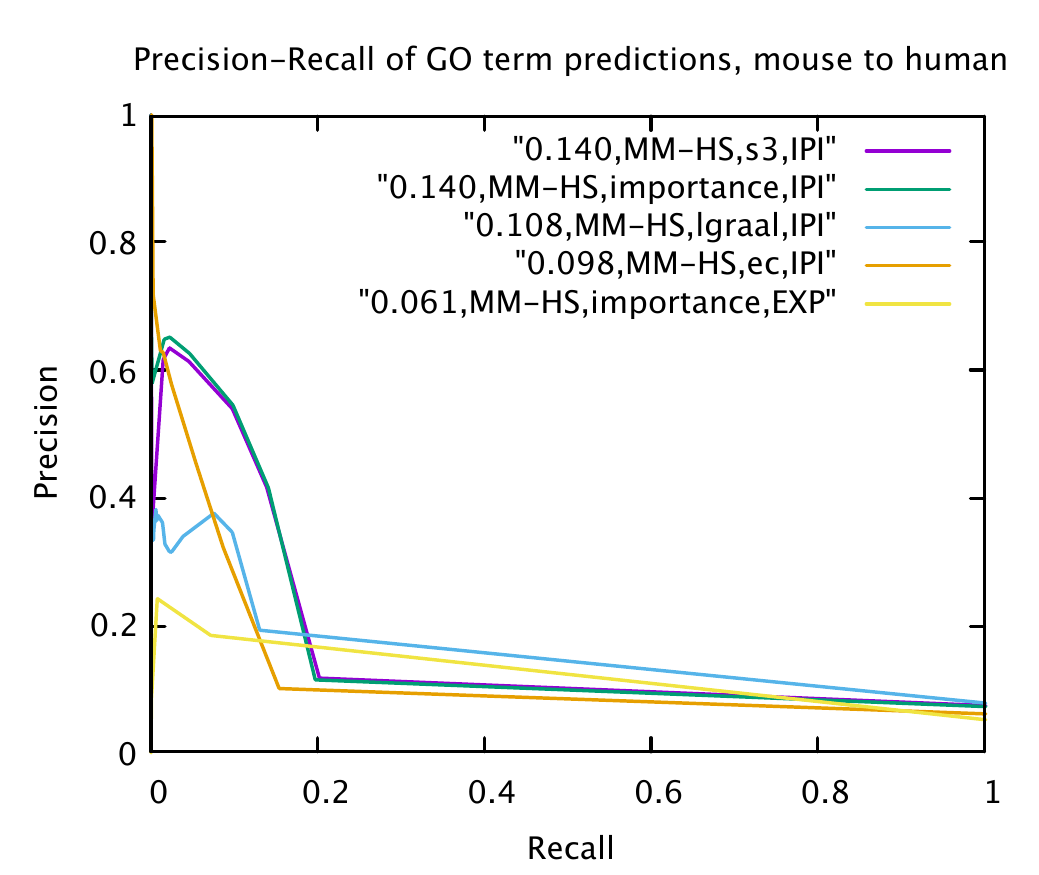}
    \caption{{\bf TOP (purple scatter plots):} Mean precision of GO term predictions {\it vs.} alignment quality between all species pairs (left, Pearson $\rho=0.19, p=0.002, n=511$) and species pairs when human was the target (right, Pearson $\rho=0.61, p=2\times 10^{-11}, n=107$). In both cases, alignment quality is measured as the product of NAF, EC, and mean degree of aligned nodes induced on the CCS with the given NAF. Predictions were made using only BioGRID networks and GO terms available as of April 2010, and validated against GO terms available a decade later (April 2020).
    {\bf BOTTOM}: Precision-Recall of 2010-based NAF predictions of GO annotations for human proteins: Predictions are made using only data available as of April 2010, validated against the GO release of April 2020. We omit any predictions in which the aligned proteins had any known orthology or detectable sequence similarity.     We plot precision vs. recall of predictions from global network alignments between the network pairs where human is the target and  which satisfy the degree-3 criterion, which are (left to right, top to bottom) {\it C. elegans, D. melanogaster, A. thaliana, S. pombe, S. cerevisiae, and M. musculus}; the six figures are ordered by species from best-to-worst by the mean AUPR of each figure. Within each figure, the legends are ordered best-to-wost by AUPR, and labeled by: AUPR, species pair, measure of topological similarity, and {\em predicting} evidence code (ie., the evidence code of the non-human protein used to source the prediction for the aligned human protein). Precision is the number of correct predictions as a fraction of all predictions arising at the threshold NAF, and the denominator of Recall is the cardinality of the set resulting from the intersection of the following two sets: {\em Predictable in Principle} pairs, and the annotations actually present in the April 2020 GO release (called the validating set---see Methods).
    See Table \ref{tab:Fstar} for $F$-scores.
    }
    \label{fig:PR-2010-by-Evidence+Pred-vs-aligQual}
%\begin{table}%
%    \begin{tabular}{|l|llllllll|}
%    \hline
%species & {\it S.cerevisiae} & {\it H.sapiens} & {\it D.melanogaster} & {\it C.elegans} & {\it A.thaliana} & {\it S.pombe} & {\it M.musculus} & {\it R.norvegicus}\\
%common name & yeast & human & fruit fly & worm & thale cress & fission yeast & mouse & rat\\
%nodes & 5611 & 8192 & 7003 & 2794 & 1607 & 1374 & 528 & 232\\
%edges & 51860 & 28117 & 22128 & 4457 & 2563 & 2323 & 475 & 178\\
%    \hline
%    \end{tabular}
%    \captionof{table}{\small Species from the April 2010 release of BioGRID (3.0.64), sorted by number of edges. %We see that yeast, human, and fly have by far the greatest number of edges, so we confine our analysis to predictions from yeast and fly to human.
%    }
%    \label{tab:networks2010}
%\end{table}
\end{figure}

In the process of studying prediction precision, we discovered that some sets of 100 alignments provided few validated predictions even with a high NAF threshold. Investigation revealed that the alignments in question had little topology in common despite the high NAF of the aligned nodes. In particular, given a set of nodes with NAF above a threshold, the {\it Common Connected Subgraph} (CCS) is the set of edges in common among the aligned nodes---cf. the purple edges emanating from purple nodes in Figure \ref{fig:NetAlign}. We found that prediction precision suffered significantly in two distinct cases. By far the most frequent case was when the mean degree of (purple) nodes of the CCS were low even with high EC or $S^3$ scores (cf. Figure \ref{fig:NetAlign})---in other words, while most edges were aligned, there simply weren't very {\em many} of them---possibly meaning the high EC and $S^3$ were due to chance. Less frequently, we found cases where the mean degree of the CCS was high, but the number of {\em non}-aligned edges was even higher, making both the EC and $S^3$ scores low. Figure \ref{fig:PR-2010-by-Evidence+Pred-vs-aligQual}(top) quantifies this effect by plotting prediction precision vs. ``alignment quality'' as measured by the product of NAF, and the EC and the mean degree of nodes in the CCS induced with that NAF. Importantly, like NAF, this measure of ``alignment qualtity'' is computable {\em a priori} as part of the alignment output. Since the low-degree case was by far the most frequent cause of low prediction precision, for the purposes of this paper we will arbitrarily apply a lower bound of 3 on the mean degree of the induced CCS to eliminate cases of low prediction precision; we call this the {\bf degree-3 threshold}, and leave to future work how to more rigorously choose such a bound.

\begin{table}
    \small
    \centering
    \begin{tabular}{|ccr|rrrr|}
    \multicolumn{7}{c}{Evidence code {\tt EXP}}\\
    \hline
    pair & measure & NAF & \#pred & PIP &\#val & precision \\
    \hline
    SC-HS& EC&    2&  383&   1903&   159&  41.5\% \\
    SC-HS& EC&    4&   73&   1903&    30&  41.1\% \\
    SC-HS& $S^3$&    2&  399&   1903&   150&  37.6\% \\
    SC-HS& $S^3$&    4&   68&   1903&    22&  32.4\% \\
    SC-HS& Importance&    2&  412&   1903&   125&  30.3\% \\
    \hline
    SP-HS& EC&    2&   49&   1906&    22&  44.9\% \\
    SP-HS& EC&    4&    7&   1906&     3&  42.9\% \\
    \hline
    \multicolumn{7}{c}{Evidence code {\tt IPI}}\\
    \hline
    SC-HS&   ec+s3+Imp&          4&      10917&  19278&  2305&   21.1\%  \\
    SC-HS&   ec+s3+Imp&          7&      19075&  19278&  6137&   32.2\%  \\
    SC-HS&   ec+s3+Imp&          11&     12341&  19278&  5436&   44.0\%  \\
    SC-HS&   ec+s3+Imp&          17&     8568&   19278&  4655&   54.3\%  \\
    SC-HS&   ec+s3+Imp&          29&     2961&   19278&  1838&   62.1\%  \\
    \hline
    AT-HS&   ec+s3+Imp&          2&      10600&  7740&   3471&   32.7\%  \\
    AT-HS&   ec+s3+Imp&          4&      6757&   7740&   3304&   48.9\%  \\
    AT-HS&   ec+s3+Imp&          7&      3416&   7740&   1945&   56.9\%  \\
    AT-HS&   ec+s3+Imp&          11&     1564&   7740&   937&    59.9\%  \\
    AT-HS&   ec+s3+Imp&          17&     610&    7740&   393&    64.4\%  \\
    AT-HS&   ec+s3+Imp&          29&     109&    7740&   79&     72.5\%  \\
    \hline
    CE-HS&   ec+s3+Imp&          2&      8652&   3703&   4200&   48.5\%  \\
    CE-HS&   ec+s3+Imp&          4&      4482&   3703&   2603&   58.1\%  \\
    CE-HS&   ec+s3+Imp&          7&      1998&   3703&   1253&   62.7\%  \\
    CE-HS&   ec+s3+Imp&          11&     637&    3703&   410&    64.4\%  \\
    CE-HS&   ec+s3+Imp&          17&     133&    3703&   88&     66.2\%  \\
    CE-HS&   ec+s3+Imp&          29&     15&     3703&   10&     66.7\%  \\
    \hline
    DM-HS&   ec+s3+Imp&          2&      18858&  8439&   5701&   30.2\%  \\
    DM-HS&   ec+s3+Imp&          4&      10119&  8439&   5473&   54.1\%  \\
    DM-HS&   ec+s3+Imp&          7&      6310&   8439&   3566&   56.5\%  \\
    DM-HS&   ec+s3+Imp&          11&     7546&   8439&   4358&   57.8\%  \\
    DM-HS&   ec+s3+Imp&          17&     3360&   8439&   2025&   60.3\%  \\
    DM-HS&   ec+s3+Imp&          29&     59&     8439&   43&     72.9\%  \\
    \hline
    SP-HS&   ec+s3+Imp&          2&      8244&   6974&   2755&   33.4\%  \\
    SP-HS&   ec+s3+Imp&          4&      3026&   6974&   1669&   55.2\%  \\
    SP-HS&   ec+s3+Imp&          7&      1677&   6974&   1054&   62.9\%  \\
    SP-HS&   ec+s3+Imp&          11&     786&    6974&   529&    67.3\%  \\
    SP-HS&   ec+s3+Imp&          17&     347&    6974&   235&    67.7\%  \\
    SP-HS&   ec+s3+Imp&          29&     71&     6974&   53&     74.6\%  \\
    \hline
    MM-HS&   ec+s3+Imp&          2&      8469&   13394&  3191&   37.7\%  \\
    MM-HS&   ec+s3+Imp&          4&      2629&   13394&  1489&   56.6\%  \\
    MM-HS&   ec+s3+Imp&          7&      849&    13394&  544&    64.1\%  \\
    MM-HS&   ec+s3+Imp&          11&     239&    13394&  148&    61.9\%  \\
    MM-HS&   ec+s3+Imp&          17&     31&     13394&  13&     41.9\%  \\
    MM-HS&   ec+s3+Imp&          29&     8&      13394&  5&      62.5\%  \\
    \hline
    \end{tabular}
    \caption{\small{\bf Prediction precision by evidence code and NAF threshold with {\it H. sapiens} as the target}. This table summarizes prediction precision as a function of NAF for species aligned with human satisfying the degree-3 criterion. The species pairs are presented in order of mean CCS degree, highest to lowest (cf. Supplementary Table \ref{tab:CCS-mean-degree}). We show predictions based on source evidence codes EXP (top section) and IPI (bottom section) available in 2010 and validated (with any evidence) in 2020. PIP means {\it Predictable In Principle}, and refers to the absolute maximum number of predictions that would be possible in principle given the information available as of April 2010 (see text). To save space in the IPI case, we have conglomerated the measures EC, $S^3$, and Importance, since all three had similar validation rates at fixed NAF (within 10\% of each other in all cases). Also to save space, not all values of NAF are listed here, but the Pearson correlation between NAF and precision across all NAF values are presented below, in Table \ref{tab:NAF-precision}.
    }
    \label{tab:HS-precision}
\end{table}

Supplementary Table \ref{tab:CCS-mean-degree} shows, for each species pair and each measure, the NAF value that achieved the highest mean degree $\overline{D}_{max}$ on the resulting induced CCS. Surprisingly, although the edge-based measures EC, $S^3$, and Importance frequently reach the degree-3 threshold, we observe that the graphlet-based measures rarely result in a mean degree above 1, and never above 3.

Table \ref{tab:HS-precision} depicts the prediction precision as a function of NAF for all species paired with human (HS), so long as the mean degree of the CCS was above 3; only RN ({\it Rattus norvegicus}) failed to satisfy the degree-3 threshold.
Observe that in {\tt IPI} section of Table \ref{tab:HS-precision}, the prediction precision generally increases with NAF. Table \ref{tab:NAF-precision} expands on this by showing that the prediction precision almost always has a strong positive correlation with NAF (though in some cases not enough distinct NAF values exist to make the correlation statistically significant, and the one case with a negative correlation is far from statistical significance). These correlations corroborate the hypothesis that higher NAF provides greater confidence that the aligned protein pair share common GO terms.

\begin{table}
    \centering
    \begin{tabular}{|c|rrrr|}
\hline
pair & rows &    $\rho$ & $p$-value & $\sigma$ \\
\hline
SC-DM&  18&     0.8436& 3.416$\times 10^{-5}$&  6.2844 \\
SC-HS&  17&     0.8741& 8.544$\times 10^{-6}$&  6.9706 \\
MM-DM&  14&     0.709&  0.0356& 3.4823 \\
SP-DM&  18&     0.5425& 0.1937& 2.5832 \\
AT-HS&  17&     0.8338& 0.0001243&      5.8490 \\
AT-DM&  17&     0.9301& 1.284$\times 10^{-8}$&  9.8053 \\
CE-HS&  16&     0.6352& 0.07426&        3.0771 \\
MM-CE&  16&     0.9252& 1.032$\times 10^{-7}$&  9.1228 \\
MM-SP&  16&     0.7642& 0.003988&       4.4333 \\
SP-AT&  18&     0.8393& 4.496$\times 10^{-5}$&  6.1739 \\
DM-HS&  16&     0.7672& 0.003627&       4.4753 \\
AT-CE&  17&     0.8804& 5.087$\times 10^{-6}$&  7.1898 \\
SP-CE&  18&     0.8864& 1.15$\times 10^{-6}$&   7.6598 \\
CE-DM&  15&     -0.2407&        0.4805& 0.8943 \\
SP-HS&  17&     0.7413& 0.005154&       4.2774 \\
MM-HS&  15&     0.2273& 0.5011& 0.8418 \\
RN-CE&  16&     0.7&    0.02165&        3.6677 \\
MM-AT&  14&     0.9327& 4.893$\times 10^{-7}$&  8.9604 \\
RN-SP&  6&      0.8793& 0.08192&        3.6916 \\
RN-DM&  6&      0.8885& 0.06864&        3.8726 \\
RN-AT&  6&      0.9643& 0.003344&       7.2840 \\
RN-MM&  6&      0.9068& 0.04537&        4.3019 \\
\hline
TOTAL (non-normalized) & 213 & 0.314& 0.00023 & 4.74 \\ 
TOTAL (normalized)     & 213 & 0.749& 7.27$\times 10^{-38}$& 15.5 \\
\hline
    \end{tabular}
    \caption{Correlation between NAF and prediction precision for each species pair, across rows similar to those in Table \ref{tab:HS-precision}. Each row represents one species pair with the NAF-precision correlation across all measures. The second-last row is the correlation between NAF and prediction precision across all species and all measures. However, as seen in Table \ref{tab:HS-precision}, the scaling between NAF and precision can differ substantially across species, which muddles the correlation. We correct for this in the final row, where we have normalized the NAF and precision to their maximum values on a per-species basis.
    }
    \label{tab:NAF-precision}
\end{table}

Armed now with the knowledge of which species pairs have ``robust'' alignments based on the mean degree-3 threshold of the CCS, Figure \ref{fig:PR-2010-by-Evidence+Pred-vs-aligQual}(bottom) presents precision-recall curves using NAF thresholds, across the 6 species aligned with human that satisfied the degree-3 threshold, broken down by predicting evidence code and measure of topological similarity used to drive the alignment.
The number of predictions are not depicted, but for example GO terms with IPI evidence in 2010 from yeast and fly produced 2959 and 2050 validated, novel GO annotations of human proteins, respectively; EXP produced 367 and 187, respectively. Other evidence codes for these species had AUPR's below 0.01, though some other species pairs had non-negligible AUPRs (see Supplementary).
Table \ref{tab:Fstar} lists the top 20 sets of predictions across all species pairs satisfying the degree-3 threshold, ranked by $F^*$ (best $F_1$ score), broken down by GO evidence code; Supplementary Table \ref{tab:Category} does the same for GO category (Biological Process, Cellular Component, Molecular Function). We see from Table \ref{tab:Fstar} that the most successful evidence code for making predictions is IPI (Inferred through Physical Interaction), while Supplementary Table \ref{tab:Category} shows that GO terms in the category Molecular Function are by far the most successfully predicted. These conclusions may change as the date of prediction moves forward.

We note that, even though these predictions are made with 10-year-old networks, our best AUPRs are competitive with the best sequence- and structure-based predictors in the 2017 CAFA3 competition as well as recent algorithms comparing themselves to CAFA3 \cite{cao2017prolango,savojardo2018busca,zhang2018metago,zhou2019cafa,kulmanov2020deepgoplus}. (It is impossible to compare directly against CAFA because no PPI network data is available for the species used in CAFA.) We emphasize again, however, that our predictions were neither made nor validated using sequence information, and so we believe our predictions are orthogonal to those that are possible from CAFA, and thus purely {\em additive} to existing prediction methods. Finally, it is interesting to note the high quality of these predictions even though Resnik-NAF correlations are much weaker in 2010 data than in Figure \ref{fig:Resnik-vs-NAF+Pearsons} (Supplementary Figure \ref{fig:Resnik-vs-NAF2010}).

\begin{table}
    \centering\small
\begin{tabular}{|r|r|r|l|l|l|r|r|r|r|}
\hline
rank & $F^*$  & NAF & pair& $M$ &EvCode& $|P_{12}\;\cap\;\Gamma'_2|$  & pred & valid & Precision \\
\hline
     1&  0.500&  2\%&      CE-HS&   $S^3$&     IPI&    3443&   3131&   1643&   52.5\% \\
     2&  0.496&  2\%&      CE-HS&   Import.&   IPI&    3443&   3146&   1634&   51.9\% \\
     3&  0.424&  16\%&     SP-AT&   Import.&   IPI&    455&    569&    217&    38.1\% \\
     4&  0.417&  16\%&     SP-AT&   $S^3$&     IPI&    455&    577&    215&    37.3\% \\
     5&  0.416&  5\%&      DM-HS&   $S^3$&     IPI&    4080&   3237&   1521&   47.0\% \\
     6&  0.404&  5\%&      DM-HS&   EC&     IPI&    5978&   3204&   1855&   57.9\% \\
     7&  0.375&  8\%&      SP-AT&   EC&     IPI&    472&    968&    270&    27.9\% \\
     8&  0.358&  2\%&      CE-HS&   EC&     IPI&    3443&   2127&   998&    46.9\% \\
     9&  0.346&  2\%&      AT-HS&   $S^3$&     IPI&    5038&   4360&   1627&   37.3\% \\
    10&  0.341&  2\%&      AT-HS&   Import.&   IPI&    5038&   4365&   1605&   36.8\% \\
    11&  0.340&  8\%&      MM-AT&   $S^3$&     IPI&    725&    768&    254&    33.1\% \\
    12&  0.309&  5\%&      DM-HS&   Import.&   IPI&    9018&   3245&   1895&   58.4\% \\
    13&  0.298&  2\%&      SP-HS&   $S^3$&     IPI&    5380&   4022&   1401&   34.8\% \\
    14&  0.298&  2\%&      SP-HS&   Import.&   IPI&    5380&   3985&   1394&   35.0\% \\
    15&  0.291&  5\%&      MM-AT&   EC&     IPI&    755&    1043&   262&    25.1\% \\
    16&  0.277&  1\%&      CE-HS&   {\sc lgraal}& IPI&    3140&   1511&   644&    42.6\% \\
    17&  0.260&  8\%&      SP-AT&   graphlet& IPI&    455&    808&    164&    20.3\% \\
    18&  0.259&  8\%&      SP-AT&   {\sc lgraal}& IPI&    472&    827&    168&    20.3\% \\
    19&  0.257&  1\%&      SC-HS&   EC&     EXP&    2111&   1158&   420&    36.3\% \\
    20&  0.249&  2\%&      AT-HS&   EC&     IPI&    5038&   3629&   1080&   29.8\% \\
\hline
\end{tabular}
    \caption{{\small\bf 2010-based predictions ranked by $F^*$.}
    %Refer to Figure \ref{fig:PR-2010-by-Evidence+Pred-vs-aligQual}(bottom) for precision-recall curves.
    {\bf Legend}:
    {\bf NAF}: threshold that achieved $F^*$;
    {\bf pair}: species pair (cf. Table \ref{tab:BioGRID+NOSEQ}(top));
    {\bf $M$}: topological measure;
    {\bf EvCode}: evidence code supporting the annotation of the source (non-human) protein that produced the predicted human protein annotation;
    {\bf $|P_{12} \cap\;\Gamma'_2|$}: intersection of the number of {\it Predictable in Principle} annotations ($P_{12}$, see Methods) with $\Gamma'_2$---all annotations available in the validation set.
    {\bf pred}: number of predicted annotations made using the specified source evidence code at the specified NAF (note this number can be bigger than the previous column since, clearly, any number of predictions can be made that may not appear in the validating GO release).
    {\bf  valid}: the number of validated predictions by any non-sequence-based evidence code.
    {\bf Precision}: number of predictions that were validated.
    }
    \label{tab:Fstar}
\end{table}

\subsubsection*{Predictions using 2018 data, validated today by literature search}
The painstaking effort required to create the Gene Ontology database by human curation of the literature necessarily means that the GO database lags behind knowledge available in the most recent, live literature. Thus, we repeated the same effort as we did for 2010, but using \href{https://downloads.thebiogrid.org/BioGRID/Release-Archive/BIOGRID-3.4.164}{BioGRID 3.4.164} (Sept. 2018, the same release as was used in Figures \ref{fig:Resnik-vs-NAF+Pearsons}--\ref{fig:Resnik-Know}), using the GO database of \href{http://release.geneontology.org/2018-09-05/}{the same month}. Our goal is to produce {\it bona fide} predictions of GO annotations to human proteins.
We expect that the relevance of inter-species GO term predictions will be highest when (a) the two species are as closely related as possible; and (b) both PPI networks are as complete as possible. Thus, we choose to align the human PPI network with that of mouse, since mouse and human are both mammals, and mouse has the most complete mammalian PPI network after human.

All below predictions of the annotation of human protein $p$ with GO term $g$ are {\it bona fide} predictions, in the sense that the annotation of $p$ with $g$ was not present in the Sept. 2018 GO release, either directly, nor by inference on the GO hierarchy. For reference, out of the approximately 150,000 GO annotations to human proteins, only 340 (0.23\%) contained the word ``cilia''; the numbers for mouse were comparable, at 285 out of 110,000 (0.26\%).

\subsubsection*{Literature validation of our top cilia-related GO term predictions}
To keep our job of manual literature curation tractable, we narrowed the scope to cilia-related predictions from mouse to human with a NAF of 8\% or greater, with cilia chosen on the advice of a senior curator of the Gene Ontology Consortium (Karen Christie, personal communication). We use cilia-related GO annotations of mouse proteins to predict the same GO annotations to human proteins lacking such annotations. We avoid all cases that could be related via sequence or orthology---in other words, we omit predictions where the aligned mouse and human proteins had any known orthology or detectable sequence similarity, even if the mouse protein had an annotation that the human one did not.
Table \ref{tab:cilia} shows that these predictions achieve a high rate of literature validation. We stopped at NAF$=8$ since lower values of NAF had dozens to hundreds of predictions, which is too many to validate manually.

\begin{table}
    \centering\small
    \begin{tabular}{|r|ll|l|l|l|}
    \hline
    NAF     &Mouse  & Human (\& aliases)& T & Predicted cilia-related GO: term(s) & Species+Validation \\
\hline
16      &MKS1   &HDAC5~CLUH     &P      &1905515, 0044458, 0060271, 0060122, & F\cite{rothschild2011camk,rothschild2018calcium} M,H\cite{winyard2011putative}\\
16      &MKS1   &               &C      &0005929, 0036064, 0035869 & \\
15      &NPHP1  &               &C      &0031514, 0005929, 0035869, 0097546, 0032391 & \\
12      &AHI1   &               &P      &1905515, 0060271 & \\
12      &AHI1   &               &C      &0005929, 0036064, 0097730 & \\
\hline
13      &SHANK3 &CAND1          &C      &0060170 & H\cite{zahid2020rapid} \\
\hline
12      &SHANK3 &RPL6~IRS4      &C      &0060170 & H?\cite{jakobsen2013centrosome} \\
\hline
12      &IQCB1  &CUL7~RAD51C    &P      &0060271 & H\cite{barraza2016two} \\
12      &IQCB1  &               &C      &0032391 &  \\
\hline
11      &NPHP4  &POLR3D~CFTR    &C      &0005929, 0036064, 0035869, 0097546, 0032391, 0097730 & H\cite{li2009generation,wong2012directed,scudieri2020ionocytes}\\
\hline
10      &APP    &CDH1           &C      &0035253 & H\cite{miyamoto2011insufficiency}?\cite{wang2014master}\\
\hline
10      &SHANK3 &EEF1A2~HNRNPU  &C      &0060170 & M?\cite{tadenev2011bardet}\\
\hline
10      &SHANK3 &RPL18~ITGA5    &C      &0060170 & R?\cite{mayer2008proteomic}\\
\hline
 9      &IQCB1  &RNF2           &P      &0060271 & H\cite{novas2015bardet,mcclure2017nuclear}\\
 9      &IQCB1  &               &C      &0032391 & \\
\hline
 9      &FAM92A &VCAM1          &P      &0060271 & M\cite{dewispelaere2015icam}\\
 9      &FAM92A &VCAM1          &C      &0036064, 0097546 & \\
 8      &IQCB1  &VCAM1          &P      &0060271 & \\
 8       &IQCB1  &VCAM1         &C      &0032391 & \\
\hline
 9      &NPHP4  &MUC4~XPO1      &C      &0005929, 0036064, 0035869, 0097546, 0032391, 0097730 & M\cite{santos2014central}\\
\hline
 8      &DYNC1H1        &RPS9   &P      &0003341 & F\cite{backfisch2014tools}\\
\hline
 8      &HTT    &CUL5~NUP50     &P      &1905505, 1902857, 0045724 & H\cite{nagai2018cullin}\\
\hline
 8      &AHI1   &C1ORF87        &P      &1905515, 0060271 & H\cite{blackburn2017quantitative} \\
 8      &AHI1   &               &C      &0005929, 0036064, 0097730 &  \\
\hline
 8      &NPHP1  &CCDC8          &C      &0031514, 0005929, 0035869, 0097546, 0032391 & M,H\cite{jackson2014regulating} \\
\hline
 8      &FAM92A &HIAT1~OBSL1    &P      &0060271 & M,H\cite{boldt2016organelle} \\
 8      &FAM92A &               &C      &0036064, 0097546 &  \\
\hline
 8      &NPHP3  &LMOD1~SOD1     &P      &1905515, 0060271 & M\cite{ma2011adenylyl}\\
 8      &NPHP3  &               &C      &0005929, 0097543, 0097546 &  \\
\hline
 8      &EPS8   &RPLP0P6        &C      &0032426, 0032420, 0032421 & not validated \\
\hline
 8      &EPS8   &CNBP~MYCL      &C      &0032426, 0032420, 0032421 & H\cite{d2015non} \\
\hline
    \end{tabular}
    \caption{\small {\bf All zero-sequence-similar cilia-related GO term predictions from BioGRID mouse to human with NAF 8\% or greater}: {\tt NAF} is the {\it Network Alignment Frequency} at which the {\tt Mouse} protein was aligned to the {\tt Human} protein. In all cases, the mouse protein was annotated with the specified GO terms but the human protein was not (even indirectly). {\tt T} is ``type'' (P = Biological Process, C = Cellular Component) of the predicted GO terms; Predicted GO terms are listed with the leading ``GO:'' and leading zeros removed; {\tt Species+Validation} lists the species (H=human, M=mouse, R=Rat, F=fish) for which cilia-related activity for that protein has been validated, along with the reference for the corroboration---however, in {\em all} cases the authors of the citations strongly implied that their results were applicable to humans, though a question-mark indicates the evidence was weak or only implicit.
    }
    \label{tab:cilia}
\end{table}

Below we provide a brief summary of each citation used in Table \ref{tab:cilia} that was used as evidence of cilia-related activity. There are 19 distinct human proteins with predicted cilia-related annotations; only 1 was entirely unvalidated; an additional 6 were validated for a non-human ortholog to the human protein without {\em explicit} mention of whether the prediction is expected to be valid for the human ortholog; and an additional 4 have what we would describe as ``weak'' human validation. The resulting validation rates are 18/19 (95\%), 12/19 (63\%), and 8/19 (42\%). In the case of ``weak'' validation, it is possible that, rather than directly transferring the specified GO term, it may be more appropriate to transfer a less specific GO term higher in the GO hierarchy. Determining when this is the appropriate action is an area of future research.

\begin{itemize}
    \item {\bf HDAC5} upregulates MEF2C; in turn, MEF2C is known to be missing during metastasis, the latter of which is necessary for ciliogenesis; conversely, inhibition of HDAC5 suppresses cyst formation that disrupt cilia formation \cite{rothschild2011camk}. HDAC5's upregulation of MEF2C also causes malformed cilia which can be rescued by knockdown of MEF2C \cite{winyard2011putative}; HDAC5 morphant Zebrafish exhibit shorter cili \cite{rothschild2018calcium}.
    
    \item {\bf CAND1} is a centrosome protein known to regulate centrosome amplification; CAND1 knockdown in mice inhibits airway ciliogenesis\cite{zahid2020rapid}.
    
    \item {\bf RPL6} is a centrosomal marker among a selection of known or candidate centrosomal proteins \cite[Figure 18.2]{jakobsen2013centrosome}.
    
    \item {\bf CUL7} Reduction in CUL7 expression is associated with defects in centrosome and cilia formation \cite{barraza2016two}.
    
    \item {\bf CFTR} at the molecular level is involved in chloride transport, but loss of function of CFTR disrupts cilia in lung tissue, causing cystic fibrosis (CF); direct delivery of CFTR to the lung is an active research area in the fight against CF \cite{li2009generation,wong2012directed,scudieri2020ionocytes}.
    
    \item {\bf CDH1}: there seems to be some controversy as to whether CDH1 does \cite{miyamoto2011insufficiency}, or does not \cite{wang2014master}, affect cilia.
    
    \item {\bf HNRNPU}: there is indirect evidence in a mouse model specifically designed to model human ciliopathy that HNRNPU interacts with SLP3 \cite{tadenev2011bardet}, a known cilia-active protein \cite{kulaga2004loss}.
    
    \item{\bf RPL18} (Ribosomal Protein L18) is one of 268 proteins identified in a rat cilia preparation \cite[Table 1]{mayer2008proteomic}; admittedly, the evidence here is weak as they make no further mention of RPL18.
    
    \item {\bf RNF2} is regulated by known BBS (cilial dysfunction) genes \cite{novas2015bardet,mcclure2017nuclear}
    
    \item {\bf VCAM1} is expressed on the ciliary body of mouse retinal cells modeled to study human autoimmune disorders \cite{dewispelaere2015icam}.
    
    \item {\bf XPO1} aids ciliary Gli2 export in mice \cite{santos2014central}.
    
    \item {\bf CUL5} knockdown weakly suppresses ciliogenesis in human epithelial cell cultures \cite{nagai2018cullin}.
    
    \item {\bf CCDC8, OBSL1, and CUL7 } form a centrosomal complex \cite{jackson2014regulating} in mice \cite{li2014cul9} and cultured human cells \cite{yan20143m}; this complex is implicated in {\it 3M Syndrome} (same references, but also as studied in human HEK293T cells \cite{boldt2016organelle}).
    
    \item {\bf SOD1} mutations are of interest because they are associated with a minority of the familial version of the muscular disease ALS; it has been shown than SOD1 mutations inhibit ciliogenesis in motor neurons in mice \cite{ma2011adenylyl}.
    
    \item {\bf RPS9} is known to be expressed in cells bearing motile cilia of model fish species \cite{backfisch2014tools}.
    
    \item {\bf C1ORF87} is found in high abundance in human airway cilia \cite{blackburn2017quantitative}.
    
    \item {\bf CNBP} integrity of the primary cilium is necessary to induce CNBP in human cancer stem cells \cite{d2015non}.
\end{itemize}
We note that, of the GO term predictions in Table \ref{tab:cilia}, 20 are {\it Cellular Component} (C), 11 are {\it Biological Process} (P), while none are {\it Molecular Function} (F). For this reason it would be misleading to label the results of this paper as ``functional prediction''. The biggest contributing factor to the lack of functional predictions is likely the fact that, of the 285 cilia-related mouse annotations, 205 are Cellular Component, 71 are Biological Process, and only 9 are Molecular Function. Thus, there is simply a dearth of truly {\em functional} annotations of cilia-related mouse proteins from which to draw predictions. A second likely contributing factor is, again, the dearth of network data which likely allows proteins to be aligned {\em close} to their ``proper'' place in the network but not directly to their functional ortholog. We hypothesize that this latter issue will become less of a problem as more PPI data are accumulated.

\subsubsection*{Detailed validation of our single highest NAF prediction}
%Looking more broadly than just cilia-related predictions, we look globally at all NAF-based predictions. 
The single highest NAF score was 82\% between mouse protein Fancd2 and human protein TRIM25. Here we provide detailed literature-based validation of all GO terms present in mouse Fancd2 but not human TRIM25 in the Sept. 2018 GO release---cf. Table \ref{tab:Fancd2-TRIM25}. Most are {\it Biological Process} GO terms, which according to recent CAFA\cite{zhou2019cafa} benchmarks is the most difficult GO category to predict. Note that in this section, we no longer restrict ourselves to cilia-related GO terms, and we arbitrarily omit validation of GO terms predicting by the IMP evidence code, due to time constraints. Thus, the text below attempts validation only of GO terms predicted by evidence codes other than IMP, though IMP-based predictions are included in Table \ref{tab:Fancd2-TRIM25}.

\begin{table}
    \centering
    \begin{tabular}{|lrlll|}
    \hline
        GO term         & freq  &Evidence     & Cat   & Description \\
    \hline
        GO:0005634      &14731  &IDA    &C      &nucleus \\
        GO:0034599      &285    &IGI    &P      &cellular response to oxidative stress \\
        GO:0000793      &100    &IDA    &C      &condensed chromosome \\
        GO:0048854      &55     &IGI    &P      &brain morphogenesis \\
        GO:0097150      &47     &IGI    &P      &neuronal stem cell population maintenance \\
    \hline\hline
        GO:0006974      &673    &IMP    &P      &cellular response to DNA damage stimulus \\
        GO:0050727      &84     &IMP    &P      &regulation of inflammatory response \\
        GO:0007129      &72     &IMP    &P      &synapsis \\
        GO:0007276      &44     &IMP    &P      &gamete generation \\
        GO:0051090      &29     &IMP    &P      &regulation of DNA-binding transcription factor activity \\
        GO:0045589      &20     &IMP    &P      &regulation of regulatory T cell differentiation \\
        GO:2000348      &4      &IMP    &P      &regulation of CD40 signaling pathway \\
    \hline
    \end{tabular}
    \caption{GO terms present in Mouse Fancd2 but not Human Trim25, along with the global frequency of the GO term, the evidence code, the GO Category (Biological Process, Cellular Component), and description. {\bf Top section}: Non-IMP-based GO term predictions, sorted from most general (high frequency in the 2nd column) to most specific (low frequency). {\bf Bottom section}: GO terms predicted by the IMP evidence code, for which we did not attempt literature validation due to time constraints.}
    \label{tab:Fancd2-TRIM25}
\end{table}

\paragraph{Biological Process GO:0048854 (brain morphogenesis)}
Formation of the brain requires differentiation of stem cells into determined cell types. 
Autophagy plays an important role in stem cell differentiation, as it allows the cell to degrade obsolete organelles to become a more specialist cell \cite{vessoni2012autophagy}. 
TRIM family proteins are emerging as important regulators of autophagy, and interact with a range of known autophagy proteins\cite{kimura2016precision}. 
A number of autophagic genes, including Ambra1, are expressed in mouse embryos during neuronal differentiation\cite{vazquez2012atg5}. Ambra1 has been shown to be a key modulator of neurogenesis\cite{fimia2007ambra1}. Recently, it has been demonstrated that TRIM25 interacts with Ambra1 to up-regulate autophagy in mouse liver cells\cite{zhang2019vivo}. Whether TRIM25 interacts with Ambra1 similarly in neural cells is not known, but two of its close relatives have been shown to promote neural differentiation by different pathways: TRIM32\cite{sato2011trim32}, and TRIM69\cite{han2016trim69}. TRIM25 has been shown to enhance transcriptional activity of the differentiator gene RAR$\alpha$ to a similar degree as TRIM32\cite{sato2011trim32}, further implicating it in this pathway for promoting neural stem cell differentiation.

\paragraph{Biological Process GO:0097150 (neuronal stem cell population maintenance)}
Understanding the functions of different TRIM proteins in this regard is an area of cutting-edge research, as discoveries that TRIM proteins have regulatory functions in neural development and maintenance have only recently been made\cite{nenasheva2020many}. As with stem cell differentiation, autophagy is an important process in stem cell maintenance \cite{vessoni2012autophagy}, and TRIM proteins have important roles in autophagy \cite{mandell2014trimJ1,mandell2014trimJ2}.
Deficiencies in autophagy can result in neuro-degenerative disorders and premature aging\cite{levine2019biological}.
TRIM25 is expressed and contributes to stem cell maintenance in mouse embryos\cite{kwon2013rna} by ensuring genomic stability following DNA replication\cite{zhao2018mouse}. A recent survey \cite{nenasheva2020many} states that TRIM25's function in stem cells appears to be the least well understood out of all TRIM family proteins, and makes no mention of a role for TRIM25 in neurological processes. The indirect evidence presented above, along with its high NAF score, suggests that TRIM25’s role in this area be further investigated.

\paragraph{Biological Process GO:0034599 (cellular response to oxidative stress (ROS))}
Oxidative stress in cells is used as a signal of protein activity and function. Viral infection can lead to oxidative stress and degradation of viral proteins via proteasomes, and the TRIM25 ubiquitylation pathway \cite{osna2014proteasome}. Viral-origin proteins, when expressed in the cell, commonly generate reactive oxygen species. The RIG-1 pathway is known to respond to ROS to trigger cellular processes as part of the innate immune system\cite{soucy2010requirement}.
Importantly, reactive oxygen species are also a known stimulus for activating autophagic processes \cite{chen2009superoxide}, providing an obvious potential link between this prediction and the autophagy ones discussed above.

\paragraph{Components GO:0000793 (condensed chromosome) and GO:0005634 (nucleus)}
TRIMs have roles in cell cycle progression\cite{venuto2019e3}. The cell cycle is composed of various different phases, one of which is mitosis (M phase). During mitosis, a number of changes occur within the cell, including the condensation of DNA into chromosomes (in prophase). While the review of Venuto \& Merla\cite{venuto2019e3} does not acknowledge TRIM25 to have a specific role in prophase mitosis, the relatively uncharacterised status of TRIM25\cite{nenasheva2020many} does not contradict our prediction. Finally, chromosome condensation occurs in the nucleus, so if TRIM25 is involved in condensing the chromosome, this additionally implies GO:0005634.

In sum, TRIM25 appears a poorly understood member of the TRIM family. Given the importance of E3 ubiquitin ligases in neurological development, disorders and degenerative conditions\cite{upadhyay2017e3} these predictions from PPI network alignment provide plausible directions for future research in the function of TRIM25.

%Finally, since the above predictions use 10-year-old data, we offer evidence that NAF applied to modern networks is likely to have substantially stronger predictive power. Table \ref{tab:cilia} provides an exhaustive list of predictions of cilia-related GO terms made using mouse-human protein pairs aligned with NAF of 8 or above on EC-driven alignments on BioGRID 3.4.164 (Sept. 2018). For each mouse protein with a cilia-related annotated with a cilia-related GO term but the other is not. To these make predictions, we chose to look for high-NAF alignments between a mouse and a human protein---i.e., pairs of proteins that appeared a highly significant number of times out of the 100 alignments between BioGRID mouse and human---where the mouse protein was annotated with any GO term mentioning ``cilia'', and the human one was not. We then performed a literature search on Google Scholar for papers that matched ``cilia'' with the name of the human protein, and then manually curated the resulting papers for evidence of cilia-related function of the human protein. Table \ref{tab:cilia} shows the top predictions, and their results.

\subsection*{Comparison with other methods that use only network topology}
At the time of writing, we are aware of only two methods in the literature that predict GO annotations using only network topology: SINaTRa\cite{jacunski2015connectivity} and Mashup\cite{cho2016compact}; neither is based on network alignment.

{\it Synthetic Lethality} (SL) refers to a pair of genes neither of which is alone essential to life, but death occurs if both are knocked out simultaneously.
SINaTRa \cite{jacunski2015connectivity} uses a vector of traditional (non-graphlet) local measures of network topology to quantify the neighborhood of a node, and then uses standard machine learning techniques to train an SL classifier on pairs of genes in one species, and then predict SL pairs in another species.
While the authors attempt no other types of prediction other than SL, and they use data from just one year (2015), the closest approximation to our results are when they train on yeast ({\it S. cerevisiae}) and test on an ``ablated'' version of the fission yeast ({\it S. pombe}) network designed to mimic the edge density of the human network. In this test (their Figure S10), they achieved AUPRs between 0.43 and 0.60\cite[p.9]{jacunski2015connectivity}. Their higher AUPRs may be attributable to their using more recent data (by 5 years).

Mashup \cite{cho2016compact} uses {\it network diffusion} to construct a compact, low-dimensional vector of features for each node in a network. They then integrate the feature vectors extracted from many different types of networks from the {\em same} species to train an off-the-shelf machine learning algorithm to learn properties of interest, such as GO terms. Using the 2013 STRING database as input, they achieve AUPRs for prediction of human GO terms in the range of about 0.15 to 0.40 (their Figures 2 and 3), depending on details of their ranking. These numbers are comparable to ours (cf. Figure \ref{fig:PR-2010-by-Evidence+Pred-vs-aligQual}(bottom)).

\section*{Discussion}

In broad outline, our main results are:

\begin{enumerate}
    \item Across many stochastically-generated inter-species network alignments with near-optimal\cite{WeBeat} topological scores, the frequency that a pair of proteins is aligned together correlates with, and has predictive value of, Resnik similarity.
    
    \item Network Alignment Frequency (NAF) exposes Resnik similarity not only in the absence of significant sequence similarity, but exposes such similarity between non-sequence similar proteins that is just as strong as the Resnik similarity between sequence-similar proteins (cf. Figure \ref{fig:Resnik-Know}).  This leads to the {\bf NAF-function Postulate} (page \pageref{eq:NFP}).
    
    \item While sequence comparison is obviously an accepted and valuable tool when predicting functional similarity, it is simply {\em not applicable} when no significant sequence similarity is detectable. Thus, sequence similarity is a sufficient but not necessary condition for inferring functional or semantic similarity (cf. Figure \ref{fig:Resnik-Know}).
     
    \item To our knowledge, NAF is the first measure based solely on topology-driven network alignment to provide GO term predictions with success that is competitive with state-of-the-art methods, whether based on sequence, structure, or topology.
\end{enumerate}

Though not depicted in any Figures, we also measured precision, recall, and AUPR of our 2010-based predictions (similar to Figure \ref{fig:PR-2010-by-Evidence+Pred-vs-aligQual}(bottom)) by validation against GO releases for every year from 2011 to 2019. We found that the number of validated predictions, sourced from 2010, increases significantly year-over-year, suggesting that many ``non-validated'' predictions may become validated at some future date. Also, though not discussed in the main text, Supplementary Figure \ref{fig:degree-vs-frequency-noSEQ} demonstrates that the ability to detect and predict semantic similarity scales with degree and, more generally, edge density (see also our companion paper\cite{WeBeat}). This leads us to predict that the following will occur as network data continue to accrue:
    \begin{enumerate}
        \item[(a)] larger regions of the networks will become robustly alignable---ie., NAF scores will increase, along with the number of protein pairs aligned with NAF above any fixed threshold.        
        \item[(b)] topology-driven network alignments will be able to discover better topological {\em agreement} between networks, resulting in more GO term predictions, and with greater confidence. This hypothesis is corroborated by the much higher prediction accuracy of our literature validation of 2018-sourced predictions than those from 2010.
        \item[(c)] in general, the biological relevance of topology-driven network alignments will increase dramatically.
    \end{enumerate}
    
Related to the above, it is important to emphasize that we are not claiming that the results expounded in this paper are of practical use---yet. The fundamental problem is dearth of PPI network data. Yeast and Human are by far the most complete species pair, and yet they do not produce the best predictions, possibly due to their great taxonomic distance. The mere fact that we had to run {\em one hundred} independent 1-hour runs of SANA {\em per species pair} in order to tease out the weak signal attests to how weak that signal is at present. The signal is just too weak, and the CPU requirements too large, for the method to be practical on existing networks. We expect, however, that as PPI networks become more complete and less noisy, a much more clear signal will appear in network alignments, allowing topology-only network alignments to more efficiently extract predictions.

One may notice that the ``good'' values of NAF and other parameters of our algorithm varies widely between species. We believe this, again, is due to the wide disparity in network densities between species.  This makes it fruitless to ``tune'' the parameters of our algorithm on one species pair and use those parameters on another pair.
We also have not accounted for multiple hypothesis testing in any of the $p$-values herein.

Clearly, our goal is to make the best novel GO term predictions using {\em today}'s data. To do that, it is important to have an estimate for the confidence level of predictions made today when no validating data is available. We intend to explore the many relationships observed in this paper to get a better handle on how to assign a confidence to each prediction made. For example, we expect that as PPI data accumulate with time, predictions will be more precise and have higher confidence; this hypothesis is supported by the literature validation rates above applied to predictions using recent PPI data. However, the more recent the PPI data, the smaller the duration between the date of the prediction, and the date of validation; thus, validation rates will {\em appear} lower simply due to the lack of passage of time. Untangling these effects in order to produce predictions with a reliable confidence level is an obvious direction for future research.

\section*{Methods}

\subsection*{Sequence similarity according to BLAST}
For all analyses other than those in Table \ref{tab:cilia}, we ran BLASTP locally with the default parameters to align all-to-all pairs of proteins between each species pair. Pairs of proteins were labeled as ``having sequence similarity according to BLAST'' if and only if BLASTP listed that pair anywhere in its output, otherwise not; the lowest observed bit score was 13.5, while E-values ranged from zero up to 1000. As a more sensitive test specifically applied to Table \ref{tab:cilia}, we visited \href{https://blast.ncbi.nlm.nih.gov/Blast.cgi?CMD=Web&PAGE=Proteins&PROGRAM=blastp&RUN_PSIBLAST=on#}{NCBI's PSI-BLAST page}, and for each row we entered the accession code for the mouse protein and used the {\tt PSI-BLAST} program choice. In all cases, many matches (dozens to hundreds) among human proteins were found with E-values ranging from 10 down to 1e-180, but in all cases we verified that none of those matches came from the protein in the Human column of Table \ref{tab:cilia}.

\subsection*{Formal definition of Pairwise Global Network Alignment}
Let $G_1,G_2$ be two undirected graphs (ie., networks), with node sets $V_1,V_2$ and edge sets $E_1,E_2$. Let $n_i=|V_i|,i=1,2$ be the number of nodes in the networks, and $m_i=|E_i|,i=1,2$ be the number of edges in each. Without loss of generality, assume $n_1\le n_2$. We define a {\em global} network alignment $a$ as a 1-to-1 function $a:V_1\rightarrow V_2$ that maps each node in $G_1$ to some node in $G_2$. (While the 1-to-1 requirement does not handle all biologically relevant cases, it is a widely adopted assumption; however, SANA's randomness effectively eliminates this restriction.) Figure \ref{fig:NetAlign} provides a simple schematic example of such a network alignment.

\subsection*{GO term prediction and automatic validation}
The following description applies only to automatic prediction and validation, not to manually literature-curated validations.

Assume we have two species $s_1,s_2$. Our goal is to use the PPI networks and GO annotations of both species available as of date $t$ to predict the existence of novel GO annotations not available at time $t$, and validate these predictions using GO terms available at some later date $t'>t$. Without loss of generality assume we are making predictions in the direction $s_1\rightarrow s_2$, ie., using GO annotations of proteins in $s_1$ to predict GO annotations of proteins in $s_2$. We refer to $s_1$ as the {\it source} species, and $s_2$ as the {\it target} species. In our case we are making predictions using networks and annotations available at $t=$ April 2010 (BioGRID 3.0.64 and GO release 2010-04, both available in April 2010), and validating those predictions using annotations available from the GO release at $t'=$ April 2020. The GO databases were retrieved from \href{ftp://ftp.ebi.ac.uk/pub/databases/GO/goa/old/UNIPROT}{the EMBL-EBI UNIPROT historical GO database}, which specifically focuses on protein (as opposed to gene) function.

Assume that on date $t$, species $s_1,s_2$ have PPI networks, $G_1,G_2$ with node sets $V_1,V_2$, and let $n_1=|V_1|, n_2=|V_2|$. Node sets consist of $V_1=\{p_i\}_{i=1}^{n_1},$ and $V_2=\{q_j\}_{j=1}^{n_2}$. For simplicity we will drop the node subscripts and refer to $p\in V_1$ and $q\in V_2$. Assume that on date $t$, $p$ is annotated with GO terms $\gamma_p$, and $q$ is annotated with GO terms $\gamma_q$. We will use the same letters for all entities at the later date $t'$, but with a prime added: for example $G'_1$ refers to the PPI network of $s_1$ at time $t'$, $p'$ refers to a protein in $V'_1$, and $\gamma_{p'}$ refers to the set of annotations to $p'$ at time $t'$. Note that $p'$ and $p$ are the same protein, but there exist proteins that may only exist in one of the two PPI networks, or one of the two GO annotation databases; thus, $p$ may exist in the PPI network at time $t$ but have no GO annotations at that time, or vice versa. (Note we do not include proteins with degree zero in our PPI networks, since they possess no useful topological information.)

We say that the association of GO term $g$ with protein $q'$ of the target species $s_2$ at time $t'$, sourced from any protein $p$ in $s_1$ at time $t$, is {\em predictable in principle} if both of the following are true:
\begin{itemize}
    \item $q\in V_2$---ie., the protein exists in the earlier PPI network of the target species $s_2$. This is because $q$ acquires annotations from proteins in the source species by being aligned to them at time $t$; $q$ cannot be aligned if it does not exist in $G_2$.
    \item $\exists p\in V_1$ such that $g\in \gamma_p$---ie., at least one protein from source species $s_1$ is annotated with $g$ at the earlier time. (Otherwise there is no place from which to source $g$ as a prediction for $q'$.)
\end{itemize}
We define $P_{12}$ as the set of all such {\it predictable in principle} annotations from species 1 to species 2; this set is derivable from information known only at the earlier time. Note, however, that its size is {\em huge}, being the product of the number of nodes in $s_2$ at time $t$ and the number of distinct GO terms annotating $s_1$ at time $t$.

Note that, although $q$ needs to be in the earlier network $V_2$, we do not demand that it exists in either of the GO term databases; those that exist in the later but not the earlier GO database, and for which we can make predictions, count as {\em completely unannotated} proteins at the earlier time, for which we may be able to make, and validate, predictions; those that also fail to exist in the later GO database may have predictions that are not yet, but may ultimately become, validated. Finally, we say that a predicted annotation $(v',g)$ is {\it validatable} if $g\in\gamma_{v'}$---that is, $g$ annotates $q'$ in the later GO database.

To measure recall, we need a maximal set of ``ground-truth'' annotations at the later date. The most obvious candidate ``ground truth'' is all GO annotations across all proteins in the target species at the later date, which we call $\Gamma'_2$. However, there are likely to exist annotations $(v',g)\in\Gamma'_2$ that are not {\it predictable in principle} as defined above, either because $g$ annotated no proteins in $V_1$, or because $q$ had no known interactions at time $t$ and thus did not exist in $G_2$. Thus, we define our maximal ``ground truth'' set as $P_{12}\cap \Gamma'_2$, and the number of elements in that set becomes the denominator in our computation of Recall.

We use AUPR rather than ROC curves because the data are {\em extremely} unbalanced: in particular, $|P_{12}|\gg |\Gamma'_2|$, directly informing us that the negative set is much larger than the positive one.
For example, in April 2010, the human BioGRID PPI network had 8192 nodes, and the other species listed above all had 3,000-10,000 GO terms, so $|P_{12}|$ is in the tens of millions, but the number of validating annotations for human in 2020 is less than 20,000, making the negative set about 1,000 times larger than the positive one.
%while the set of positives is
%$$ \mathbf{P} = \{(\gamma,v') | \gamma\in \gamma_{v'}, \;\; \forall v'\in V_2'\}.$$
%Thus, the negative set of predictions is $\mathbf{N=R\setminus P}$, where $\setminus$ means ``set subtraction''. The problem is that $|\mathbf{R}|$ is {\em huge}---it's the number of nodes in $G_2$ (thousands) multiplied by the entire set of GO terms observed across all proteins in $G_1$ (generally tens of thousands). The set of positives is miniscule in comparison (essentially only the second term above), so our dataset is extremely {\em imbalanced}---with negatives outnumbering positives by approximately a factor of $|V_2|$---making the ROC statistic inappropriate.
% {\bf I don't think we can do a ROC statistics, because our data set is heavily imbalanced: anything not predicted is a negative prediction, and there's an almost unlimited number of things we {\em could} predict but don't; and similarly, the number of {\em actual} negatives---ie., functions that a protein does {\em not} perform---is also huge. Thus, the ROC statistic is inappropriate for our dataset.} 

We make every attempt to eliminate any prediction that could have been made {\em or validated} using sequence analysis. In particular, we
\begin{itemize}
    \item eliminate any protein pairs $(p,q)$, regardless of NAF, which have sequence similarity according to BLAST (bit score threshold of 13), or those with known (even distant) orthology according to NCBI Homologene \cite{ncbi2016database}, InParanoid 8 \cite{sonnhammer2015inparanoid}, or the 2019 release of EggNog 5 \cite{huerta2019eggnog};
    \item eliminate any GO terms of $p$ possessing evidence codes from Table \ref{tab:BioGRID+NOSEQ}(bottom), {\em even if they also possess non-sequence-based evidence.}
    \item discard any ``predicted'' annotations that were already known at time $t$ between $q$ and GO terms with {\em any} evidence code (including those in Table \ref{tab:BioGRID+NOSEQ}(bottom));
    \item discard any predicted annotations for which sequence evidence had been produced by time $t'$.
\end{itemize}
We are left with predictions of GO terms annotating $q'$ that were entirely unknown at time $t$, that came from GO annotations of $p$ at time $t$ that did not possess {\em any} sequence-based evidence, and that {\em still} lacked sequence-based evidence as of time $t'$, even when including orthology based on the best homology methods of time $t'$. Note that for consistency, when we remove any predictions coming from a pair of proteins $(p,q)$ using the above criteria, we also remove the predictions from $P_{12}$ unless the same prediction can be sourced from another protein $\hat p$ in $s_1$ that is not eliminated based on the above criteria. (That is, we eliminate it from both the numerators and denominators of precision and recall.)

Using these criteria, we feel confident that the majority of (possibly all) predictions discussed in this paper were unattainable by any other means using data or methods available as of $t=$ April 2010, and additionally had still not been discovered by any sequence or homology based method available as of $t'$ = April 2020.

\section*{Additional Information}

\subsection*{Data Availability}
BioGRID networks are available from BioGRID.org; GO term releases are available at GeneOntology.org.
The output alignments, including alignment frequency, Resnik score, and paired proteins used to generate all Figures in the manuscript are available for the EC measures in the paper at \href{http://sana.ics.uci.edu/Topo-Function-2019-alignments-EC.7z}{http://sana.ics.uci.edu/Topo-Function-2019-alignments-EC.7z}, and for a longer list of objectives (many of which were used in our companion paper\cite{WeBeat}) at \href{http://sana.ics.uci.edu/Topo-Function-2019-alignments-all.7z}{http://sana.ics.uci.edu/Topo-Function-2019-alignments-all.7z}

\subsection*{Code Availability}
\href{https://GitHub.com/waynebhayes/SANA}{Source code for SANA is available at our GitHub repository};
the Resnik Python script, which uses the FastSemSim library \cite{guzzi2012semantic}, is part of the SANA repo.

\subsection*{Acknowledgements}
We thank Rishi Desai and William Longabaugh for creating the schematic depiction of network alignment in Figure \ref{fig:NetAlign}, Gary Bader and Brian Law of the University of Toronto for insightful comments that significantly improved our presentation, and Karen Christie of the Gene Ontology consortium for suggesting the use of cilia for making the predictions listed in Table \ref{tab:cilia}.

\subsection*{Author contributions}

SW conducted most computational experiments, including computing NAFs, Resnik scores and producing figures related to NAF-vs-Resnik scores. GRSA performed the detailed literature validation of TRIM25's predictions. Both worked under the direction of WBH, who conceived of the project and produced the $p$-value, precision-recall data and figures, as well as the GO predictions and cilia-related validations. All authors contributed to and approved the manuscript.

\subsection*{Competing interests}

The authors declare no \textbf{Competing interests}.

%The corresponding author is responsible for submitting a \href{http://www.nature.com/srep/policies/index.html#competing}{competing interests statement} on behalf of all authors of the paper. This statement must be included in the submitted article file.

\bibliography{wayne-new}

%\noindent LaTeX formats citations and references automatically using the bibliography records in your .bib file, which you can edit via the project menu. Use the cite command for an inline citation, e.g.

%For data citations of datasets uploaded to e.g. \emph{figshare}, please use the \verb|howpublished| option in the bib entry to specify the platform and the link, as in the \verb|Hao:gidmaps:2014| example in the sample bibliography file.

\ifanswers
\documentclass[fleqn,10pt]{wlscirep}
\usepackage[utf8]{inputenc}
\usepackage[T1]{fontenc}
\usepackage{pdflscape}

%%%%% Allowing cross-source-file references in Overleaf
\usepackage{xr}
\makeatletter
\newcommand*{\addFileDependency}[1]{% argument=file name and extension
  \typeout{(#1)}
  \@addtofilelist{#1}
  \IfFileExists{#1}{}{\typeout{No file #1.}}
}
\makeatother
\newcommand*{\myexternaldocument}[1]{%
    \externaldocument[#1:]{#1}%
    \addFileDependency{#1.tex}%
    \addFileDependency{#1.aux}%
}
\myexternaldocument{main}

\title{SUPPLEMENTARY MATERIAL for \\
SANA: Cross-Species Prediction of Gene Ontology GO Annotations via Topological Network Alignment}

\author[1]{Siyue Wang}
\author[1]{Giles R. S. Atkinson}
\author[1,*]{Wayne B. Hayes}
\affil[1]{Department of Computer Science, University of California, Irvine, CA 92697-3435, USA}

\affil[*]{whayes@uci.edu}

%\affil[+]{these authors contributed equally to this work}

%\keywords{Keyword1, Keyword2, Keyword3}

\begin{abstract}
We present supplementary material for the paper.
\end{abstract}

\begin{document}
%\flushbottom
\maketitle
\thispagestyle{empty}
%The main text (not including abstract, Methods, References and figure legends) is limited to 5,000 words. Figure legends are limited to 350 words each. As a guide, references should not exceed 70. Footnotes are not used.
%Display items: maximum about 1 per 500 words of text, max 10 for 5,000 word paper.

\fi

\newpage
\section*{Supplementary Info}

\subsubsection*{The effect of edge density and degree}
\begin{figure}
%     $\;\;\;\;\;\;\;\;\;\;\;\;\;\;\;\;\;\;\;\;\;\;\;\;\;\;\;$MMHS-EC
     %$\;\;\;\;\;\;\;\;\;\;\;\;\;\;\;\;\;\;\;\;\;\;\;\;\;\;\;\;\;\;\;\;\;\;\;\;\;\;\;$ graphlets
     %$\;\;\;\;\;\;\;\;\;\;\;\;\;\;\;\;\;\;\;\;\;\;\;\;\;\;\;\;\;\;\;\;\;\;\;\;\;\;\;\;\;\;\;\;\;\;\;$SCHS-EC
%     %$\;\;\;\;\;\;\;\;\;\;\;\;\;\;\;\;\;\;\;\;\;\;\;\;\;\;\;\;\;\;\;\;\;\;\;\;\;\;$ graphlets
\centering
%\rotatebox{90}{$\;\;\;\;\;\;\;$DM-HS}
%\includegraphics[width=0.19\linewidth]{Figures/sd-DMHS-ec-noseq2-averageDegree.pdf}
%\includegraphics[width=0.19\linewidth]{Figures/sd-DMHS-s3-noseq2-averageDegree.pdf}
%\includegraphics[width=0.19\linewidth]{Figures/sd-DMHS-importance-noseq2-averageDegree.pdf}
%\includegraphics[width=0.19\linewidth]{Figures/sd-DMHS-f100-graphlet-noseq2-averageDegree.pdf}
%\includegraphics[width=0.19\linewidth]{Figures/sd-DMHS-f100-lgraal-noseq2-averageDegree.pdf}\\
\rotatebox{90}{$\;\;\;\;\;\;\;$MM-HS-degree}
\includegraphics[width=0.40\linewidth]{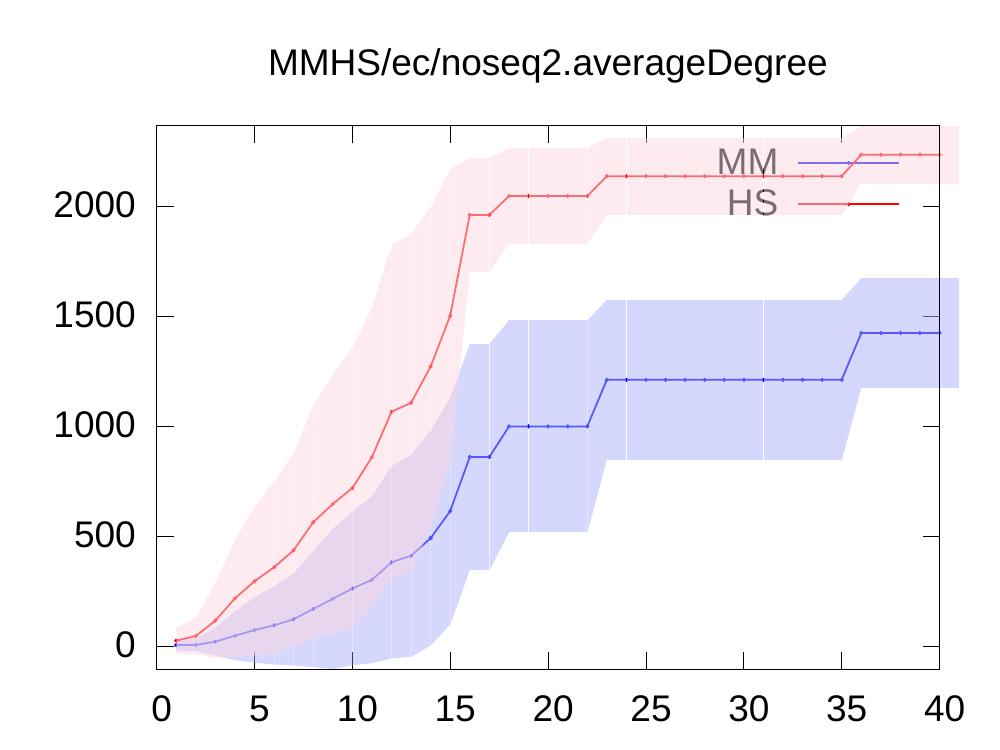}
%\includegraphics[width=0.19\linewidth]{Figures/sd-MMHS-s3-noseq2-averageDegree.pdf}
%\includegraphics[width=0.19\linewidth]{Figures/sd-MMHS-importance-noseq2-averageDegree.pdf}
%\includegraphics[width=0.23\linewidth]{Figures/sd-MMHS-f100-graphlet-noseq2-averageDegree.pdf}
%\includegraphics[width=0.19\linewidth]{Figures/sd-MMHS-f100-lgraal-noseq2-averageDegree.pdf}\\
%\rotatebox{90}{$\;\;\;\;\;\;\;$RN-HS}
%\includegraphics[width=0.19\linewidth]{Figures/sd-RNHS-ec-noseq2-averageDegree.pdf}
%\includegraphics[width=0.19\linewidth]{Figures/sd-RNHS-s3-noseq2-averageDegree.pdf}
%\includegraphics[width=0.19\linewidth]{Figures/sd-RNHS-importance-noseq2-averageDegree.pdf}
%\includegraphics[width=0.19\linewidth]{Figures/sd-RNHS-f100-graphlet-noseq2-averageDegree.pdf}
%\includegraphics[width=0.19\linewidth]{Figures/sd-RNHS-f100-lgraal-noseq2-averageDegree.pdf}\\
\rotatebox{90}{$\;\;\;\;\;\;\;$SC-HS-degree}
\includegraphics[width=0.40\linewidth]{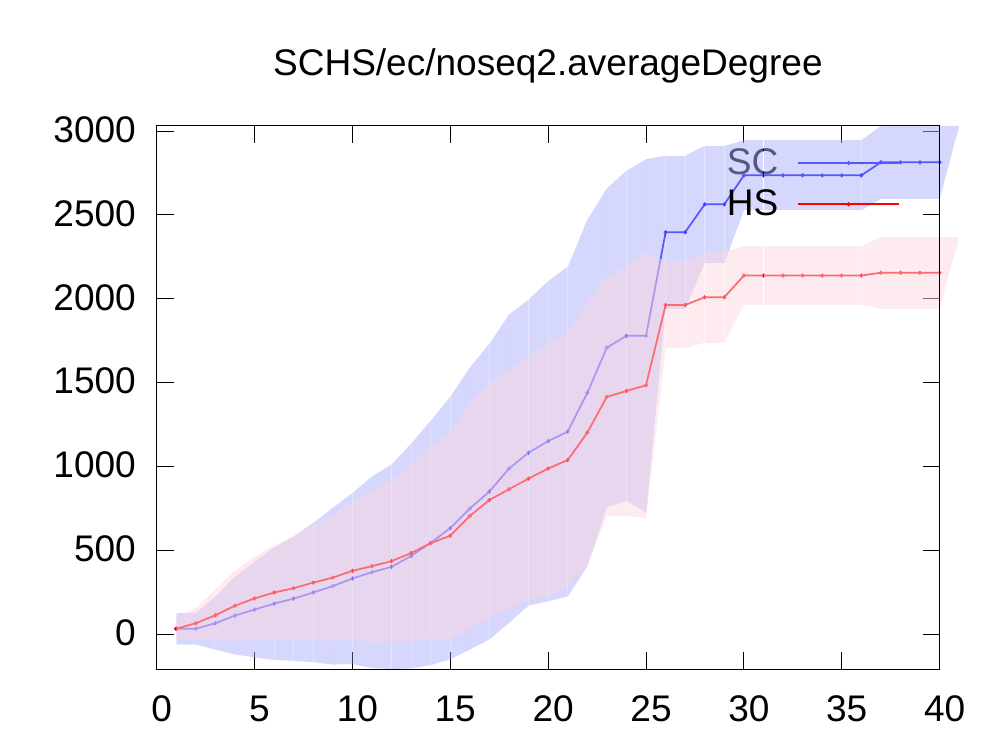}
\caption{{\bf Mean and standard deviation of degree of aligned protein pairs, as a function of NAF (\%)}, for the same parameters and species as Figure (main manuscript) \ref{fig:Resnik-vs-NAF+Pearsons}. In every plot, red is human, and blue is the other species.
We observe that nodes with higher degree in the original networks tend to be more frequently aligned. The depicted mouse-human Pearson correlation is 0.92, while yeast's is 0.65; both $p$-values are below $10^{-300}$.
%However, note that the edge-based measures are only able to ``correctly'' align nodes with degrees in the thousands, even if the actual degrees differ by up to a factor of 3--8 (for example, in MM-HS, all three of EC, $S^3$, and Importance are able to align nodes with very high semantic similarity but degrees 2,500 and 300).%, while the graphlet-based measures seem to work better on more closely-related species, and in addition are able to ``correctly'' align nodes with much smaller degree.
}
\label{fig:degree-vs-frequency-noSEQ}
\end{figure}

We expect that edge density strongly affect alignment robustness, simply because the more edges we have, the more topological information we have to align similar regions\cite{WeBeat}. We can quantify this at the node level by observing how node degree correlates with NAF.
Figure \ref{fig:degree-vs-frequency-noSEQ} depicts the relationship between node degree and NAF. We can clearly see that higher degree nodes are more reliably aligned than lower degree ones. %We also see that the degrees of aligned pairs are {\em much} higher for the edge-based measures than for the graphlet-based ones; we suspect this is because the graphlet-based measures are aligning regions of lower density with almost perfectly-matching topology, whereas the edge-based measures are able to “cope” with greater diversity of degrees—a hypothesis supported by the observation here that, for the edge-based measures, degrees of aligned nodes can differ by factors of 3--8, while for graphlet-based measures the degree curves are much closer together.
%We also note that, for the edge-based measures, the mouse-human pair have both (a) the highest Resnik scores (cf. Figures (main manuscript) \ref{fig:Resnik-vs-NAF+Pearsons}[top],(main manuscript) \ref{fig:Resnik-vs-NAF+Pearsons}[bottom]) and (b)  the highest degrees (Figure 6). This is consistent with Table 1, where we see that human and mouse are the top 2 mammals in terms of number of edges, far outpacing rat.

\begin{figure}
\centering
    $\;\;\;\;$
    BP $\;\;\;\;\;\;\;\;\;\;\;\;\;\;\;\;\;\;\;\;\;\;\;\;\;$
    CC $\;\;\;\;\;\;\;\;\;\;\;\;\;\;\;\;\;\;\;\;\;\;\;\;\;$
    MF $\;\;\;\;\;\;\;\;\;\;\;\;\;\;\;\;\;\;\;\;\;\;\;\;\;$
    $\;\;\;\;$
    BP $\;\;\;\;\;\;\;\;\;\;\;\;\;\;\;\;\;\;\;\;\;\;\;\;\;$
    CC $\;\;\;\;\;\;\;\;\;\;\;\;\;\;\;\;\;\;\;\;\;\;\;\;\;$
    MF
     \\
%\rotatebox{90}{\footnotesize DMHS-edges}
%\includegraphics[width=.15\linewidth]{Figures/DMHS-ecs3imp-seq2-netGO-resnik-BP-allGO.pdf}
%\includegraphics[width=.15\linewidth]{Figures/DMHS-ecs3imp-seq2-netGO-resnik-CC-allGO.pdf}
%\includegraphics[width=.15\linewidth]{Figures/DMHS-ecs3imp-seq2-netGO-resnik-MF-allGO.pdf}
\rotatebox{90}{$\;\;\;\;\;\;\;$MM-HS}
\includegraphics[width=.14\linewidth]{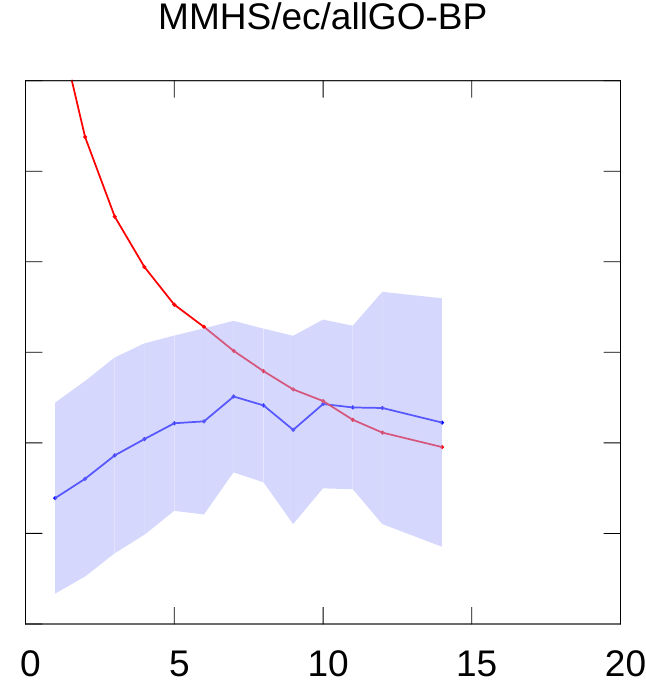}
\includegraphics[width=.14\linewidth]{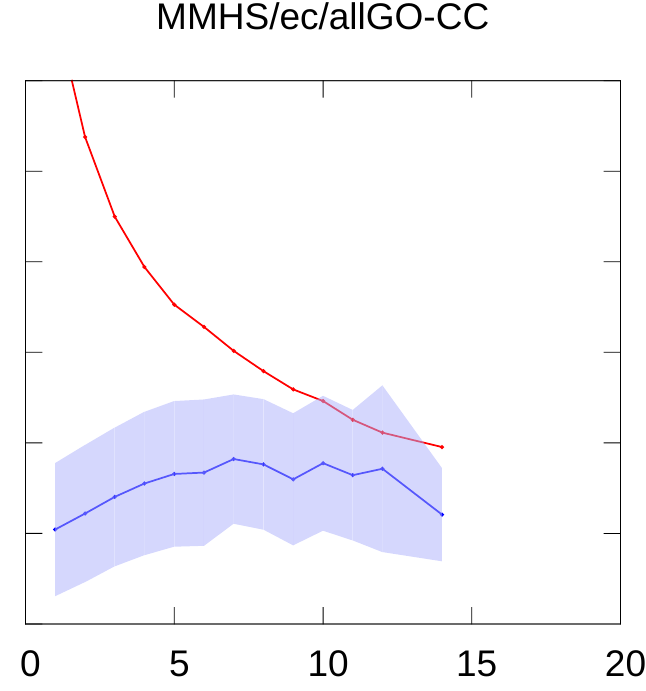}
\includegraphics[width=.14\linewidth]{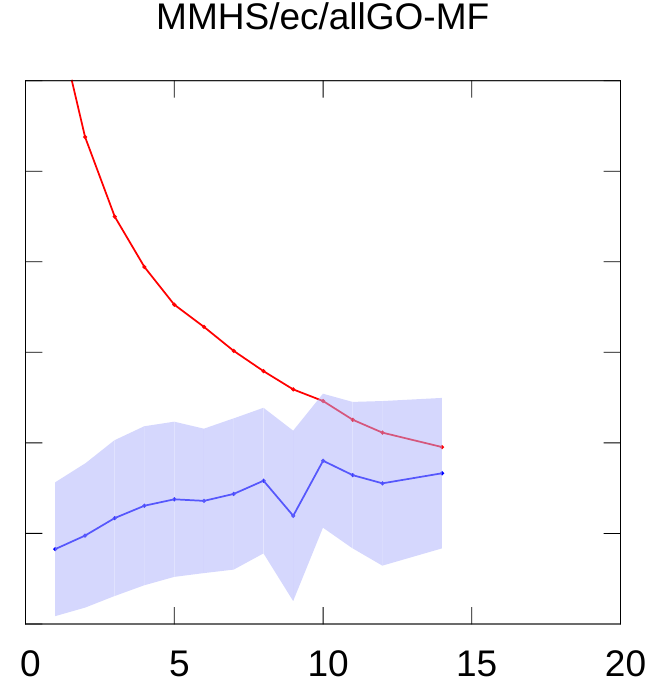}
\rotatebox{90}{$\;\;\;\;\;\;\;$SC-HS}
\includegraphics[width=.14\linewidth]{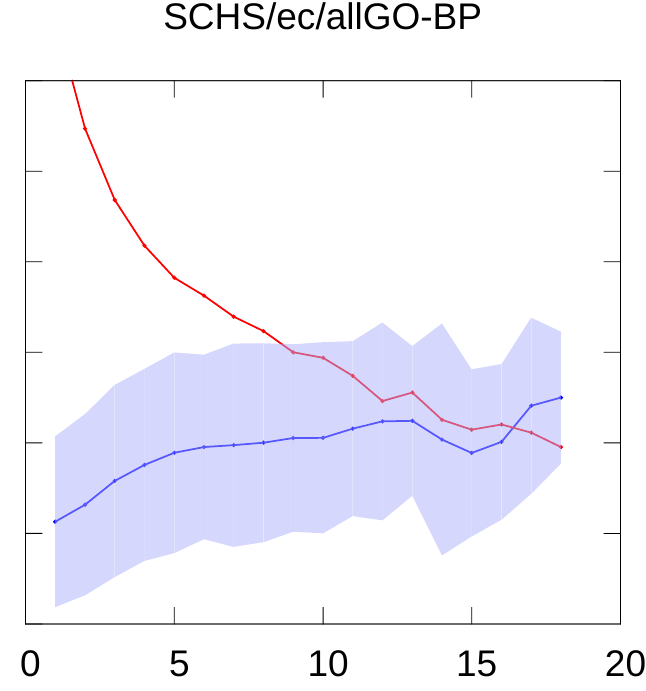}
\includegraphics[width=.14\linewidth]{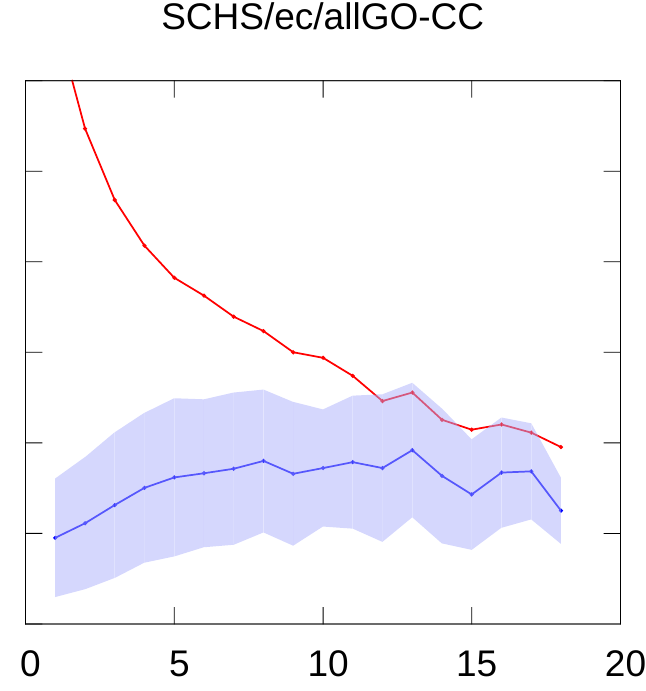}
\includegraphics[width=.14\linewidth]{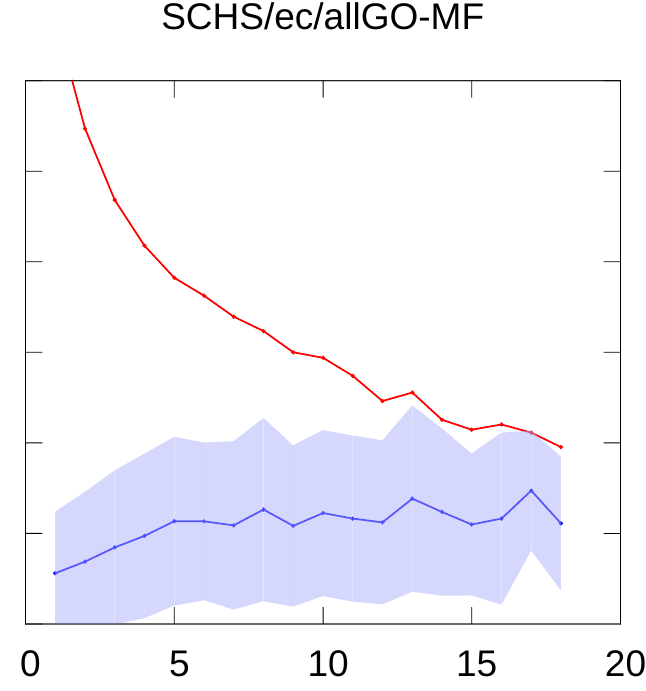}
\caption{
    {\bf Splitting GO into Biological Process, Cellular Component, and Molecular Function:} We plot mean (blue line) and standard deviation (blue shade) of Resnik scores vs. Network Alignment Frequency (NAF\%), split into the three GO Categories. The left three panels depict mouse-human, and the right three are yeast-human. All alignments are driven to optimize EC. We use the same axes as Figure (main manuscript) \ref{fig:Resnik-vs-NAF+Pearsons}. To reduce clutter, we only plot the version corresponding to Figure (main manuscript) \ref{fig:Resnik-vs-NAF+Pearsons}'s second sub-plot (all aligned pairs, all GO terms). The relationships all have Pearson correlations in the range $[0.05,0.10]$ and $p$-values below $10^{-30}$.
}
\label{fig:Resnik-vs-NAF-BP-CC-MF}
\end{figure}

One of the referees pointed out that since both Resnik and mean degree correlate with NAF, there is the possibility that NAF simply aligns node pairs with high degree; since both degree and GO annotation level are correlated with ``popularity'' of proteins in the research literature \cite{luck2017proteome}, increased popularity would result in both higher node degree and more GO annotations, potentially contributing to the Resnik-NAF correlations observed in Figures (main manuscript) \ref{fig:Resnik-vs-NAF+Pearsons} and \ref{fig:Resnik-vs-NAF-BP-CC-MF}.
This hypothesis can be tested in at least three ways.
First, assume we list {\em all} pairs of proteins $p\in G_1,q\in G_2$. %If $G_1$ and $G_2$ have $n_1$ and $n_2$ nodes, respectively, the number of possible pairs is $n_1n_2$. This is {\em far} larger than number of protein pairs across all of our alignments (eg., $n_1n_2\approx 10^8$ for mouse-human, which is 182 times as many pairs as occur across all 100 of our mouse-human alignments). Assume we
Then sort the list of pairs by degree---for example sorted by the arithmetic or geometric mean degree of the two, or by the minimum or maximum degree of the two. We have computed the mean Resnik similarity across all mouse-human protein pairs, and indeed, we find an {\em enormously} powerful correlation between degree and Resnik: degree-1 node pairs have mean Resnik scores of about 0.5, which rises dramatically up to a Resnik value of about 3.5 at degree 30. Unfortunately, this trend halts abruptly at that point: there are over 600,000 pairs of mouse-human proteins in which {\em both} have degrees above 30, which is higher by many orders of magnitude than the number of high-Resnik pairs in Figures (main manuscript) \ref{fig:Resnik-vs-NAF+Pearsons} and \ref{fig:Resnik-vs-NAF-BP-CC-MF}; however, their mean Resnik score remains constant at about 3.5 for all degree thresholds above 30. Thus, degree alone cannot be responsible for the results of Figures (main manuscript) \ref{fig:Resnik-vs-NAF+Pearsons} and \ref{fig:Resnik-vs-NAF-BP-CC-MF}.
A second argument against the Resnik-degree hypothesis is by our companion paper\cite{WeBeat}, which clearly shows that our alignments recover orthologous protein pairs---essentially the strongest definition of a ``correctly'' aligned pair between species---at a rate {\em far} higher than random, which again eliminates high degree as the sole cause of good alignments (though of course they {\em enable} good alignments by providing more information). Third, many of the correctly recovered orthologs in our companion paper\cite{WeBeat} do not have particularly high node degree: there were 16 orthologs in which the mouse protein has a degree less than the median, 10 for which the human one does, and 7 for which both do, for a grand total of 19 instances of correctly aligned orthologs with degree below the global median. Since a randomly chosen protein has a 50\% probability of having degree below the median, these low-degree ortholog alignments have a collective probability of $2^{-19}$ (one in half a million). Again we conclude that higher degree {\em enables} good alignments but does not, in and of itself, cause them.

%We conclude that NAF is genuinely matching protein pairs with high Resnik similarity. In other words, Figure \ref{fig:degree-vs-frequency-noSEQ} shows that this is easier to do when both nodes have higher degree, the higher degree {\em alone} provides no predictive value in and of itself.

%The conclusion is simply that the higher the edge density, the easier it becomes to detect the common interaction topology that arises from semantic similarity; and conversely, the greater our ability to {\em predict} semantic similarity based on similar interaction topology as detected by NAF.

\begin{landscape}
\begin{table}
\small
\centering
\begin{tabular}{|cr|crr|crr|crr|crr|crr|}
\hline
&&\multicolumn{3}{c|}{EC} &\multicolumn{3}{c|}{$S^3$} &\multicolumn{3}{c|}{Importance} &\multicolumn{3}{c|}{graphlet} &\multicolumn{3}{c|}{lgraal} \\
species&$\max(\overline{D})$ &NAF&nodes&$\overline{D}$ &NAF&nodes&$\overline{D}$ &NAF&nodes&$\overline{D}$ &NAF&nodes&$\overline{D}$ &NAF&nodes&$\overline{D}$ \\
\hline
SC-DM&  28.99&  5& 874& 28.84&  5& 870& 28.99&  5& 774& 28.16&  4& 21135& 0.14& 4& 20466& 0.15 \\
SC-HS&  14.10&  4& 1291& 12.91& 4& 1186& 14.10& 4& 1235& 12.84& 5& 18504& 0.28& 4& 19593& 0.26 \\
MM-DM&  12.70&  7& 101& 10.93&  6& 89& 11.26&   8& 46& 12.70&   19& 453& 0.44&  11& 757& 0.38 \\
SP-SC&  12.58&  3& 2116& 7.96&  6& 67& 9.64&    9& 24& 12.58&   80& 287& 0.08&  4& 3525& 0.20 \\
SP-DM&  12.23&  8& 108& 11.57&  8& 60& 12.23&   8& 63& 12.19&   7& 2843& 0.30&  12& 2171& 0.14 \\
CE-SC&  11.38&  4& 1093& 11.38& 4& 348& 8.30&   5& 219& 10.09&  90& 430& 0.04&  5& 11027& 0.04 \\
AT-HS&  10.25&  4& 879& 8.56&   4& 501& 8.89&   5& 340& 10.25&  19& 1872& 0.52& 13& 2631& 0.32 \\
AT-DM&  10.23&  4& 831& 7.85&   3& 1421& 6.63&  9& 69& 10.23&   30& 1281& 0.32& 12& 2922& 0.18 \\
CE-HS&  9.77&   4& 808& 9.04&   5& 272& 9.77&   5& 312& 9.26&   30& 2147& 0.27& 13& 4438& 0.28 \\
MM-CE&  9.48&   6& 313& 6.93&   10& 112& 9.45&  9& 142& 9.48&   12& 649& 0.68&  17& 513& 0.75 \\
MM-SP&  9.33&   7& 246& 9.33&   9& 94& 7.98&    8& 117& 8.34&   13& 638& 0.53&  6& 1554& 0.50 \\
SP-AT&  9.22&   7& 1022& 6.91&  6& 978& 8.01&   7& 704& 9.22&   13& 1920& 0.85& 12& 2096& 0.54 \\
AT-SC&  9.15&   4& 896& 9.15&   4& 565& 8.00&   4& 514& 8.21&   30& 1027& 0.08& 4& 6094& 0.14 \\
DM-HS&  9.11&   3& 1696& 5.33&  3& 1548& 5.00&  4& 690& 9.11&   8& 21033& 0.23& 5& 27210& 0.36 \\
AT-CE&  8.32&   6& 786& 6.79&   7& 513& 7.52&   11& 194& 8.32&  12& 2773& 0.54& 7& 4988& 0.39 \\
MM-SC&  7.73&   3& 388& 2.84&   3& 337& 5.49&   5& 60& 7.73&    90& 47& 0.09&   40& 180& 0.06 \\
SP-CE&  7.52&   7& 438& 7.32&   12& 84& 7.52&   9& 137& 7.30&   13& 1953& 0.49& 4& 4075& 0.37 \\
CE-DM&  6.23&   4& 538& 5.80&   4& 374& 5.14&   5& 175& 6.23&   9& 5962& 0.26&  15& 3811& 0.31 \\
SP-HS&  5.62&   3& 1991& 5.38&  4& 188& 5.31&   7& 47& 5.62&    15& 1634& 0.41& 15& 1700& 0.23 \\
MM-HS&  5.02&   3& 533& 3.67&   4& 139& 5.02&   4& 157& 4.47&   13& 687& 0.93&  14& 609& 0.51 \\
RN-CE&  4.78&   6& 71& 3.77&    13& 23& 4.78&   17& 15& 4.67&   20& 146& 0.81&  12& 245& 0.69 \\
MM-AT&  4.47&   4& 927& 4.47&   20& 19& 4.21&   5& 497& 3.81&   8& 1137& 1.18&  11& 763& 0.60 \\
RN-SP&  4.33&   7& 24& 4.33&    4& 297& 2.34&   5& 145& 2.07&   30& 118& 0.66&  40& 87& 0.55 \\
RN-DM&  3.93&   4& 61& 3.93&    3& 152& 2.54&   4& 63& 2.83&    17& 203& 0.60&  13& 245& 0.57 \\
RN-AT&  3.12&   4& 191& 3.12&   3& 663& 1.80&   3& 665& 1.96&   30& 104& 0.90&  11& 273& 0.70 \\
RN-MM&  3.00&   6& 471& 3.00&   9& 114& 2.32&   8& 188& 2.41&   14& 6& 1.33&    16& 206& 0.87 \\
RN-SC&  1.32&   2& 1116& 1.32&  2& 745& 0.99&   2& 714& 1.02&   100& 26& 0.08&  50& 56& 0.07 \\
RN-HS&  1.15&   2& 936& 1.14&   2& 661& 1.13&   3& 96& 1.15&    30& 122& 0.85&  12& 258& 0.53 \\
    \hline
    \end{tabular}
    \caption{Mean degree of the {\it Common Connected Subgraph} (CCS, cf. purple edges in Figure (main manuscript) \ref{fig:NetAlign}) induced on the aligned node pairs with NAF above the threshold in the NAF column. For each species pair and measure, we list only the set of nodes with the highest mean degree and the value of NAF that gave it. The table includes all ${8 \choose 2}=28$ species pairs from Table (main manuscript) \ref{tab:BioGRID+NOSEQ}(top), and is sorted by the first column (the maximum mean degree across all 5 measures). Note that RN-HS (rat-human) has the lowest mean degree across all 28 pairs.}
    \label{tab:CCS-mean-degree}
\end{table}
\end{landscape}

\begin{figure}
         $\;\;\;\;\;\;\;\;\;\;\;\;\;\;\;\;\;\;\;\;\;\;\;\;$
     (a) $\;\;\;\;\;\;\;\;\;\;\;\;\;\;\;\;\;\;\;\;\;\;\;\;\;\;\;\;\;\;\;\;\;\;\;\;\;\;\;\;\;\;\;\;\;\;\;\;\;\;$
     (b) $\;\;\;\;\;\;\;\;\;\;\;\;\;\;\;\;\;\;\;\;\;\;\;\;\;\;\;\;\;\;\;\;\;\;\;\;\;\;\;\;\;\;\;\;\;\;\;\;\;\;$
     (c) $\;\;\;\;\;\;\;\;\;\;\;\;\;\;\;\;\;\;\;\;\;\;\;\;\;\;\;\;\;\;\;\;\;\;\;\;\;\;\;\;\;\;\;\;\;\;\;\;\;\;$
     (d)
     
    \includegraphics[width = 0.24\linewidth]{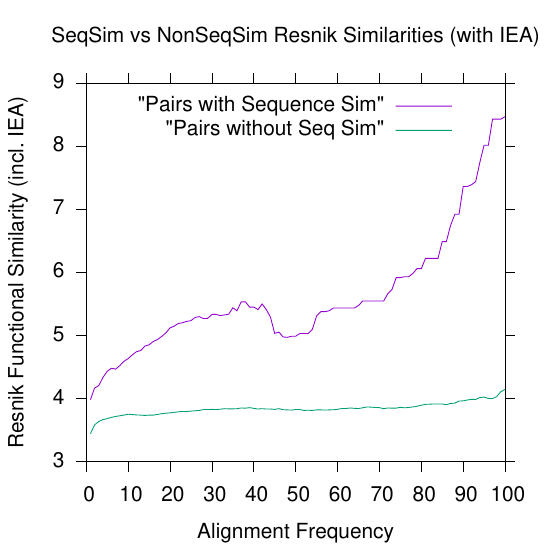}
    \includegraphics[width = 0.24\linewidth]{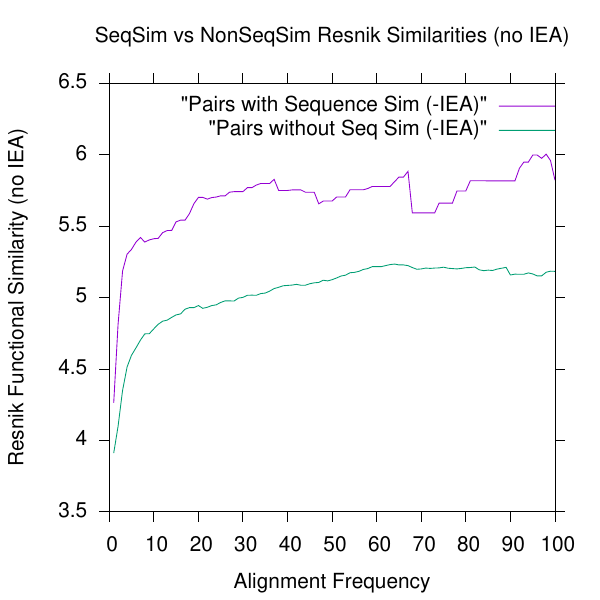}
    \includegraphics[width = 0.24\linewidth]{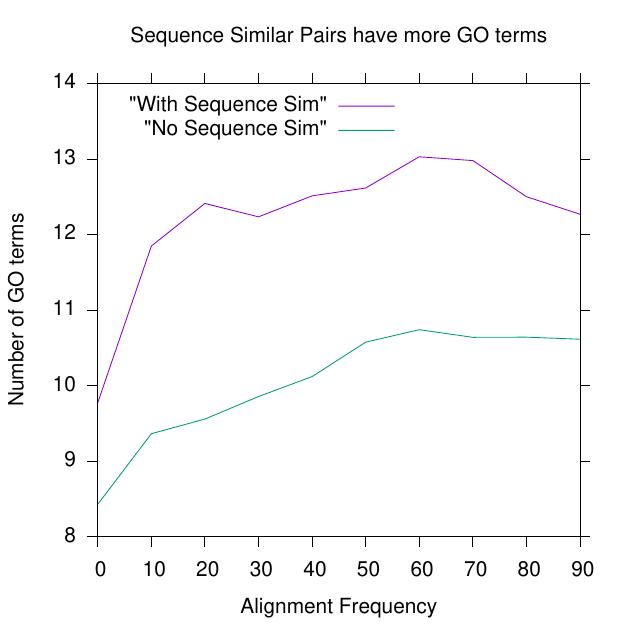}
    \includegraphics[width = 0.25\linewidth]{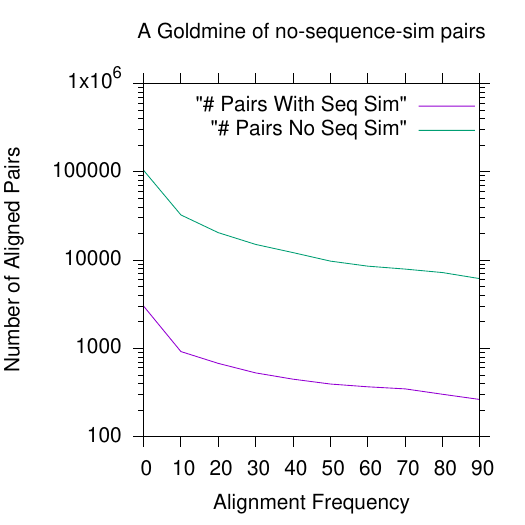}
    \caption{
    {\bf The effect of BLAST scores and sequence-based GO terms}
    A closer study of the allGO vs. NOSEQ aspect of Figure (main manuscript) \ref{fig:Resnik-vs-NAF+Pearsons}.
    We use purple curves to depict aligned pairs with {\em any} sequence similarity, and green for those that do not, both
    according to BLAST.
    {\bf (a)}: We re-plot on the same graph the mean mouse-human Resnik scores, with core sizes and
    standard deviations removed for clarity. Clearly the scores are far better for pairs with sequence similarity.
    {\bf (b): Removing sequence from the evaluation entirely:} Same as (a) but we remove {\em all} GO evidence codes
    based on sequence---including human curated ones: IEA, ISS, ISO, ISA, ISM, IGC, and RCA were all removed.
    Though diminished significantly, the effect persists.
    {\bf (c): Mean number of experimental GO terms for pairs with and without sequence similarity}
    We find that pairs with sequence similarity have more {\em experimental} GO terms. This is likely a social / ``popularity'' bias similar to the known bias in PPI edge selection towards ``interesting'' proteins\cite{rolland2014proteome,luck2017proteome,luck2019reference}. For example, PIs may intentionally repeat the same experiment on known ortholog for confirmation/validation, or observe someone else's experiment on protein p in species X, and attempt to repeat it on a ortholog of p in species Y.
    {\bf (d): Non-sequence-similar pairs dominate:} At a given network alignment Frequency, NAF produces far more protein pairs without
    sequence similarity than with. Finally, we note that this Figure is not in conflict with Figure (main manuscript) \ref{fig:Resnik-Know} since, in the latter, we have {\em forced} a comparison betwneen protein pairs with equal annotation levels (ie., the disparity in part (c) of this Figure has been removed).
}
\label{fig:seq-vs-noseq-GOk}
\end{figure}

\begin{table}
    \centering \small
\begin{tabular}{|r|r|r|l|l|l|r|r|r|r|}
\hline
rank & $F^*$  & NAF & pair& $M$ &Category&$|P_{12}\cap\Gamma'_2|$  & pred & valid & Precision \\
\hline
1&      0.413&  8\%&      DM-HS&   $S^3$&     Func&   3510&   3207&   1386&   43.2\%\\
2&      0.410&  8\%&      DM-HS&   Import.&   Func&   3510&   3248&   1387&   42.7\%\\
3&      0.403&  2\%&      CE-HS&   $S^3$&     Func&   4572&   3593&   1644&   45.8\%\\
4&      0.401&  3\%&      CE-HS&   Import.&   Func&   2743&   2351&   1021&   43.4\%\\
5&      0.400&  7\%&      DM-HS&   EC&     Func&   3510&   3231&   1348&   41.7\%\\
6&      0.400&  16\%&     SC-HS&   $S^3$&     Func&   3458&   3367&   1364&   40.5\%\\
7&      0.398&  16\%&     SC-HS&   Import.&   Func&   3458&   3479&   1382&   39.7\%\\
8&      0.396&  16\%&     SP-AT&   Import.&   Func&   480&    645&    223&    34.6\%\\
9&      0.384&  16\%&     SP-AT&   $S^3$&     Func&   480&    656&    218&    33.2\%\\
10&     0.336&  16\%&     SC-HS&   EC&     Func&   3458&   2752&   1043&   37.9\%\\
11&     0.336&  16\%&     SP-AT&   EC&     Func&   480&    699&    198&    28.3\%\\
12&     0.330&  16\%&     MM-AT&   Import.&   Func&   558&    539&    181&    33.6\%\\
13&     0.321&  16\%&     MM-AT&   $S^3$&     Func&   558&    538&    176&    32.7\%\\
14&     0.278&  3\%&      AT-HS&   $S^3$&     Func&   5196&   4455&   1341&   30.1\%\\
15&     0.276&  3\%&      AT-HS&   Import.&   Func&   5196&   4404&   1326&   30.1\%\\
16&     0.276&  2\%&      CE-HS&   EC&     Func&   3041&   2691&   791&    29.4\%\\
17&     0.266&  3\%&      SP-HS&   Import.&   Func&   2872&   3358&   830&    24.7\%\\
18&     0.264&  3\%&      SP-HS&   $S^3$&     Func&   2872&   3438&   833&    24.2\%\\
19&     0.234&  32\%&     SP-AT&   Import.&   Comp&   1718&   3439&   604&    17.6\%\\
20&     0.226&  32\%&     SP-AT&   $S^3$&     Comp&   1718&   3427&   582&    17.0\%\\
\hline
\end{tabular}
    \caption{{\bf 2010-based predictions, by GO Category.} Similar to Table (main manuscript) \ref{tab:Fstar} but for Categories.
    }
    \label{tab:Category}
\end{table}

\subsubsection*{Sequence-similar pairs have more {\em non}-sequence evidence}
\begin{figure}
    \centering
    %\vspace{11cm}
    %\includegraphics[width=0.5 \textwidth]{Supp-Figures/prot-GOs-Resnik-all.pdf}
    \includegraphics[width=0.9 \textwidth]{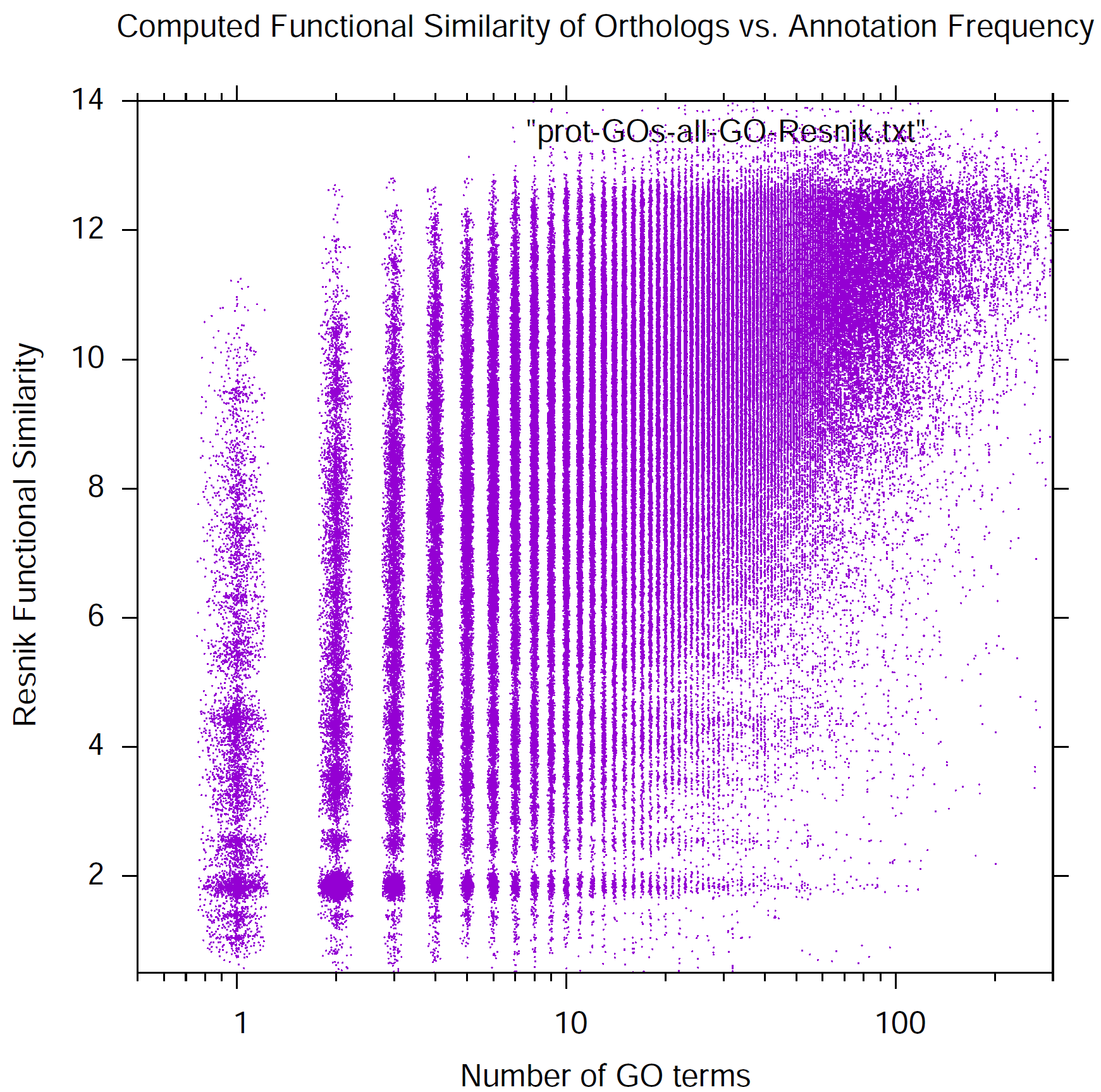}
    \caption{{\bf Computed Resnik similarity between orthologs, as a function of GO annotation count} Each dot represents a pair of orthologous proteins between pairs of BioGRID species from Table (main manuscript) \ref{tab:BioGRID+NOSEQ}(top).
    The vertical axis is the computed Resnik semantic similarity between a pair of orthologs, while the horizontal axis is the GO term count of whichever of the pair has fewer GO annotations. Since they are orthologs, we expect them usually to have high Resnik similarity.
    We see this is true as long as both proteins are well-annotated, but fails when at least one is poorly annotated. (A small amount of random ``jitter'' has been added in both the horizontal and vertical directions to more clearly depict the density of points across the surface.)}
    \label{fig:ortholog-Resnik-vs-GOcount}
\end{figure}

Supplementary Figure \ref{fig:seq-vs-noseq-GOk} provides a more in-depth analysis of the difference observed between the ``allGO'' vs. ``NOSEQ'' parts of Figure (main manuscript) \ref{fig:Resnik-vs-NAF+Pearsons}.
Part (a) plots the curves of mouse-human mean NetGO-weighted mean Resnik scores as a function of frequency, with core sizes and standard deviations removed for clarity. We observe more clearly the stark difference
between scores of pairs with and without sequence.
The most obvious possible explanations are (i) pairs with sequence similarity are genuinely closer
in Resnik similarity than those without, or (ii) the two sets have comparable Resnik similarity but the computed Resnik score is higher for those with sequence similarity simply because there is more information available for such pairs.
Part (b) of the figure tries to resolve the ambiguity by removing {\em all} sequence-based GO terms, including those
that have been human curated---because human curation does not change the fact that nobody would even have {\em looked}
at the possibility of functional or semantic similarity unless an automated method first uses sequence to suggest the possibility.
Surprisingly, even after removing all sequence-based GO terms, sequence-similar pairs {\em still} have a slight edge
in mean Resnik similarity.
In part (c), we find a potential explanation: we find that pairs of proteins that exhibit sequence similarity have, on
average, two more {\em experimental} GO terms than pairs that do not have sequence similarity. Observe from Figure \ref{fig:ortholog-Resnik-vs-GOcount}
that the mean {\em pairwise} Resnik score tends to increase with number of annotations; we hypothesize that perhaps there is a publication
bias in GO terms similar to the publication bias that results in PPI network edges being biased towards pairs of highly
studied proteins\cite{rolland2014proteome,VidalY2H01}: perhaps protein pairs that share sequence similarity tend to have more experimental tests for
functionality than those that do not. In any case, we have at the very least demonstrated that the difference between
``allGO'' and ``NOSEQ'' {\em vastly} diminished if we remove sequence-related GO terms. Though not entirely satisfactory,
we think it is sufficient grounds to continue to our next point, which is Part (d) of Figure \ref{fig:seq-vs-noseq-GOk},
which is to point out that there are vastly more non-sequence-similar pairs at a given NAF alignment frequency than there
are sequence-similar pairs.
Circling back to part (a),
recall that the separation of the green and purple curves was done after-the-fact---the {\em only} independent variable
was network alignment frequency. Thus, since FastSemSim cannot use sequence similarity that doesn't exist, and if we trust the
high Resnik similarity computed for those pairs that {\em do} share sequence similarity, it follows that all pairs
at a given network alignment frequency should be drawn from the same distribution of Resnik similarity, and so NAF has the
potential to fundamentally alter the landscape of protein functional predictions.

\begin{figure*}
     $\;\;\;\;\;\;\;\;\;\;\;\;\;\;\;\;\;\;\;\;\;$EC-seq
     $\;\;\;\;\;\;\;\;\;\;\;\;\;\;\;\;\;\;\;$EC-allGO
     $\;\;\;\;\;\;\;\;\;\;\;\;\;\;\;$EC-NOSEQ
     $\;\;\;\;\;\;\;\;\;\;\;\;\;\;\;\;\;\;\;$graphlets-seq
     $\;\;\;\;\;\;\;\;\;$graphlets-allGO
     $\;\;\;\;\;\;$graphlets-NOSEQ$\;\;\;\;\;\;\;$
     
\centering
\rotatebox{90}{$\;\;\;\;\;\;\;$MM-HS}
\includegraphics[width=0.155\linewidth]{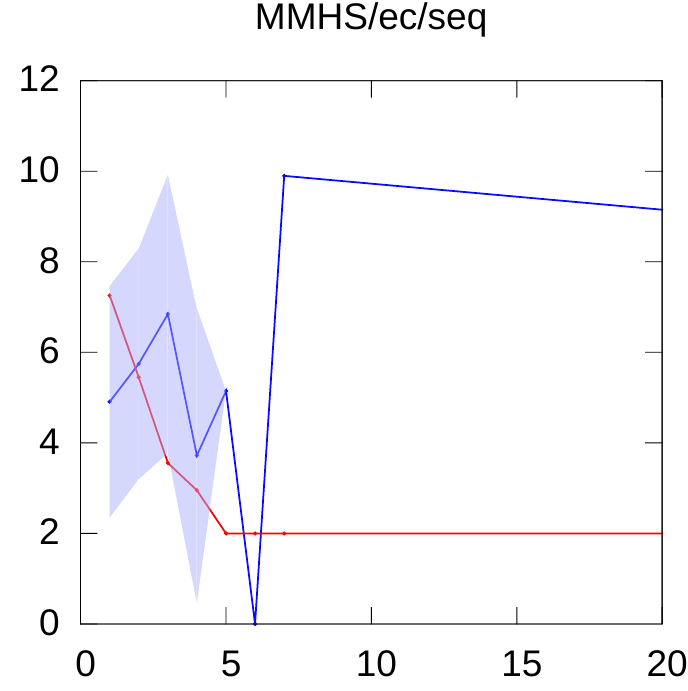}
\includegraphics[width=0.145\linewidth]{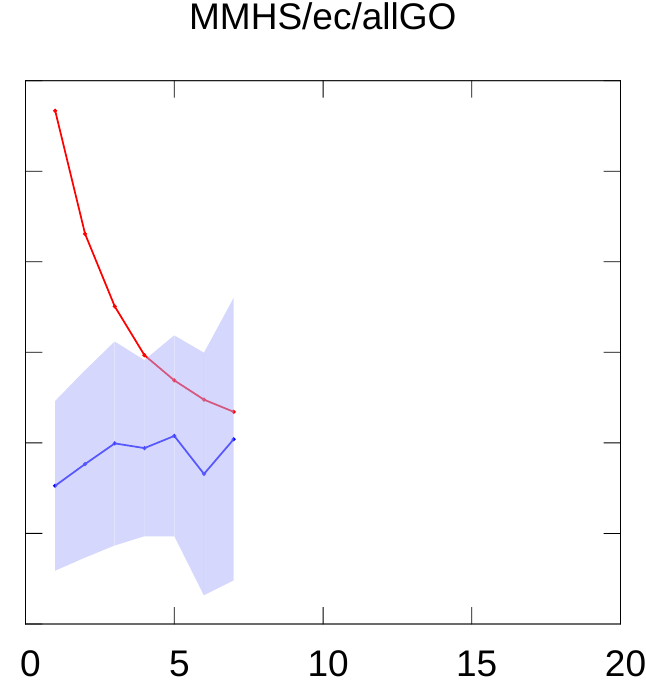}
\includegraphics[width=0.17\linewidth]{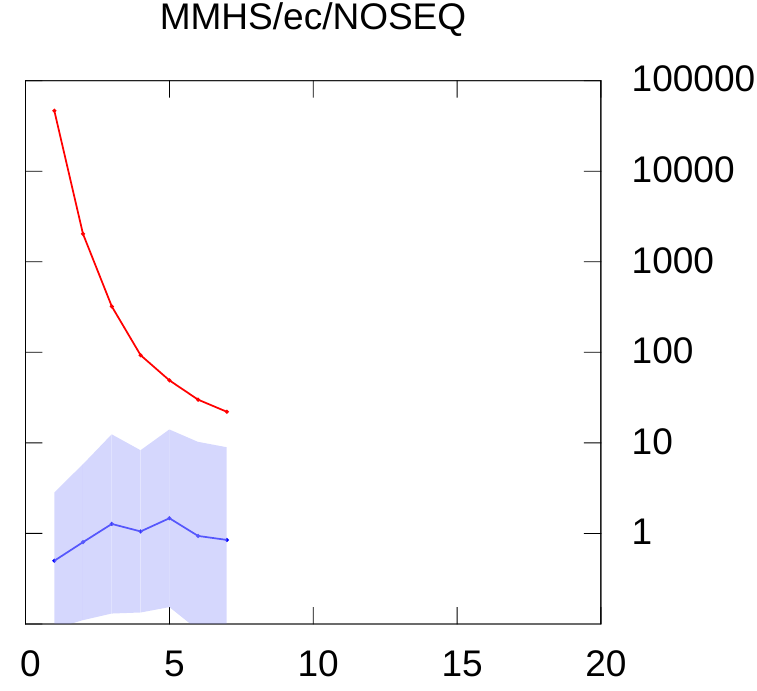}
\includegraphics[width=0.155\linewidth]{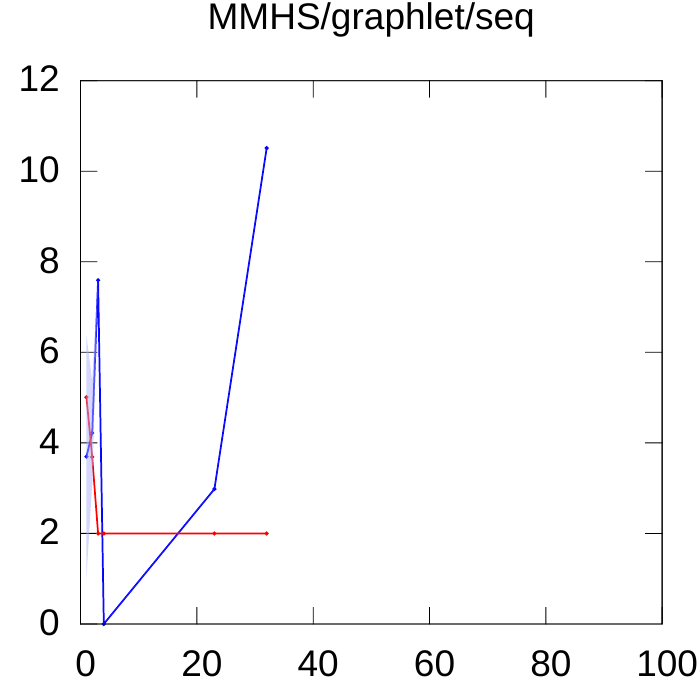}
\includegraphics[width=0.145\linewidth]{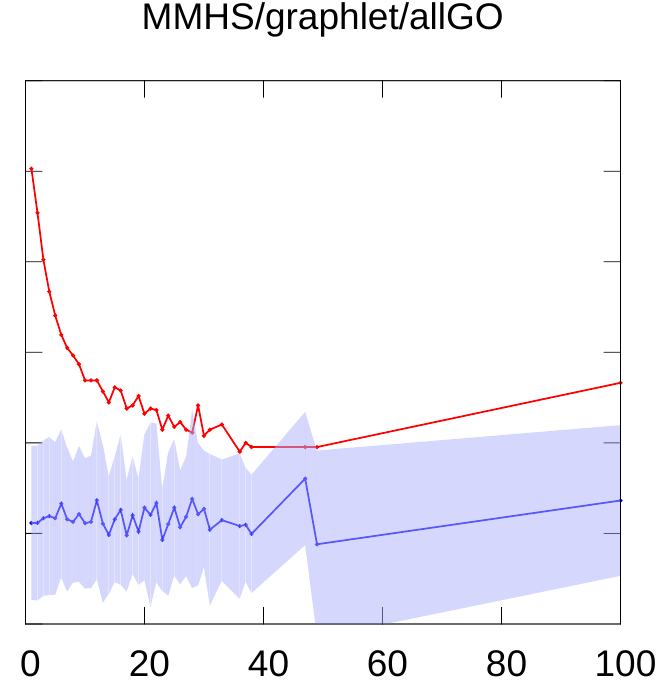}
\includegraphics[width=0.17\linewidth]{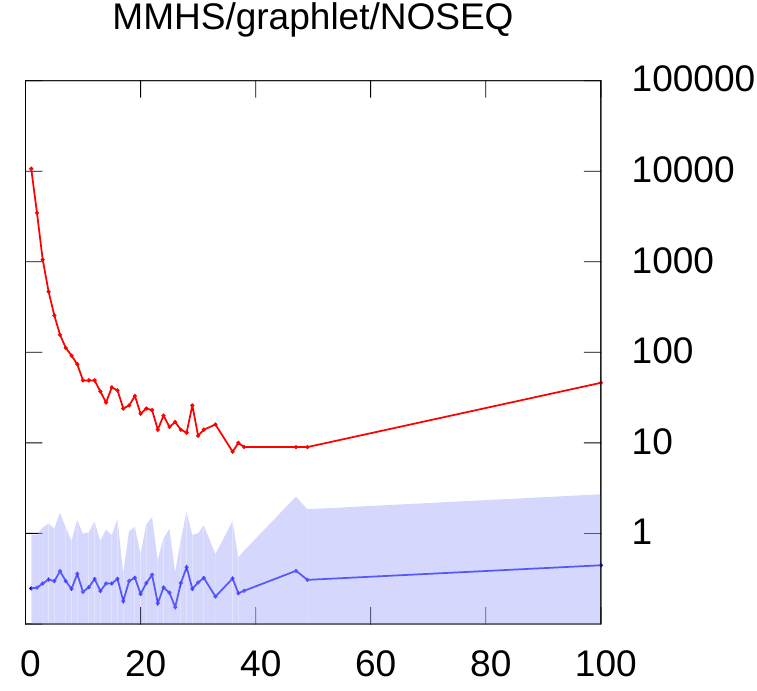}\\
\rotatebox{90}{$\;\;\;\;\;\;\;$SC-HS}
\includegraphics[width=0.155\linewidth]{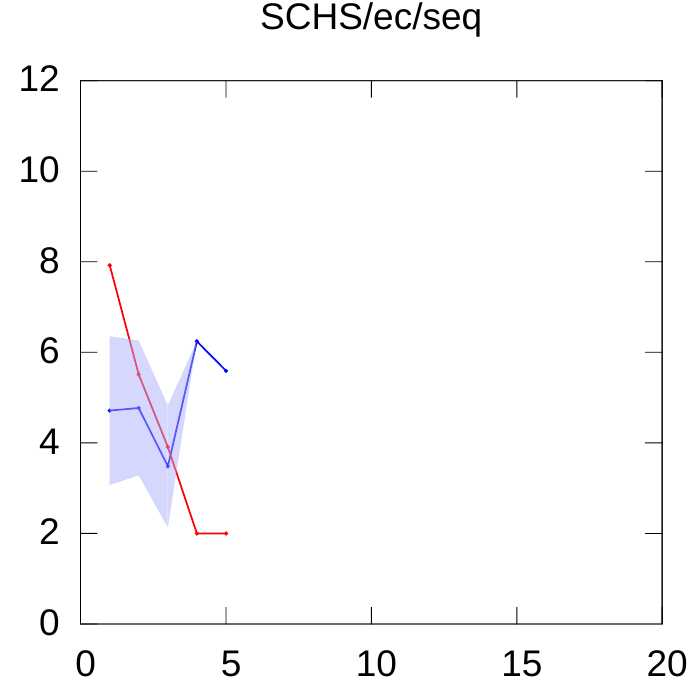}
\includegraphics[width=0.145\linewidth]{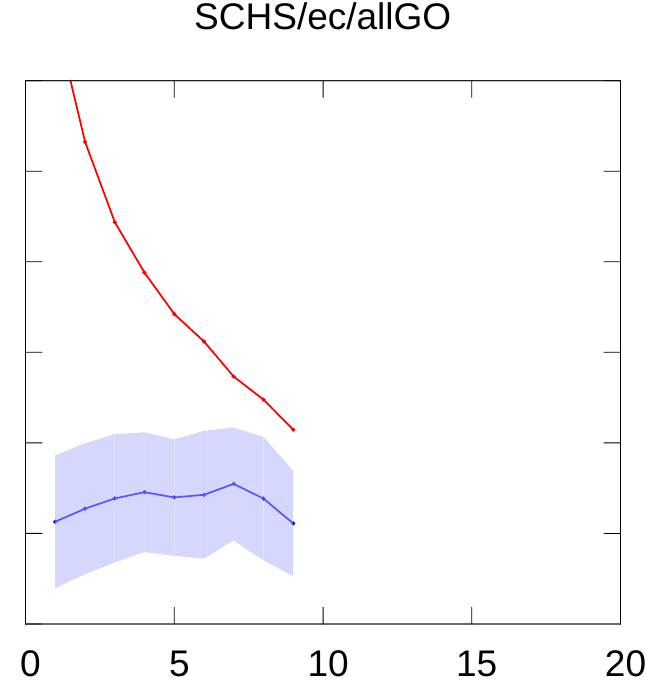}
\includegraphics[width=0.17\linewidth]{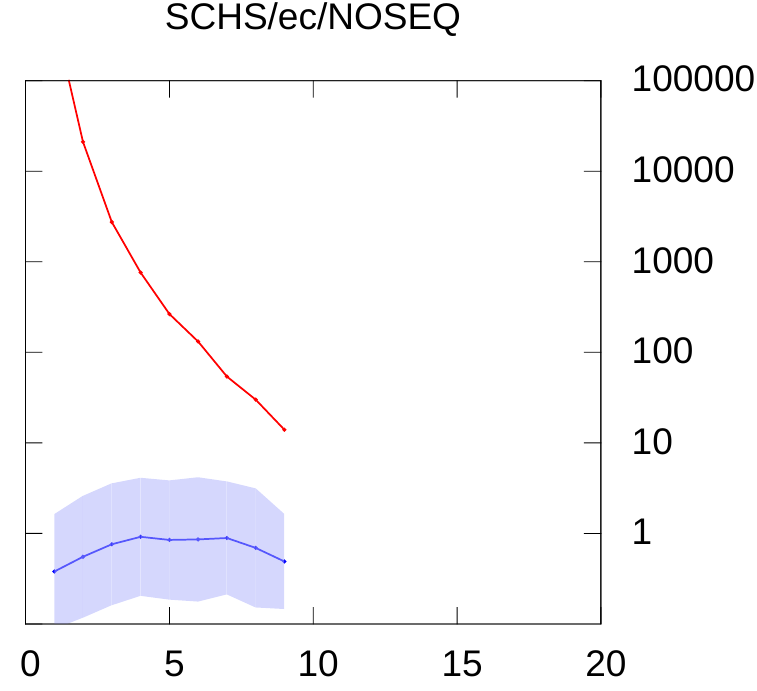}
\includegraphics[width=0.155\linewidth]{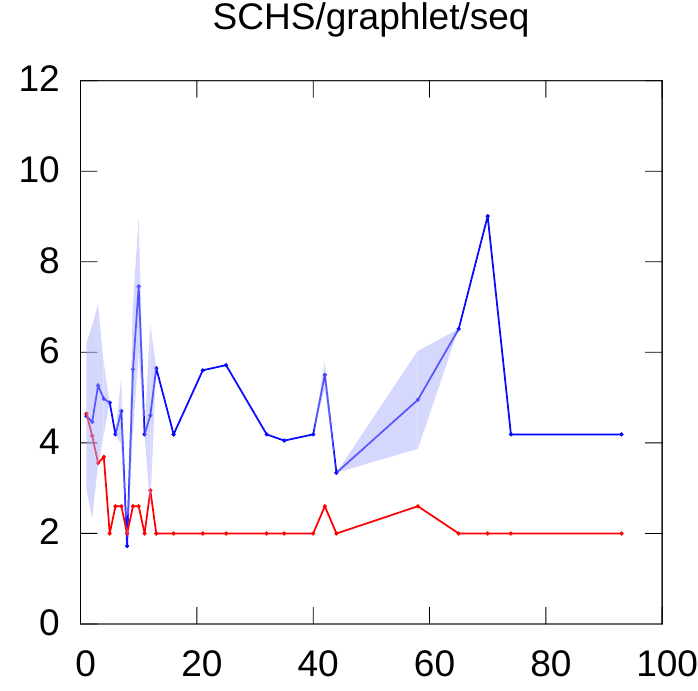}
\includegraphics[width=0.145\linewidth]{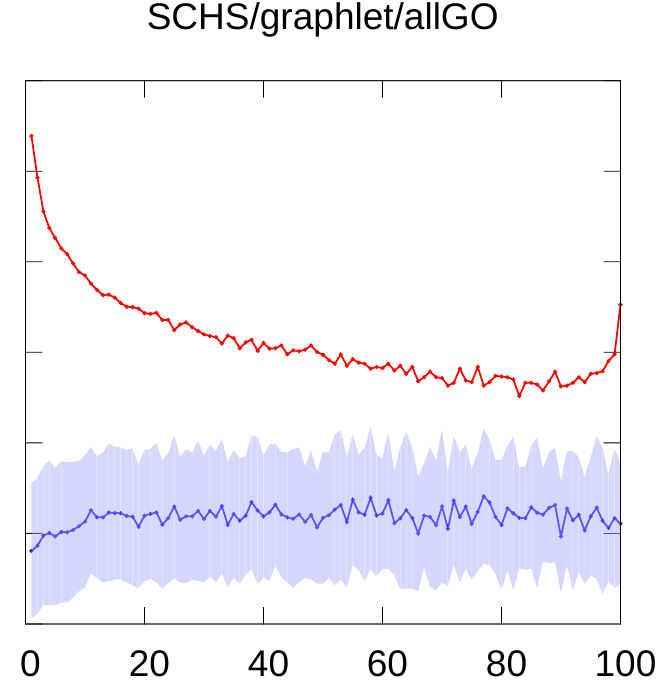}
\includegraphics[width=0.17\linewidth]{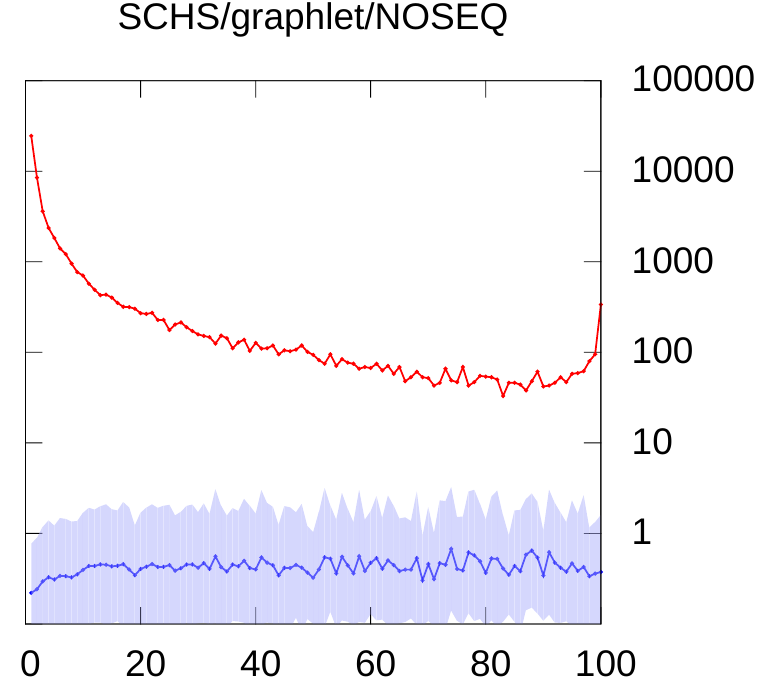}
\caption{\small
    {\bf Resnik-vs.NAF between mouse-human (top) and yeast-human (bottom) in the year 2010} Similar to Figure (main manuscript) \ref{fig:Resnik-vs-NAF+Pearsons} except using BioGRID 3.0.64 (released Apr. 23, 2010) and GO terms released the same month).
    }
\label{fig:Resnik-vs-NAF2010}
\end{figure*}

\ifanswers
\bibliography{wayne-new}
\end{document}
\fi

\end{document}